\documentclass[aps,prd,twocolumn,superscriptaddress,floatfix]{revtex4-2}

\usepackage[switch]{lineno}
\usepackage{graphicx}
\usepackage{enumitem}
\usepackage{comment}
\usepackage{siunitx}
\usepackage{amsfonts}
\usepackage{nicefrac}
\usepackage{lipsum}
\usepackage{multirow}
\usepackage{color}
\usepackage{xspace}
\usepackage{soul}
\usepackage{amsmath}
\usepackage{mathtools}
\usepackage{multirow}
\usepackage{listings} 
\usepackage{color}
\usepackage{import}
\usepackage{bm}
\usepackage{booktabs}
\usepackage{makecell}
\usepackage{hyperref} 
\newcommand{\ra}[1]{\renewcommand{\arraystretch}{#1}}

\makeatletter
\def\lst@makecaption{%
  \def\@captype{table}%
  \@makecaption
}
\makeatother

\definecolor{kw}{cmyk}{1, 0.50, 0, 0}

\begin{document}

\title{Anomaly preserving contrastive neural embeddings for end-to-end model-independent searches at~the~LHC} \author{Kyle Metzger}
\affiliation{ETH Z\"{u}rich, Z\"{u}rich, Switzerland}

\author{Lana Xu}
\affiliation{Massachusetts Institute of Technology, Cambridge, MA, USA}

\author{Mia Sodini}
\affiliation{Massachusetts Institute of Technology, Cambridge, MA, USA}

\author{Thea K. \AA rrestad}
\affiliation{ETH Z\"{u}rich, Z\"{u}rich, Switzerland}

\author{Katya Govorkova}
\affiliation{Massachusetts Institute of Technology, Cambridge, MA, USA}

\author{Gaia Grosso}
\affiliation{Massachusetts Institute of Technology, Cambridge, MA, USA}
\affiliation{NSF AI Institute for Artificial Intelligence and Fundamental Interactions, Cambridge, MA, USA}
\affiliation{School of Engineering and Applied Sciences, Harvard University, Cambridge, MA, USA}

\author{Philip Harris}
\affiliation{Massachusetts Institute of Technology, Cambridge, MA, USA}
\affiliation{NSF AI Institute for Artificial Intelligence and Fundamental Interactions, Cambridge, MA, USA}

\email{thea.aarrestad@cern.ch, ekaterina.govorkova@cern.ch, gaia.grosso@cern.ch, philip.coleman.harris@cern.ch}

\begin{abstract}
Anomaly detection — identifying deviations from Standard Model predictions — is a key challenge at the Large Hadron Collider due to the size and complexity of its datasets. This is typically addressed by transforming high-dimensional detector data into lower-dimensional, physically meaningful features. We tackle feature extraction for anomaly detection by learning powerful low-dimensional representations via contrastive neural embeddings. This approach preserves potential anomalies indicative of new physics and enables rare signal extraction using novel machine learning-based statistical methods for signal-independent hypothesis testing. We compare supervised and self-supervised contrastive learning methods, for both MLP- and Transformer-based neural embeddings, trained on the kinematic observables of physics objects in LHC collision events. The learned embeddings serve as input representations for signal-agnostic statistical detection methods in inclusive final states. We achieve significant improvement in discovery power for both rare new physics signals and rare Standard Model processes across diverse final states, demonstrating its applicability for efficiently searching for diverse signals simultaneously. We study the impact of architectural choices, contrastive loss formulations, supervision levels, and embedding dimensionality on anomaly detection performance. We show that the optimal representation for background classification does not always maximize sensitivity to new physics signals, revealing an inherent trade-off between background structure preservation and anomaly enhancement. We demonstrate that combining compression with domain knowledge for label encoding produces the most effective data representation for statistical discovery of anomalies.
\end{abstract}

\maketitle
\thispagestyle{plain}
\pagestyle{plain}

\section{Introduction}
The Large Hadron Collider (LHC) at CERN is the world’s largest particle accelerator, designed to explore the fundamental properties of matter by colliding protons and heavy ions at high energies. The LHC allows physicists to probe the deep structure of the universe, searching for new particles and phenomena that could extend our understanding of the Standard Model of particle physics. Anomaly detection in high-energy physics is crucial for identifying events that might indicate the presence of new unexpected phenomena beyond the ones described by the Standard Model. Given the immense volume of data generated by the LHC, advanced techniques for anomaly detection are essential for sifting through the data to find rare signals that could point to groundbreaking discoveries. Such methods must be highly sensitive and capable of distinguishing true anomalies from background noise, making them a powerful tool in the ongoing search for new fundamental particles and forces.

A crucial challenge for anomaly detection on LHC data is the step of \textit{feature learning}, constructing a low-dimensional set of informative variables from the high-dimensional set of observables that describes each collision event. Raw collision data at the LHC consists of millions of sensor measurements, which are processed using clustering algorithms to reconstruct individual particles. The data is then further reduced to physically meaningful summary statistics, such as particle energy, momentum, mass, and direction. These summary statistics serve as inputs to data analyses aimed at detecting deviations from expected backgrounds, which in turn could hint at new physics. The process of transforming high-dimensional detector data into compact, but informative representations is therefore well established in high-energy physics. Machine learning (ML) has demonstrated high effectiveness in extracting powerful representations from high-dimensional data~\cite{openai2024gpt4technicalreport}. Such neural embeddings can similarly be leveraged for data compression, mapping raw data to lower-dimensional spaces that retain essential features for downstream tasks~\cite{CGV-107}. In this work, we study whether neural embeddings, when trained appropriately, can outperform classical feature representations commonly used in LHC data analysis, particularly in the context of anomaly detection.

When solving supervised problems, the information loss due to feature reduction can be curated to minimally impact the specific objective of the training task. Conversely, for semi-supervised tasks, like anomaly detection at the LHC, the absence of a priori knowledge about the signal increases the risk of performing a disruptive feature reduction.

Contrastive learning, when used for feature reduction in the context of anomaly detection, offers a compelling approach to learn a meaningful form of data organization in an unsupervised or semi-supervised fashion.
In the context of anomaly detection, the goal of contrastive learning is to create a latent space where common data points are grouped closely together. Meanwhile, anomalies, which are points that deviate from the bulk of the data, are pushed further apart. By training the model to maximize the similarity between representations of normal data and minimize it for data that is distinct or unusual, the model learns to highlight the features that distinguish anomalies from the regular pattern. This process is particularly advantageous in high-dimensional and complex datasets, such as those encountered in high-energy physics or other scientific domains, where anomalies may be subtle or rare. Once the latent space is structured in this way, unsupervised clustering techniques can be employed to further isolate and identify potential anomalies. In this context, contrastive learning helps construct well-defined clusters in the latent space, easing the tasks of outliers or statistical anomaly detection.

This paper presents a comprehensive study focused on the development and evaluation of neural embeddings for anomaly detection, particularly in the context of high-energy physics. Section~\ref{sec:related_work} provides an overview of existing methods and approaches in anomaly detection and contrastive learning. In Section~\ref{sec:neural_embeddings}, we go into the details of how different models and loss functions are employed to pre-train embeddings that can effectively represent the underlying data structure. 
In Section \ref{sec:anomaly_detection} we apply these pre-trained embeddings to the task of identifying anomalies, with a focus on evaluating how well the learned latent spaces can separate anomalous events from the normal data distribution. In Section~\ref{sec:discussion}, we analyze the results, comparing the performance of different training strategies and architectures, and discussing the implications of our findings for future research in both machine learning and high-energy physics. Finally, Section~\ref{sec:conclusion} summarizes the key findings from our work and outlines potential directions for further investigation, emphasizing the importance of robust anomaly detection techniques in advancing our understanding of complex LHC datasets.

\section{Related Work}\label{sec:related_work}

Machine learning has long been an integral component of anomaly detection in high-energy physics. Traditionally, anomaly detection in this field has relied on the formulation of a theory model for the signal, used by physicists to define specific criteria or regions of interest within the data space (signal-dependent searches). The advent of machine learning has enabled the development of flexible methods, capable of learning optimal signal to background multi-dimensional discriminants directly from data and simulation. 
These approaches mainly include supervised learning techniques, where models are trained on labeled datasets to distinguish between normal and pre-specified anomalous events. 
These methods, while effective in targeted scenarios, are limited by their dependency on predefined signal models.

On the other hand, looking for new physics in the absence of a signal model (signal-agnostic searches) is extremely hard due to the lack of information on how to build relevant summary statistics. Following the steps of the D0 and CDF collaborations at Tevatron and the H1 collaboration at HERA~\cite{D0:2000dnz,D0:2001mmn, H1:2004rlm,CDF:2007iou,H1:2008aak, CDF:2008voc,D0:2011ccx}, both the ATLAS and CMS experiments at CERN have developed strategies to scan the experimental phase space searching for anomalies prior to the advent of deep learning~\cite{ATLAS:2014sxa,CMS:2020zjg}. In recent years, semi-supervised, weakly-supervised, and unsupervised learning methods have emerged as powerful tools to enhance these approaches.  
Deep learning techniques have shown great promise in identifying anomalies in complex, high-dimensional datasets and have been recently deployed in both the CMS and ATLAS experiments~\cite{ATLAS:2020iwa,CMS:2024lwn, ATLAS:2023azi,ATLAS:2023ixc,CMS:2024lwn,ATLAS:2025obc}. 
Techniques such as autoencoders, generative adversarial networks (GANs), and contrastive learning have been explored for their ability to learn meaningful representations of data and detect deviations from expected patterns, without the need of strict a priori assumptions on the nature of the anomalous signal (see~\cite{Belis:2023mqs} for a review on the topic). Despite the recent advancements, there remain significant challenges. Particularly, the interpretability of these ML methods, as well as their practical application in the real-world context of LHC experiments, that requires the handling of vast and complex datasets. Challenges in this field include: (1) mitigating sensitivity degradation caused by the curse of dimensionality~\cite{backtoroots,Freytsis:2023cjr}, (2) speeding up training schemes to handle extremely large sample sizes~\cite{Letizia:2022xbe,Grosso:2024nho}, and (3) uncertainty quantification~\cite{dAgnolo:2021aun,Khot:2025kqg}.

In this work we tackle the first of the aforementioned challenges by exploring neural embeddings for anomaly-preserving data compression using contrastive learning techniques. Several recent results in particle physics have highlighted the success of contrastive learning and neural embedding strategies to train models capable of performing various tasks~\cite{Park:2022zov,Harris:2024sra,Dillon:2021gag} and foundation models providing data representations suitable for many tasks at one~\cite{Mikuni:2024qsr,Leigh:2024ked,Birk:2024knn, Wildridge:2024yeg,Bardhan:2025icr}. Focusing on anomaly detection applications, Refs.~\cite{Golling:2024abg,Li:2024htp,PhysRevD.106.056005} showed that pre-trained foundation models can enhance the performance of weakly-supervised classifiers for anomaly detection. In Ref.~\cite{Favaro:2023xdl}, physics-informed self-supervised strategies are used to defined a contrastive space for anomaly detection. Whereas, in Ref.~\cite{Dillon:2023zac}, self-supervised strategies are explored in the context of real-time anomaly detection in jets as a way to perform feature extraction for an autoencoder. 
In this work, we take a further step by exploring the application of contrastive learning methods to a different data representation, namely an event-level representation of proton-proton collisions at the LHC, where inclusive final states are characterized by the 3-momenta of particles reconstructed from the collision products. Furthermore, we compare two different contrastive approaches, supervised and self-supervised, and we demonstrate that supervised contrastive learning yields significantly greater performance gains compared to self-supervised approaches, as it leverages available domain knowledge to guide the organization of the representation space. To the best of our knowledge, this study presents the first application of supervised contrastive learning to high energy physics data, and provides the first evidence that the availability of labeled background data can effectively steer discovery within unlabeled datasets.
Moreover, whereas weakly supervised methods rely on defining a signal sideband and thus incorporate some signal assumptions, we propose an end-to-end, fully signal-agnostic technique applicable to generic final states.

\section{Contrastive-based neural embeddings}\label{sec:neural_embeddings}

Meaningful data representations play an essential part in how performant the ML algorithm is \cite{representations}. Therefore, a lot of emphasis has been put on processing data before giving it to an ML model. 
Representation learning describes the extraction of features using deep learning methods. A neural network $\phi_{\theta}$ parameterized by a set of trainable parameters $\theta$ maps the events from an original high-dimensional space $\mathcal{X}$ to a lower-dimensional latent space $\mathcal{Y}$. This mapping is formalized as: 
\begin{equation}
\phi_{\theta}:\left(\mathcal{X}, d_{\mathcal{X}}\right) \xrightarrow{\phi} \left(\mathcal{Y}, d_{\mathcal{Y}}\right),
\end{equation}
where $d_{\mathcal{X}}$ and $d_{\mathcal{Y}}$ are metrics in the original space and the latent space respectively. 
The learned representation can then be used for downstream tasks, such as classification. While representation learning is usually accompanied by dimensionality reduction, the defining aspect is the extraction of features that best describe the input data with regards to the downstream task. This step is often referred to as \textit{feature learning}. 
When the downstream task is signal-agnostic 
anomaly detection, the optimal representation for the signal cannot be known a priori and, ultimately, it would depend on the composition of the real data, namely which signal, if any, is present. 




The approaches to learn neural embeddings considered in this work are based on contrastive learning, originally designed to solve problems of computer vision. The methods are constructed according to the Siamese principle, where the input to a weight-sharing model consists of two (or more) views of the same data point, or in this case physics events~\cite{simsiam}. The outputs are subsequently compared by a measure of similarity, such that similar instances are pulled together in the latent space, whereas dissimilar ones are pushed apart.  

The real LHC collision dataset is several orders of magnitude larger than its simulated counterpart. As a result, it may be valuable to pre-train neural embeddings in a self-supervised manner using unlabeled data. 
In a self-supervised setting the views of an event are constructed using \textit{augmentations}, namely applying to the event representation a set of transformations that leave the physics nature of the event invariant. A collision event is augmented twice and the resulting views are grouped together in the latent space by the loss objective. The rest of the events in the batch are seen as ``negative" examples, and therefore they are pushed apart from the pair of augmented views.
As a consequence, even when they represent the same hard core physics, different collision events are pushed apart. This is potentially a problem, as we would like to build a neural embedding where data are organized according to ``concepts", namely the hard core physics underlying the events.
One way to circumvent this problem is adopting the supervised extension of the self-supervised approach outlined above, where 
similar instances (e.g. ``positive" examples) are defined using their physics process labels. This variant is referred in the literature as \textit{supervised contrastive learning}~\cite{supcon}. It should be noticed that this approach can only be applied if the physics labels are available, namely with simulated datasets, thus preventing the use of large size real LHC data in the training stage.
Given the agnostic nature of the downstream task considered in this work, our trainings in this setup will rely only on simulated data of the known background physics processes, endowed with labels. 

The dynamics of contrastive learning rely on the interplay of attractive forces and repulsive forces, subject to the constraints set by a form of regularization. We consider two loss functions serving this purpose: the SimCLR loss~\cite{simclr} and the VICReg loss~\cite{vicreg}. 
The two loss functions mainly differ in how they approach the issue of mode collapse, namely when the inputs are mapped to a constant. 
SimCLR uses a contrastive loss based on the cosine similarity metric ${\rm sim}(x_i, x_j) = \frac{x_i^T x_j}{\|x_i\| \|x_j\|}$, resulting in
\begin{equation}\label{ex:simclr-loss}
    l_{\rm SimCLR}(x_i, x_j)=-\log \frac{e^{{\rm sim}(\phi_\theta\left(x_i\right), \phi_\theta\left(x_j\right))/\tau}}{\sum_{k=1}^N e^{{\rm sim}(\phi_\theta\left(x_i\right), \phi_\theta\left(x_k\right))/\tau}}.
\end{equation}
 The loss temperature $\tau$ is a tunable hyper-parameter for regularization. In Eq.~\ref{ex:simclr-loss}, the sum at the denominator includes all possible pairs of the $x_i$ example with the $N$ examples in the batch. The presence of negative pairs act against mode collapse, eventually preventing it if the size of the batch is large enough. 
On the other hand, the similarity metric in VICReg has three terms
\begin{equation}\label{ex:vicreg-loss}
\begin{aligned}
    l_{\rm VICReg}(x_i, x_j) =& \lambda \,s(\phi_\theta\left(x_i\right), \phi_\theta\left(x_j\right))\\
                 & + \mu\,[v(\phi_\theta\left(x_i\right))+v(\phi_\theta\left(x_j\right))] \\
                 & + \nu\,[c(\phi_\theta\left(x_i\right))+c(\phi_\theta\left(x_j\right))] .\\
\end{aligned}
\end{equation}
The first term, $s$, is a mean squared error measure that encourages invariance between positive pairs; whereas the variance term $v$ and covariance term $c$ act as regularizers against mode collapse. The variance term holds the standard deviation along the batch size for every feature above a certain threshold; the covariance term decorrelates the embedding variables between pairs along the batch size. The hyper-parameters $\lambda,\, \mu$ and $\nu$ determine the relative contribution of each term to the overall loss function and are optimized via grid search.
In both cases the similarity metric is averaged across all positive pairs $P(i)$ of the $i$-th example in the batch, and summed over all of the $N$ training examples in the batch:
\begin{equation}
     L_{\rm CL}[\phi_{\theta}] = \sum_{i=1}^{N}\frac{1}{\lvert P(i)\rvert}\sum_{x\in P(i)}^{}l_{\rm CL}(x_i, x)
\end{equation}
with $\text{CL}\in[\text{SimCLR, VICReg}]$.
In the self-supervised case, each example $x_i$ in the mini-batch has exactly one positive pair, denoted $x_{j(i)}$, which corresponds to the other augmented view generated from the same underlying event. Accordingly, the set of positive pairs is defined as \[P(i) \coloneqq \{\,x_{j(i)}\,\}.\] In the supervised extension, all instances $x$ within the mini-batch with the same class labels $c$ as $x_i$ are considered positive pairs to $x_i$, such that \[P(i) \coloneqq \{\,x\mid c_x=c_{x_i}\,\}.\]
\\

\section{Anomaly detection on neural embeddings}\label{sec:anomaly_detection}
In this work, we consider the detection of anomalous signal events as the downstream task of the feature learning step. A sample of experimental data is passed through the neural embedding map and its latent representation is analyzed to detect and statistically quantify data departures from their expected behavior. 
We assume that the expected behavior is faithfully represented by means of a mixture of simulated background processes ($\rm B$) depicting the Standard Model of particle physics. In realistic scenarios, the model representing the background processes is not known exactly, but rather affected by systematic uncertainties that can impact on the robustness of the anomaly detection task. While a way to incorporate systematic uncertainties for the method considered in this work has been proposed~\cite{dAgnolo:2021aun}, in this work we don't address this problem and focus on the impact of feature learning on the anomaly detection task. We leave the study of systematic uncertainties to future work. We assume that the signal ($\rm S$), if present, manifests itself as a mild over-density in the data density distribution. The exact location and shape of the signal contribution is a priori unknown, giving rise to the challenging problem of engineering the right statistical test for novelty discovery. In a $d$-dimensional latent space, the density of the embedded data $y$ under their true ($\rm T$) generative model can be modeled as the sum of the signal and background components
\begin{equation}
    n(y|{\rm T}) = n(y|{\rm B}) + n(y|{\rm S}).
\end{equation}
The aim of the anomaly detection task boils down to recognizing the presence of signal events and estimating their significance. In other words, determining $n(y|\rm{S})$ to maximize the likelihood of the observed data.

To address this problem, we consider the signal-agnostic Neyman-Pearson goodness-of-fit test performed by the NPLM algorithm~\cite{Letizia:2022xbe,Grosso:2023scl}.

New Physics Learning Machine (NPLM) is a semi-supervised approach to agnostic signal detection that aims to quantify the statistical significance of an observed deviation in the data~\cite{DAgnolo:2018cun}.
Inspired by the Neyman-Pearson approach to hypothesis testing based on the log-likelihood-ratio~\cite{Neyman:1933wgr}, NPLM compares the background hypothesis, $\rm B$, and an unknown alternative $\rm H_{\bold{w}}$, which is learned from the data $\cal D$ in a machine learning fashion.
A model $f_{\bold{w}}(y)$ with trainable parameters $\mathbf w$ is defined on the space of the data $y$ to parametrize  the density distribution of the data in the family of alternatives
\begin{equation}\label{eq:n_w}
    n(y|{\rm H_{\mathbf w}})= n(y|{\rm B}) e^{f_{\mathbf w}(y)}\,.
\end{equation}

Since the total number of events in the experimental dataset from the LHC is a Poissonian random variable whose expectation depends on the theory model, the Neyman-Pearson test statistic is computed as the extended log-likelihood-ratio, which in terms of $f_{\mathbf w}$ is written as
\begin{equation}\label{eq:t_np2}
t({\cal D})=2\max\limits_{\mathbf w}\left[ {\rm N(B)}-{\rm{N}}({\rm H_{\mathbf w}}) +\sum\limits_{x\in{\cal D}} f_{\mathbf w}(y)
\right]\,,
\end{equation}
with $\rm N(B)$ and ${\rm{N}}({\rm H_{\mathbf w}})$ the expectation values of the total number of events in $\cal D$ under the reference and the alternative hypotheses respectively.
The maximization problem described by Eq.~\ref{eq:t_np2} can be solved by a machine learning task that learns the optimal values of the trainable parameters of $f$, $\bold{\widehat w}$, to approximate the log-density-ratio of the two samples
\begin{equation}
    f_\bold{\widehat w}(y) = \log\frac{n(y|{\rm H_\bold{\widehat w}})}{n(y|{\rm R})}
    \approx \log\frac{n(y|{\rm T})}{n(y|{\rm R})}
\end{equation}

In this work we adopt the fast version of the NPLM algorithm originally introduced in~\cite{Letizia:2022xbe} which relies on the \textsc{Falkon} package~\cite{falkon}. The latter is based on kernel methods whose coefficients $\mathbf w$ are trained with an L2 regularized weighted binary cross-entropy loss function
\begin{equation}\label{eq:loss_bce}
\begin{aligned}
L_{\rm FLK}[f_{\mathbf w}]= 
&\sum\limits_{y\in {\cal R}}w_{\cal R}\log(1+e^{f_{\mathbf w}(y)}) \\
&- \sum\limits_{y\in{\cal D}} \log(1+e^{-f_{\mathbf w}(y)}) \\
& + \lambda\sum ||{\mathbf w}||^2, \\
\end{aligned}
\end{equation}
where the weights $w_{\cal R}$ is introduced to match the size of sample of reference $\cal R$ with the expected experimental size in absence of signal, $\rm N(B)$.
The test statistic is then computed as in Eq.~\ref{eq:t_np2}, using the learned model $f_\bold{\widehat w}$.

As for any test statistic, the value of the test obtained for $\cal D$ has to be compared to the distribution of the test under the background hypothesis, $p(t|{\rm R})$, and the $p$-value is finally used as a metric of significance.
In the results presented in Section~\ref{sec:experiments}, $p(t|{\rm R})$ is estimated both empirically, using 1000 pseudo-experiments, and analytically relying on the asymptotic $\chi^2$ behavior of the test.
To help the interpretation, $p$-values resulting from our experiments will be reported in their standardized version of $Z$-scores, defined as the quantile $\Phi^{-1}$ of the normal distribution at the $p$ complement to $1$:
\begin{equation}
\label{eq:zscore}
    Z_{p} = \Phi^{-1}(1-p).
\end{equation}\\

\section{Neural embeddings on LHC data}\label{sec:experiments}
In this Section we apply the representation learning approaches outlined in  Section~\ref{sec:neural_embeddings} to a synthetic dataset representing realistic LHC collision events.
We train various neural embeddings with different learning schemes in a signal agnostic way, namely only using background data. We evaluate the quality of their representation as a backbone for both the classification of background physics processes and for the detection of anomalies. For the first task, we use a simple one layer linear evaluation as a downstream task. For the second problem, we consider the anomaly detection approach described in~\ref{sec:anomaly_detection}. For both tasks we compare the performances among different learning strategies and rates of compression.
 
\subsection{Dataset}
We evaluate our models on the \textit{ADC2021} dataset, specifically designed for anomaly detection in high energy particle physics~\cite{delphesdataset}. This dataset emulates realistic proton-proton collisions at the LHC with a center-of-mass energy of 13 TeV, applying an event selection criterion that mimics an online filter, requiring at least one electron ($e$) or muon ($\mu$) with transverse momentum $p_T > 23$ GeV. Standard Model background processes are generated using \textsc{Pythia}~\cite{Sjostrand:2014zea}, with their detector responses simulated via \textsc{Delphes}~\cite{deFavereau:2013fsa}. 
For each collision event at most 19 objects are reconstructed: the missing transverse energy (MET), the four most energetic electrons and muons and the ten most energetic jets. For Each object the kinematics variables $(p_T, \,\eta,\,\phi)$ are provided. Data are provided in a $19\times3$ tabular format where zero padding is used for events with less than 19 reconstructed objects.
The dataset is partitioned into training and validation sets using 5-fold cross-validation. During neural embedding training, each fold serves as the validation set once, while the remaining four folds are combined to form the training set. Additionally, 20\% of the Standard Model background processes are withheld from the embedding training and reserved as the test dataset for anomaly detection. The Standard Model events are labeled according to the hard core physics process generating them. The dataset contains four Standard Model (background) classes in different proportions representing the expected relative abundance: W boson production (59.2\%), Z boson production (6.7\%), $t\bar t$ production (0.3\%) and QCD multijet production (33.8\%). Alongside the Standard Model dataset, four new physics scenarios are also simulated and used for evaluating the anomaly detection performance. These include a neutral scalar boson ($A$) with a mass of 50 GeV decaying to two off-shell $Z$ bosons that further decay to leptons ($A\rightarrow 4l$), a leptoquark (LQ) with a mass of 80 GeV decaying into a $b$-quark and a $\tau$ lepton, a scalar boson $h^0$ with a mass of 60 GeV decaying to two $\tau$ leptons ($h^0\rightarrow \tau\tau$), and a charged scalar boson $h^\pm$ with a mass of 60 GeV decaying to a $\tau$ lepton and a neutrino ($h^\pm\rightarrow\tau\nu$). At test time, these are injected into the Standard Model test dataset such that they make up a fraction ranging from 0.1-1\% of the total dataset size.

\subsection{Embedding models}
We trained various contrastive based neural embeddings, varying one of the following aspects at a time: the model architecture (MLP- or Transformer-based), the loss function (SimCLR or VICReg), the training scheme (supervised or self-supervised), and the input data composition (balanced background classes or a realistic background ``cocktail").

\begin{figure}[htbp]
    \centering
    \includegraphics[width=.8\linewidth]{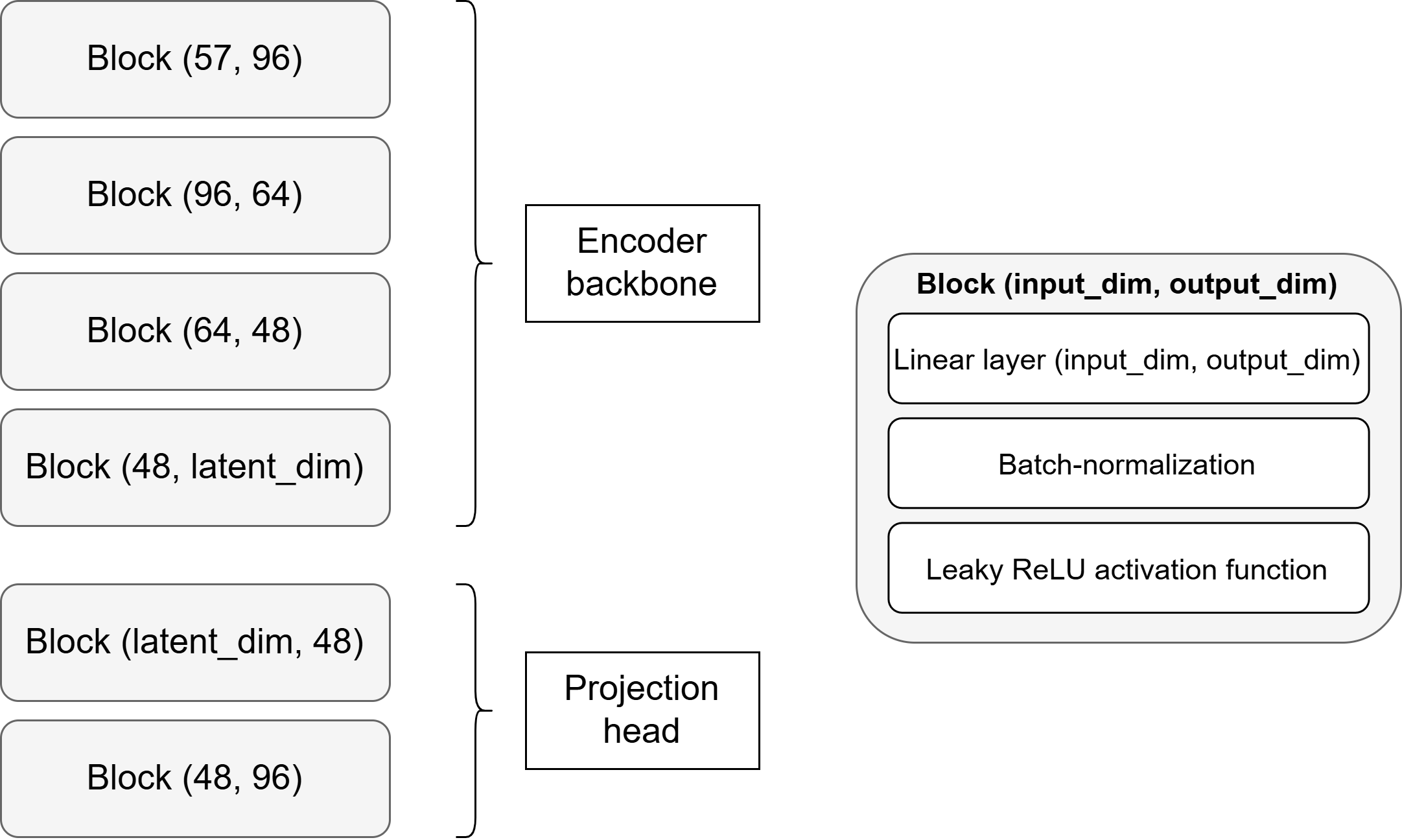}
    \caption{MLP-based neural network architecture used for training the supervised and self-supervised embeddings with SimCLR. Illustration made with Ref.~\cite{drawio}.}
    \label{fig:scheme-model-mlp-simclr}
\end{figure}

With SimCLR loss we trained both MLP- and transformer-based architectures and a realistic composition of input background data.
The network architecture used for both the supervised and self-supervised MLP-based models is depicted in Figure~\ref{fig:scheme-model-mlp-simclr}.
On the right side of the figure, we show the main building block of the architecture, namely a batch-normalized linear layer activated by Leaky ReLU. On the left side we represent the backbone encoder and the projection head used for training the neural embeddings. The former is composed of four blocks, whereas the latter is composed of two blocks. As per~\cite{simclr}, adding a projection head during training leads to improved embeddings. For inference (i.e., background classification and anomaly detection), the projection head is dropped.

In contrast, the transformer-based supervised SimCLR implementation employs a standard transformer encoder \cite{transformer, pytorch_transformer} (see dimensions in Table~\ref{tab:transformer-simclr}), and its projection head is based on the DINO head \cite{dino_github} with parameters detailed in Table~\ref{tab:dinohead}. The 19 particles in the source representation are input as 19 different tokens with the kinematic variables $(p_T,\,\eta,\, \phi)$ as features.
In this architecture, an additional classification token is used to aggregate information from all the tokens before feeding it to the projection head. The training hyperparameters are summarized in Table~\ref{tab:mlp-simclr}. The learning rate follows a linear warm up for the first ten epochs and a subsequent cosine decay to 1/1000 of the base learning rate. The MLP-based training uses the LARS optimizer \cite{lars}, while the Transformer-based training utilizes the Adam optimizer \cite{adamw} for applying the gradient updates.
\begin{table*}[htbp]
\small
\ra{1.2}
\centering
\makebox[\textwidth]{
\begin{tabular}{l|llll}
\toprule
 Method & Batch-size & Epochs & Learning rate & Weight decay  \\
\midrule
self-supervised SimCLR MLP & 1'024 & 50 & 0.8 & 1e-6\\
supervised SimCLR MLP & 1'024 & 50 & 1.2 & 1e-6\\
supervised SimCLR Transformer & 1'024 & 50 & 0.001 & 1e-6\\
supervised VICReg Transformer & 64 & 50 & 0.001 & 1e-2\\
\bottomrule
\end{tabular}
}
\caption{Training hyperparameters for the contrastive representation.}\label{tab:mlp-simclr}
\end{table*}

The SimCLR loss objective allows tuning for using the temperature parameter $\tau$ in Eq.~\ref{ex:simclr-loss}. While in the supervised implementation  a temperature parameter of $\tau=0.1$ works well for both architectures, in the self-supervised implementation the optimal temperature depends of the latent space dimension D. The temperatures used in this case are reported in Table~\ref{tab:simclr-temperatures}. It should be noted that an optimal choice of hyper-parameters is not generally possible for anomaly detection tasks, since the target is not known a priori. In this work we optimize the neural embedding modeling using background classification accuracy as a figure of merit (more details in Section~\ref{subsec:lin}). While this is not guaranteed to be optimal in the broader sense of anomaly detection we consider the background classification task to be closely related to the anomaly detection task and therefore provide a set of near to optimal hyper-parameters.
\begin{table}[htbp]
\centering
\makebox[0.5\textwidth]{
\begin{tabular}{l|llllll}
\toprule
\textbf{D} & 4 & 6 & 8 & 16 & 32 & 64 \\
$\boldsymbol{\tau}$ & 0.9 & 0.8 & 0.6 & 0.5 & 0.4 & 0.3 \\
\bottomrule
\end{tabular}
}
\caption{Loss temperatures ($\boldsymbol{\tau}$) and latent space dimension (\textbf{D}) for different models trained with self-supervised SimCLR.}\label{tab:simclr-temperatures}
\end{table}

With VICReg loss, we focused on transformer-based architectures \cite{transformer}. Each transformer block is made of MLPs with 2 layers and 16 expansions, and attention layers with number of heads ($h$) that varies with the latent space dimension D (see Table~\ref{tab:vicreg-heads}). The model consists of 4 transformer blocks stacked together. The batch size is 64, and the learning rate follows a CosineAnnealing schedule with initial value of 0.001 and weight decay of 0.01 (values reported in Table~\ref{tab:mlp-simclr}). The loss function used for the training is a combination of the VICReg loss in Eq.~\ref{ex:vicreg-loss} and a cross entropy (CE) loss, and the relative contribution of the two terms is regulated by a scale parameter $\alpha$:
\begin{equation}
    L = \alpha L_{\rm VICReg} + (1-\alpha) L_{\rm CE}.
\end{equation}
The scale parameter $\alpha$, as well as the batch size, number of layers, expansions, learning rate and weight decay are selected after a grid search is performed to optimize the accuracy of the background classification obtained by the model.

\begin{table}[htbp]
\centering
\makebox[0.5\textwidth]{
\begin{tabular}{l|llllll}
\toprule
\textbf{D} & 2 & 4 & 8 & 16 & 24 & 32  \\
$\boldsymbol{h}$ & 2 & 4 & 4 & 8 & 8 & 16 \\
\bottomrule
\end{tabular}
}
\caption{Number of heads ($\boldsymbol{h}$) and latent space dimension (\textbf{D}) for different models trained with supervised VICReg.}\label{tab:vicreg-heads}
\end{table}

\begin{table*}[htbp]
\small
\centering
\makebox[\textwidth]{
\begin{tabular}{lllllllll}
\toprule
\makecell{Model \\dimension} & \# attention heads & \makecell{Feed-forward \\dimension}  & \# layers & \makecell{positional \\encoding} & masking & \makecell{aggregate \\strategy} & dropout \\
\midrule
64 & 8 & 256 & 4 & True & False & CLS-token & 0.025\\
\bottomrule
\end{tabular}
}
\caption{Model parameters for the vanilla Transformer encoder \cite{transformer, pytorch_transformer} used as the backbone in the supervised SimCLR training.}
\label{tab:transformer-simclr}
\end{table*}

\begin{table*}[htbp]
\small
\centering
\makebox[\textwidth]{
\begin{tabular}{llllll}
\toprule
\makecell{Output \\dimension} & \makecell{hidden \\dimension} & \makecell{bottleneck \\dimension}  & \# layers & use batchnorm & \makecell{normalize \\last layer}\\
\midrule
64 & 256 & 64 & 3 & False & False\\
\bottomrule
\end{tabular}
}
\caption{Model parameters for the DINO head \cite{dino_github} used as the projection head.}
\label{tab:dinohead}
\end{table*}


\paragraph{Augmentations for self-supervised contrastive learning}
In order to achieve meaningful feature compression with self-supervised learning a suitable augmentation pipeline is crucial. We explored domain-specific augmentations for particle physics, where the underlying physics should remain unchanged under each transformation. Naive masking, where random features are zero-masked with probability \( p \). 
Gaussian resampling perturbed features with a normal distribution. A variant of Gaussian resampling involving only the transverse momentum (\( p_T \)). Rotation around the beamline randomly rotated events by \( \delta\phi \in [0, 2\pi) \), enforcing rotational invariance. Detector cropping retained only constituents within a radial cutoff ( $R^* =3$ was found to be the value performing the best on the background classification task used for evaluation). Despite these well-motivated variations, naive masking with probability \( p=0.5 \) consistently yielded the best representations and was adopted as the standard augmentation.
This augmentation strategy is inspired by the masking techniques used in language models training and can be viewed as a drop out technique used to increase the model's robustness to corrupted inputs. 
We define positive pairs as the original collision event and its augmented views 
 while the rest of the events within a batch are treated as negative pairs.
It's worth noticing that the adopted masking strategy modifies the overall kinematics of the event producing a completely different collision outcome as a view. The latter could be non physical or populate a region of the phase space that is not allowed by the original physics process. The augmented view could also resemble events coming from a different background process or an anomaly, with a potentially negative impact on the model performances. 

In addition to the augmentations defined for our experiments, we consider the two sets of physics-inspired and anomaly-inspired augmentations introduced in~\cite{Dillon:2023zac} (and defined in \url{https://github.com/bmdillon/AnomalyCLR/blob/main/EventLevelAnomalyAugmentations.py}). As in~\cite{Dillon:2023zac}, we perform two sets of experiments, the first one using only the physics-inspired augmentations and in the second one combining physics inspired and anomaly-inspired augmentations.
The effect of self-supervised approaches on both background classification and anomaly detection performance is discussed in Section~\ref{subsec:ad}.

\subsection{Baseline comparison of neural embeddings}\label{subsec:lin}
\begin{figure}[htbp]
    \centering
    \includegraphics[width=0.99\linewidth]{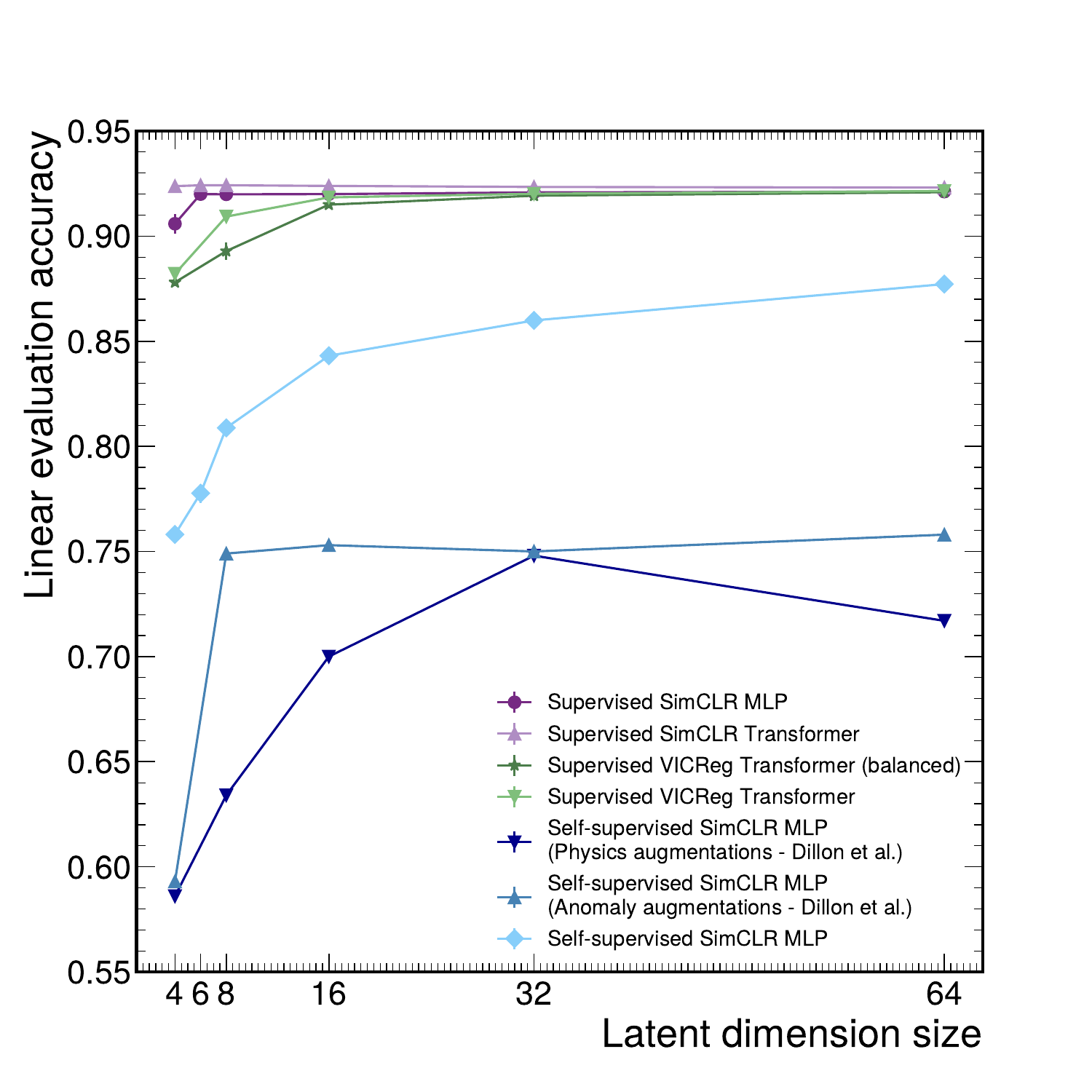}
    \caption{Linear evaluation accuracy on background classes for the MLP- and Transformer-based models trained supervised with a SimCLR loss (dark and light purple, respectively), the Transformer-based model trained supervised with a VICReg loss using either balanced background classes (dark green) or background classes weighted to match the composition expected in data (light green), the MLP-based model trained self-supervised with a SimCLR loss with 50\% random masking (light blue), physics-inspired augmentations from~\cite{Dillon:2021gag} (medium blue), or physics- and anomaly-inspired augmentations from~\cite{Dillon:2021gag} (dark blue). }
    \label{fig:lin_eval}
\end{figure}

After pre-training the different neural encoders, the learned latent space is used as an input to a simple linear evaluation classifier to determine its capability to retain and encode sufficient information for distinguishing between background classes. The linear evaluation is performed using a single linear layer on top of the neural embedding. The layer is trained using a supervised cross-entropy objective for classifying the Standard Model background events. The weights of the neural encoders are frozen during linear evaluation. We compute the standard accuracy as the mean across the 5-fold evaluation datasets, along with its uncertainty (i.e., the standard error of the mean). We use this quantity as a figure of merit to perform a grid search in the space of hyper-parameters of the considered models.. The values obtained with the selected hyper-parameters for different sizes of the latent dimension are presented in Figure~\ref{fig:lin_eval}, showcasing the effectiveness of the pre-trained latent spaces in the background classification task.
As expected, exploiting background supervision in the contrastive learning approach helps reach a high level of accuracy in the background classification task.
Moreover, we observe that higher dimensional latent spaces reach higher levels of accuracy in the classification task.
While these results give us an insight on which encoder better captures the discriminative features for an efficient background classification, this doesn't guarantee that the relevant features for detecting anomalies are preserved.

\subsection{Anomaly detection on neural embeddings}\label{subsec:ad}
To assess the power of different neural encoders to retain critical information for anomaly detection we apply the NPLM test for anomaly detection outlined in Section~\ref{sec:anomaly_detection}. While fine-tuning the embedding at the stage of testing would be possible, in this work we freeze the embedding and run the NPLM test as a downstream task. 
To test the sensitivity to previously unseen signals we consider the four signal benchmarks provided in the data challenge.
To compute the NPLM test we define a reference sample $\cal R$ composed of $1\,000\,000$ background data points, and a data sample $\cal D$ with number of expected events in the background hypothesis $\rm N(B)=100\,000$. To inspect the performances for progressively harder tasks we consider three scenarios of signal injection, $\rm N(S)=100,\,500,\,1000$, corresponding to signal fractions of $0.1, \,0.5$ and $\,1\%$.
The implementation of the NPLM test adopted in this work relies on a kernel method, for which the number of Gaussian kernels ($M$), the kernels' width ($\sigma$) and the L2 regularization coefficient ($\lambda$) are hyper-parameters to be specified.
Following the heuristic in Ref.~\cite{Letizia:2022xbe}, we take a number of kernels equal to square root of the training set size, $M=1\,000$; a value of the kernels' width equal to the $90\%$ quantile of the pair-wise distribution of the Euclidean distance between reference data points; and a small value of the L2 coefficient that still allows for a stable algorithm convergence, $\lambda=10^{-6}$. While $M$ and $\lambda$ are fixed for all experiments, the value of $\sigma$ varies with the embedding, since different representations are characterized by different typical distance scales. The only exception to the hyper-parameter setup just described is made for the 6D VAE baseline that will be introduced in the following. In this case we set $M=10\,000$ to obtained a better agreement of the test statistic distribution in absence of signal with the asymptotic $\chi^2$ used as a figure of merit for hyper-parameter selection in setting up the NPLM algorithm (see~\cite{Letizia:2022xbe} for details).
\begin{figure*}[htbp]
    \centering
    \includegraphics[width=0.45\linewidth]{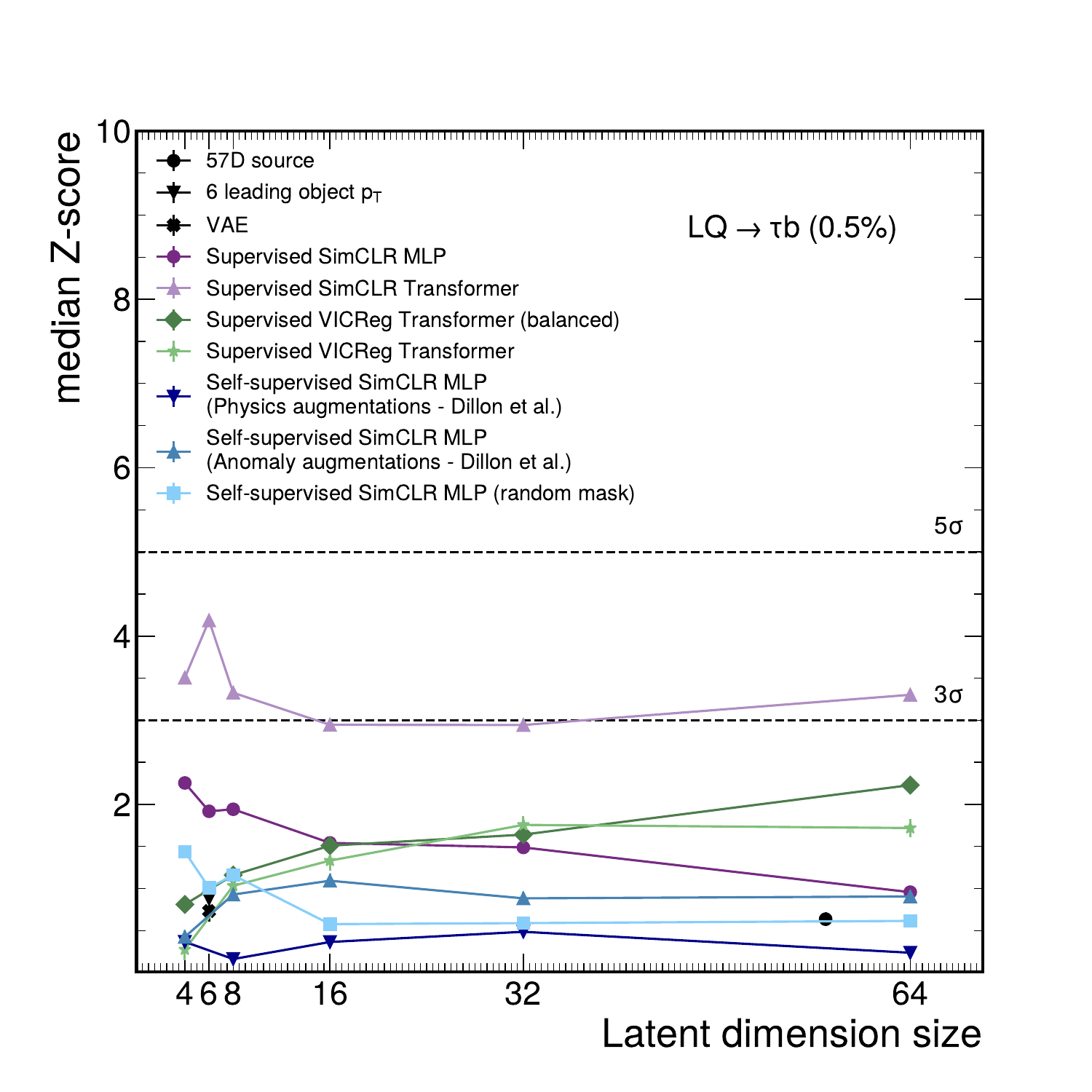}
    \includegraphics[width=0.45\linewidth]{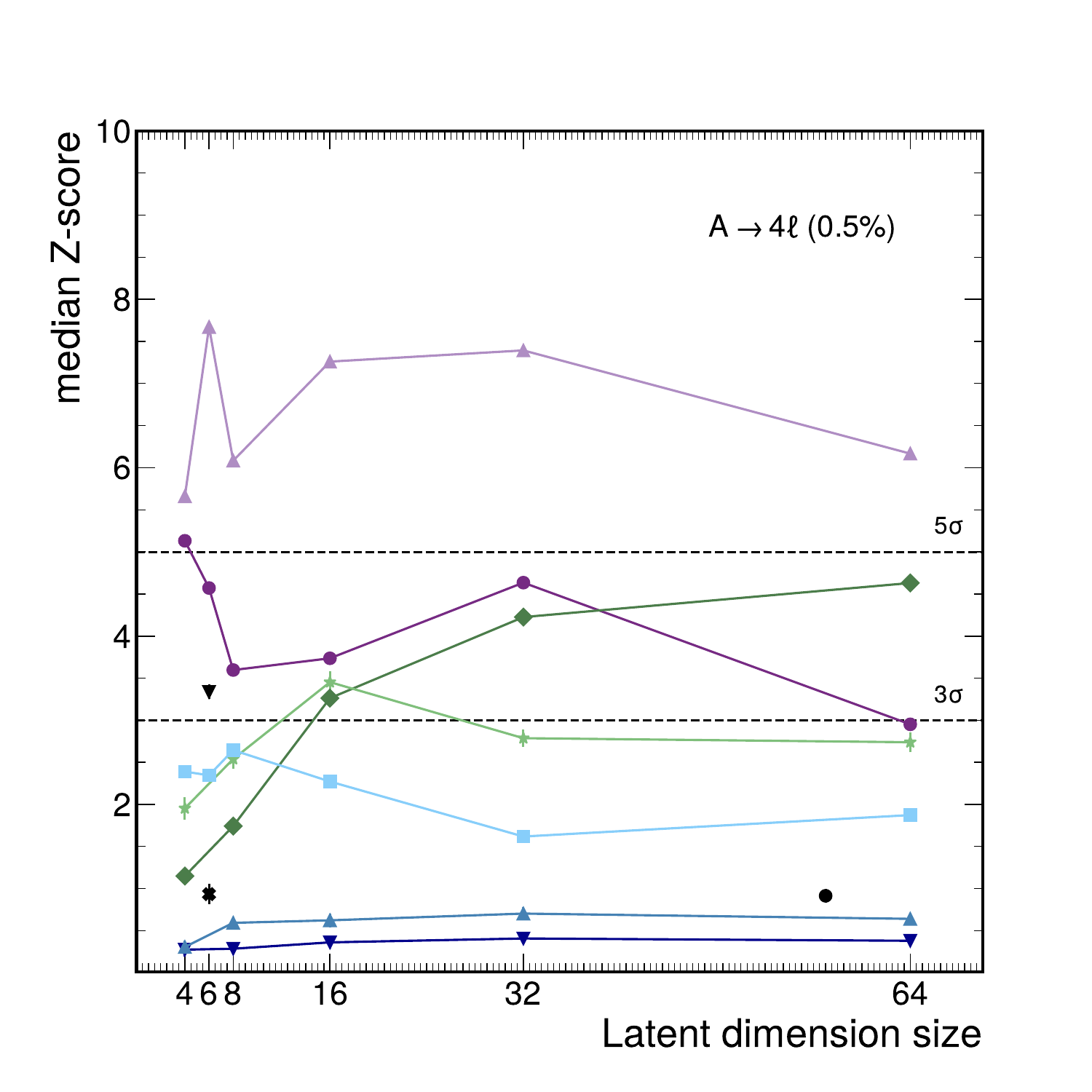}\\
    \hspace{0.02cm}
    \includegraphics[width=0.45\linewidth]{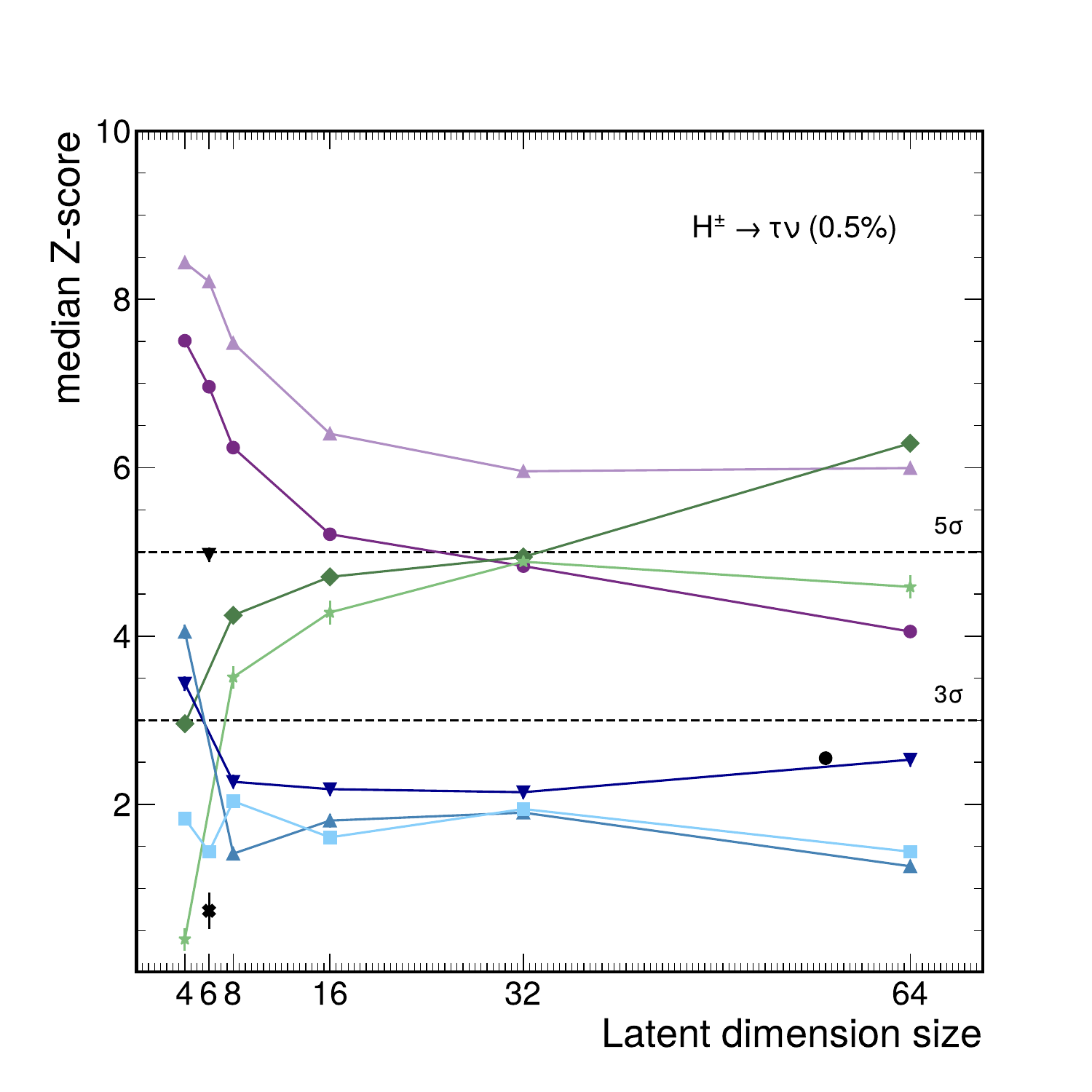}
    \includegraphics[width=0.45\linewidth]{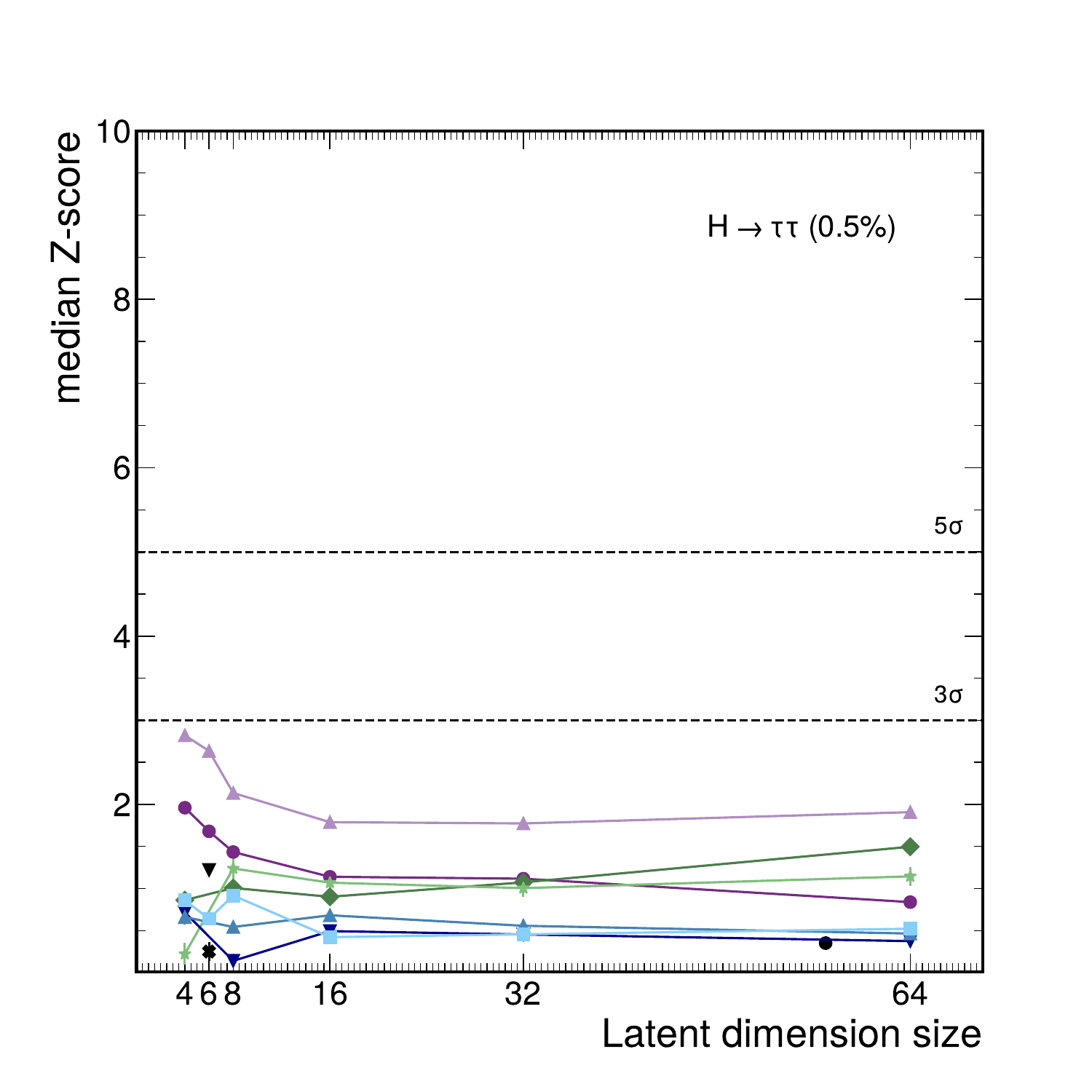}
    
    \caption{The median observed $Z$-score as a function of the feature embedding size after injecting a signal into the SM background pseudo-dataset, corresponding to $0.5\%$ of the total integral. Results are shown for the $LQ \rightarrow \tau b$ (upper left), $A \rightarrow 4\ell$ (upper right), $H^{\pm} \rightarrow \tau \nu$ (bottom left), and $H \rightarrow \tau \tau$ (bottom right) signals. The median $Z$-score is presented for different embedding models: the MLP- and Transformer-based models trained with supervision and SimCLR loss (dark and light purple, respectively),the Transformer-based model trained with supervision and VICReg loss using either balanced background classes (dark green) or background classes weighted to match the composition expected in data (light green), the MLP-based model trained self-supervised with a SimCLR loss with 50\% random masking (light blue), physics-inspired augmentations from~\cite{Dillon:2021gag} (medium blue), or physics- and anomaly-inspired augmentations from~\cite{Dillon:2021gag} (dark blue).
    Additionally, we compare these results with three baseline embeddings: 57D source (black circles), the 6D leading object $p_T$ (black triangles), and the 6D VAE (black crosses). The typical $3\sigma$ level for evidence and $5\sigma$ level for discovery are reported in dashed lines. Where empirical Z-scores cannot be computed ($Z>3\sigma$) we rely on the asymptotic formula.}
    \label{fig:pvals}
\end{figure*}

Figure~\ref{fig:pvals} reports the median $Z$-score obtained using the NPLM test over different signals and different embeddings as a function of the dimensionality of the embedding for a 0.5\% signal injection. 
Similar results on a 0.1\% and 1\% signal injections can be found in Appendix~\ref{app:1}. 
In the figure, the neural embeddings based on contrastive learning are compared to three baseline representations: (1) the original embedding, based on 57 kinematic variables describing the 19 physics objects (``57D source"); (2) a 6-dimensional embedding formed of the missing transverse momentum, the transverse momenta of the two highest momentum muons, the two highest momentum electrons and the highest momentum jet (``6D leading object $p_T$"); and (3) a 6-dimensional neural embedding formed in the latent space of a variational autoencoder (``6D VAE")~\cite{ad_nmi}.
Autoencoders are widely used in many fields of data analysis, high energy physics included, as a way to perform data compression and anomaly detection. In this work we consider the variational autoencoder developed in~\cite{ad_nmi} as a baseline for anomaly detection on the same dataset. The model learns a probabilistic latent representation of the input data. The encoder maps the 57-dimensional input to a 3-dimensional latent space, parameterized by a mean and log-variance, resulting in six latent parameters. It consists of two fully connected layers with 32 and 16 units, followed by batch normalization and LeakyReLU activations. The decoder, mirroring the encoder, reconstructs the input using a final dense layer with 57 output units. The model is trained with a loss combining reconstruction error and Kullback–Leibler divergence, weighted by a scaling factor of 0.8, ensuring a structured latent space through the reparameterization trick.

\begin{figure*}[hbt!]
    \centering
\includegraphics[width=0.45\linewidth]{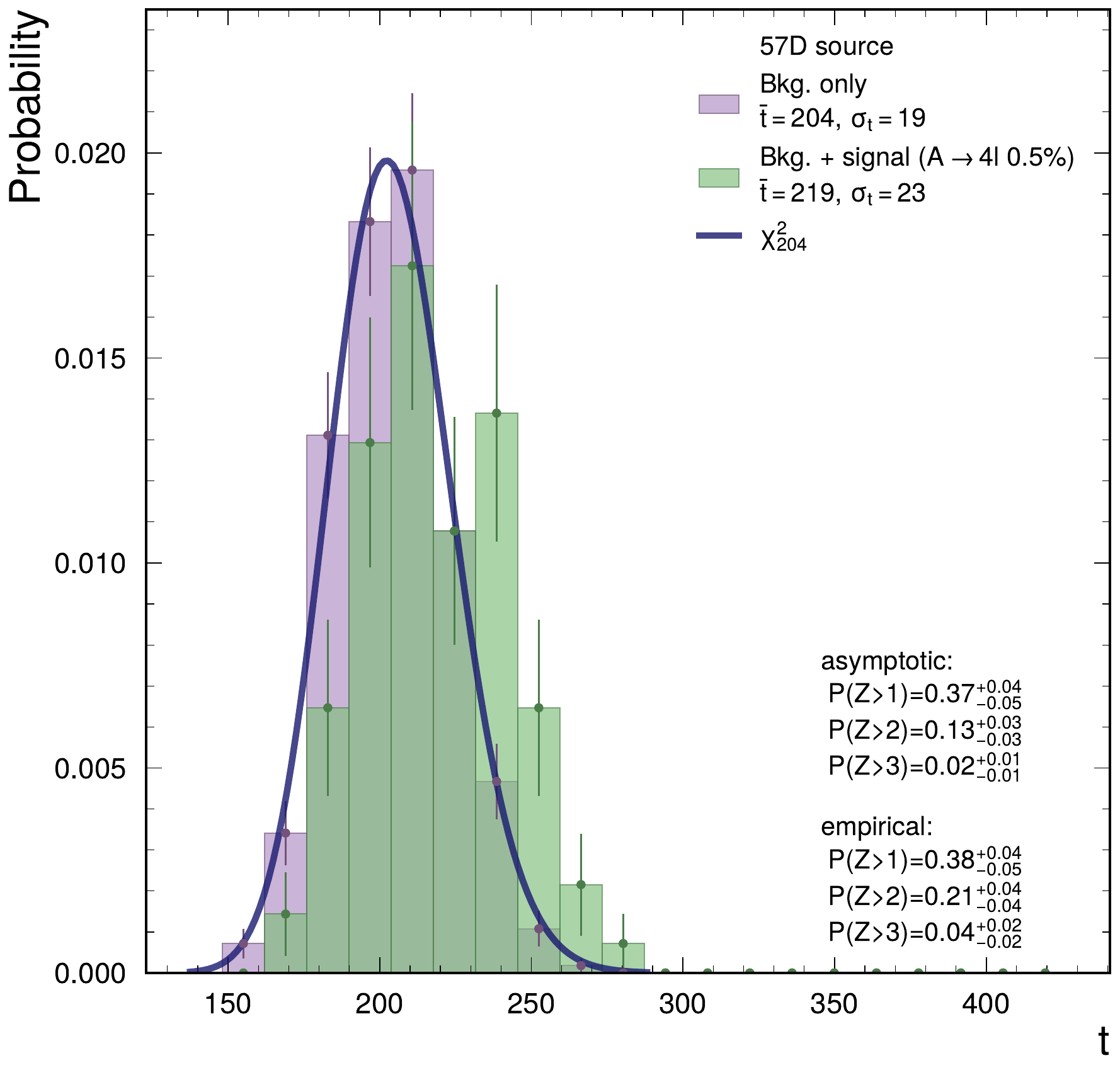}
\includegraphics[width=0.45\linewidth]{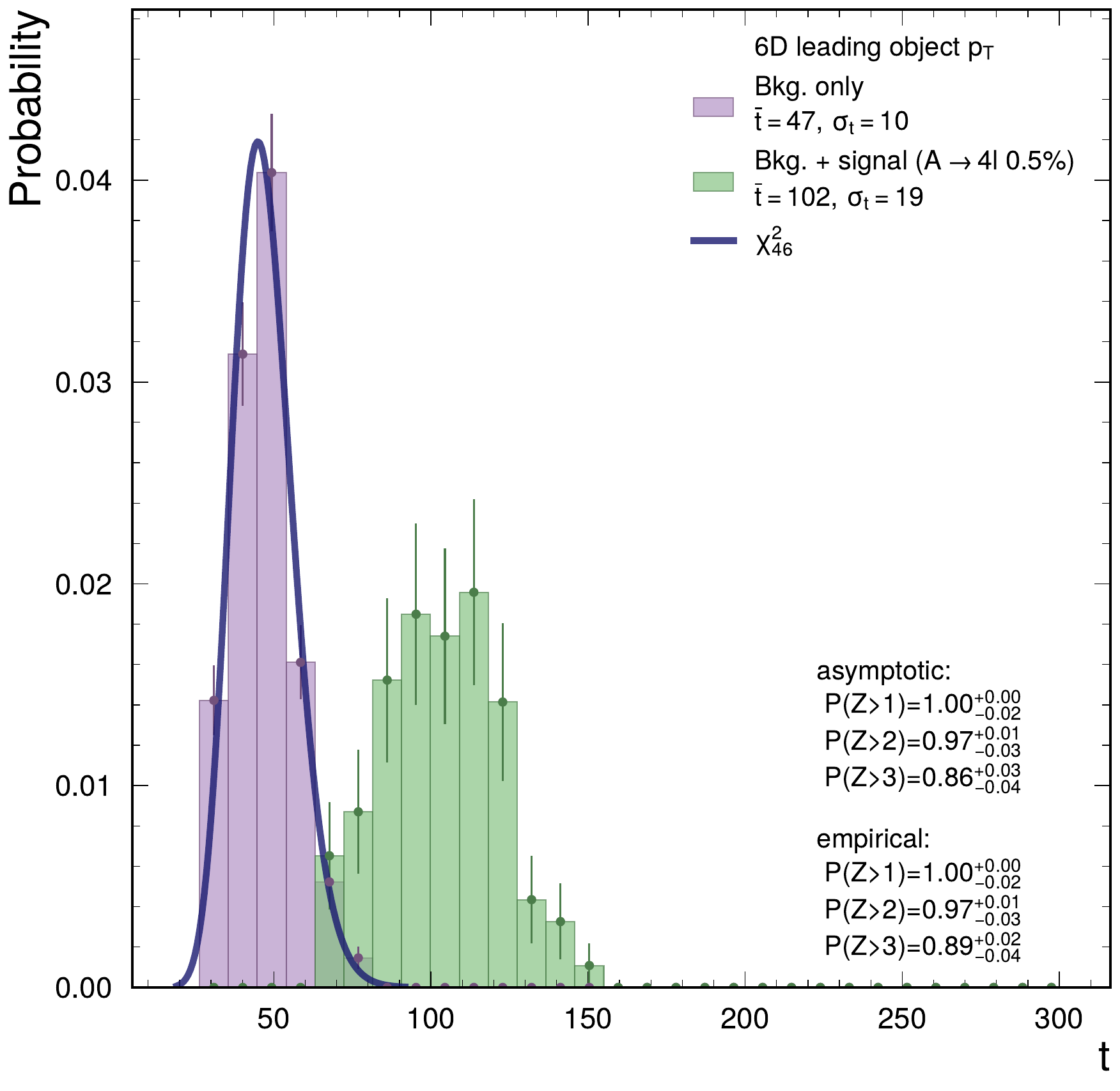}\\
\includegraphics[width=0.45\linewidth]{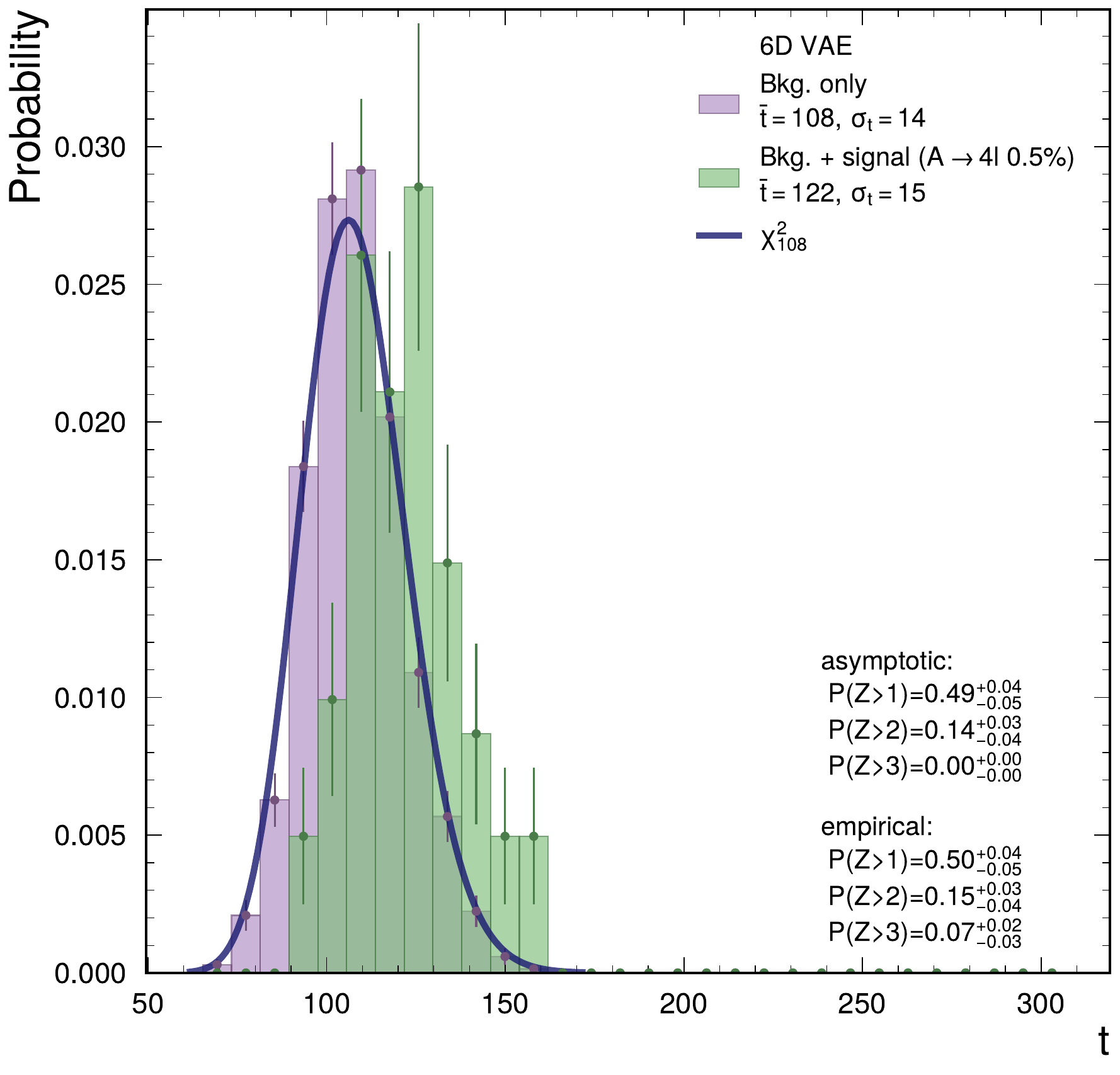}
\includegraphics[width=0.45\linewidth]{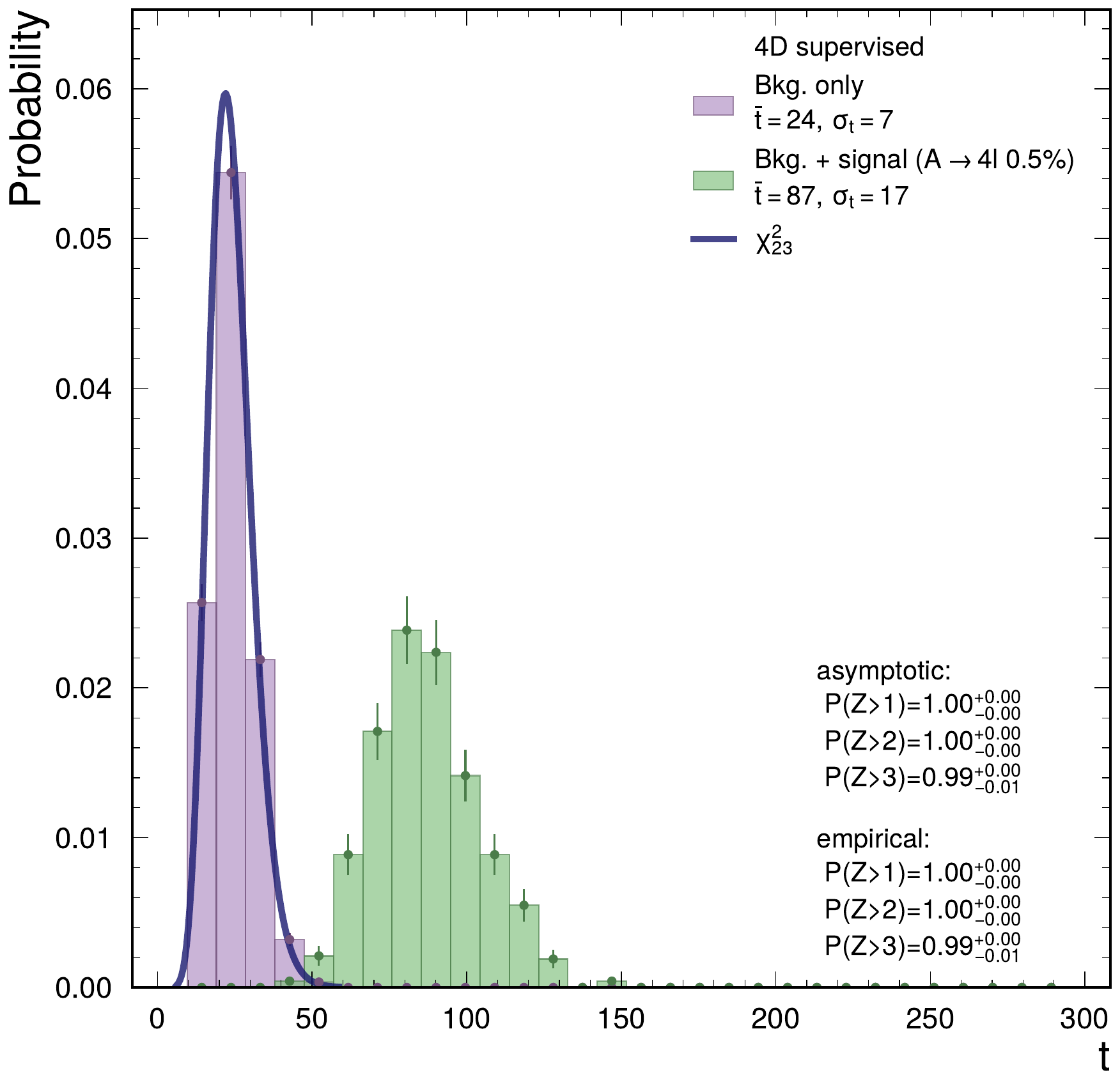}
    \caption{\textbf{Anomaly detection applied to $A\rightarrow4l$ signal benchmark.} The outcome of the NPLM test statistic for 300 toy experiments with $0.5\%$ signal injection (green histogram) are compared to 1000 experiments in absence of signal (purple histogram), representing the empirical null hypothesis. The four panels report the results for different data embeddings: the original 57-dimensional representation (top left); the 6-dimensional representation given by the six highest $p_T$ within the 19 objects in the event (top right); the 6-dimensional neural embedding given by the variational autoencoder (bottom left); and the 4-dimensional neural embedding given by the Transformer-based architecture trained with supervised Sim-CLR loss (bottom right). In each panel we report the power of the test at 1, 2, and 3 $\sigma$ level of discovery obtained from the empirical distribution or the asymptotic $\chi^2$ showed in solid blue line.}
    \label{fig:tstat}
\end{figure*}

Results reported in Figure~\ref{fig:lin_eval} showed that, generally, the performance of a linear classifier for background classes tends to be better with larger embeddings. Conversely, results reported in Figure~\ref{fig:pvals} suggest that anomaly detection benefits from smaller embedding sizes. For two signals, $A \rightarrow 4\ell$ (upper right) and $H^{\pm} \rightarrow \tau \nu$ (bottom left), SimCLR-based models reach $5\sigma$ discovery for embedding dimensionality lower than 16, with a clear increasing trend as the dimensionality is reduced. 
Moreover, transformers show superior performance, particularly relevant to detect more challenging signals, indicating that they might be better equipped to recognize complex patterns in the data. Methods based on supervised contrastive learning significantly outperform traditional approaches in all benchmarks. Morevoer supervised contrastive learning approaches perform generally better than self-supervised ones, suggesting that the physics information carried by the labels, and used to structure the background processes in the latent space, is crucial to enable anomaly detection at the level of new physics processes. 
In addition, we report in Fig~\ref{fig:tstat} the distribution of the test statistic for 1000 background-only toys (purple) and for 300 toys of background with 0.5\% signal injected (green) using the three baseline representations and the best contrastive approach, given by the 4D supervised SimCLR transformer. In each panel we report the empirical and asymptotic detection power at 1, 2 and 3$\sigma$ level. The neural embedding based on contrastive-learning is the only representation that achieves almost 100\% discovery power at a $3\sigma$ level. Numerical results are summarised in Table~\ref{tab:zscore_pvalue}.

\begin{table*}[htbp]
\renewcommand{\arraystretch}{1.5}
    \centering
    \makebox[\textwidth]{
    \begin{tabular}{c|cccc}
    \toprule
         & \multicolumn{4}{c}{0.5\% signal injection}\\
         Embedding type         & $LQ \rightarrow \tau b$ & $A \rightarrow 4\ell$ & $H^{\pm} \rightarrow \tau \nu$ & $H \rightarrow \tau \tau$ \\
         \hline
         57D source                & $0.57_{0.08}^{0.08}$  & $0.93_{0.09}^{0.09}$& $2.47_{0.09}^{0.09}$  &  $0.41_{0.09}^{0.09}$\\
         6D leading object $p_T$   &  $0.96_{0.09}^{0.09}$& $3.42_{0.09}^{0.09}$ & $4.95_{0.09}^{0.09}$ & $1.19_{0.09}^{0.09}$ \\
         6D VAE                    &  $0.7_{0.1}^{0.1}$& $0.9_{0.1}^{0.1}$ &$0.7_{0.1}^{0.1}$  & $0.2_{0.1}^{0.1}$  \\
         4D supervised Transformer & $\boldsymbol{3.66_{0.09}^{0.09}}$ & $\boldsymbol{5.61_{0.09}^{0.09}}$ &  $\boldsymbol{8.51_{0.09}^{0.09}}$& $\boldsymbol{2.81_{0.09}^{0.09}}$ \\
    \bottomrule
    \end{tabular}
    }
    \caption{Z-scores for each embedding type with $0.5\%$ signal injection across different signals: $LQ \rightarrow \tau b$, $A \rightarrow 4\ell$ , $H^{\pm} \rightarrow \tau \nu$, and $H \rightarrow \tau \tau$.}
    \label{tab:zscore_pvalue}
\end{table*}

\begin{figure}[htbp]
    \centering
\includegraphics[width=0.99\linewidth]{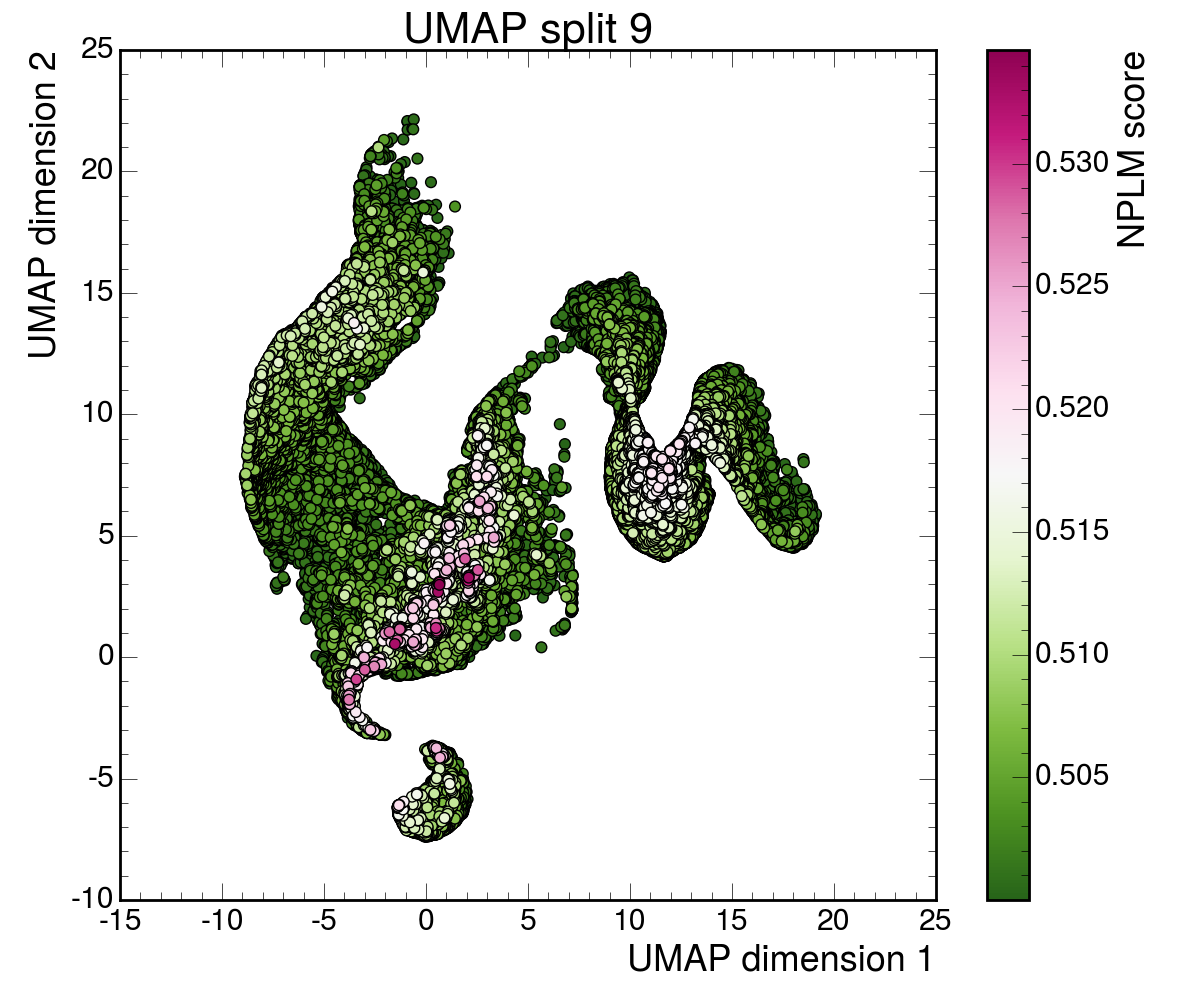}
\caption{\textbf{UMAP visualization of a ``black box" batch.} We inspect the most significant batch (split 9) by means of a two-dimensional UMAP, where the color code represents the sigmoid activated score output by NPLM. The plot only shows data with score grater than 0.5. Data points are ordered in score so that higher score data points are on top and always visible.}
\label{fig:umap}
\end{figure}
\begin{figure*}[htbp]
    \centering
\includegraphics[width=0.45\linewidth]{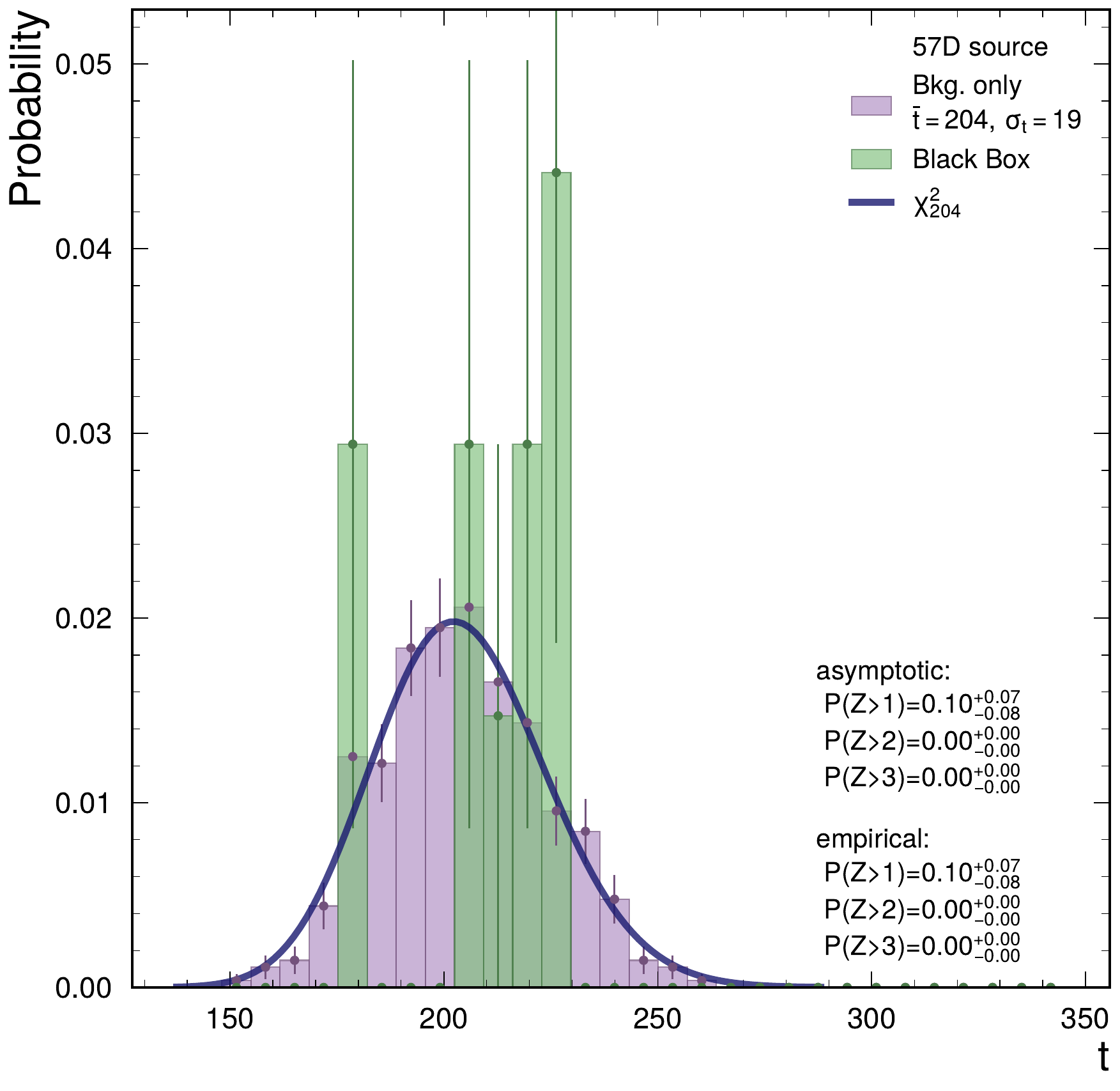}
\includegraphics[width=0.45\linewidth]{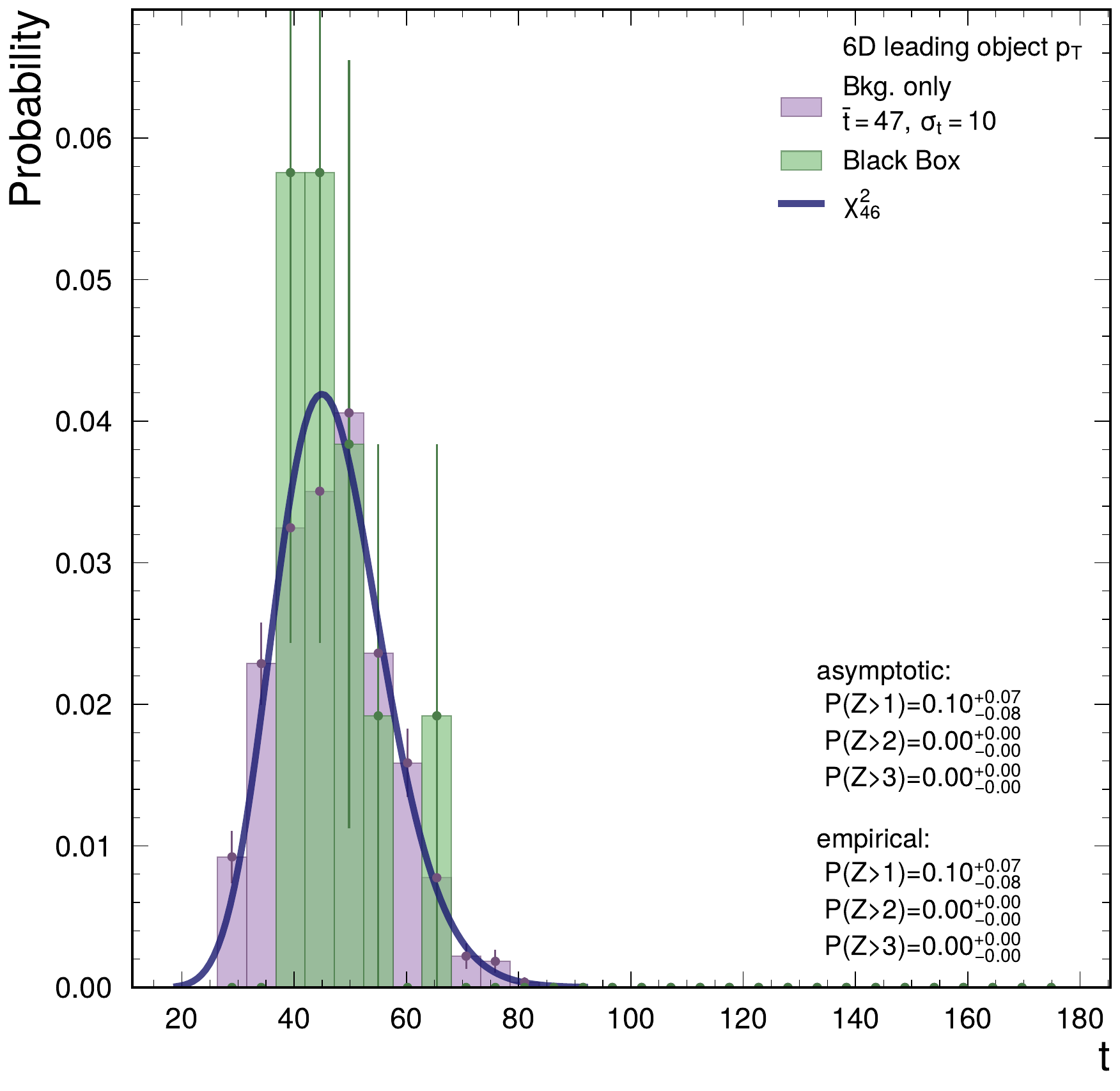}\\
\includegraphics[width=0.45\linewidth]{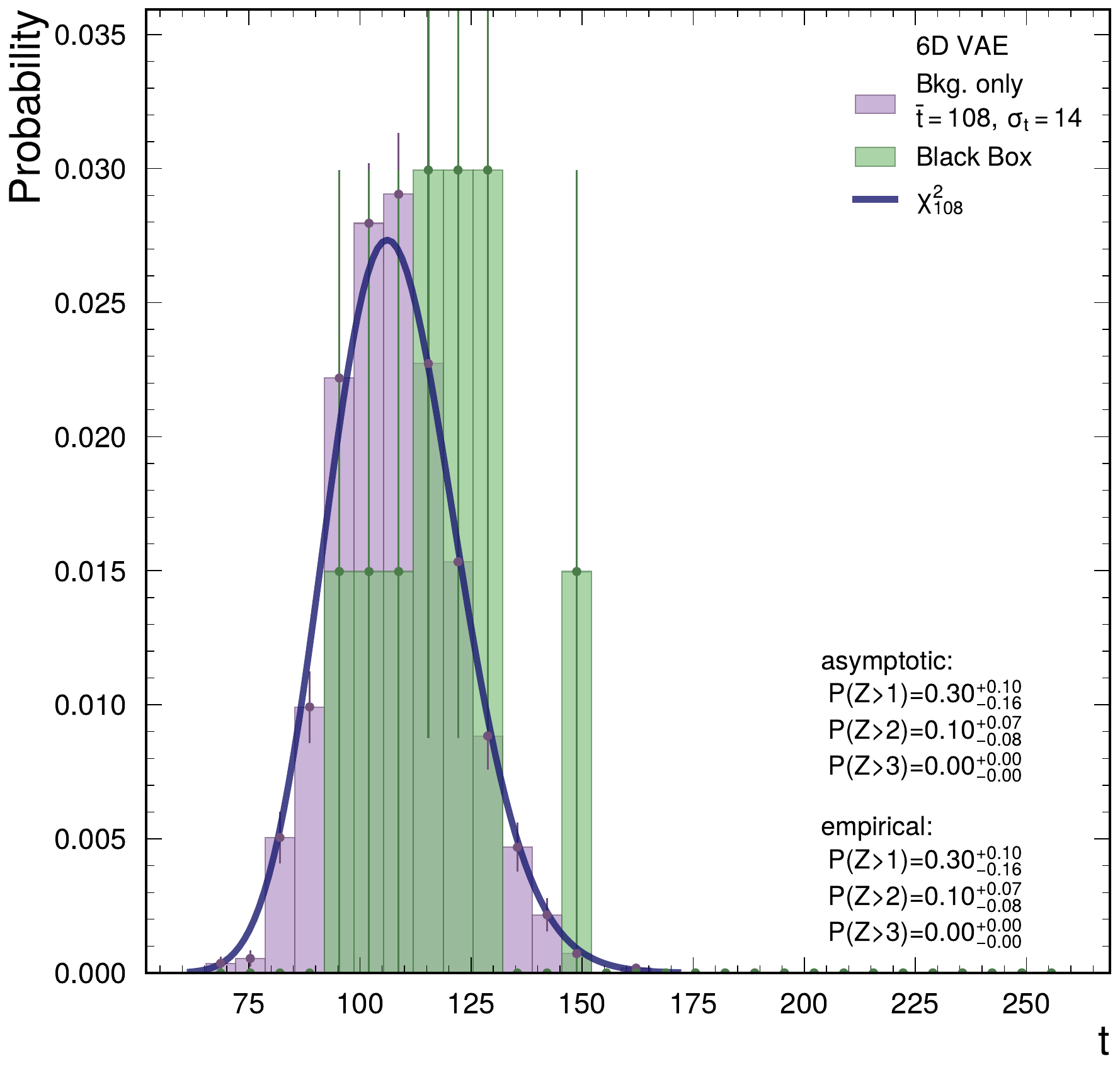}
\includegraphics[width=0.45\linewidth]{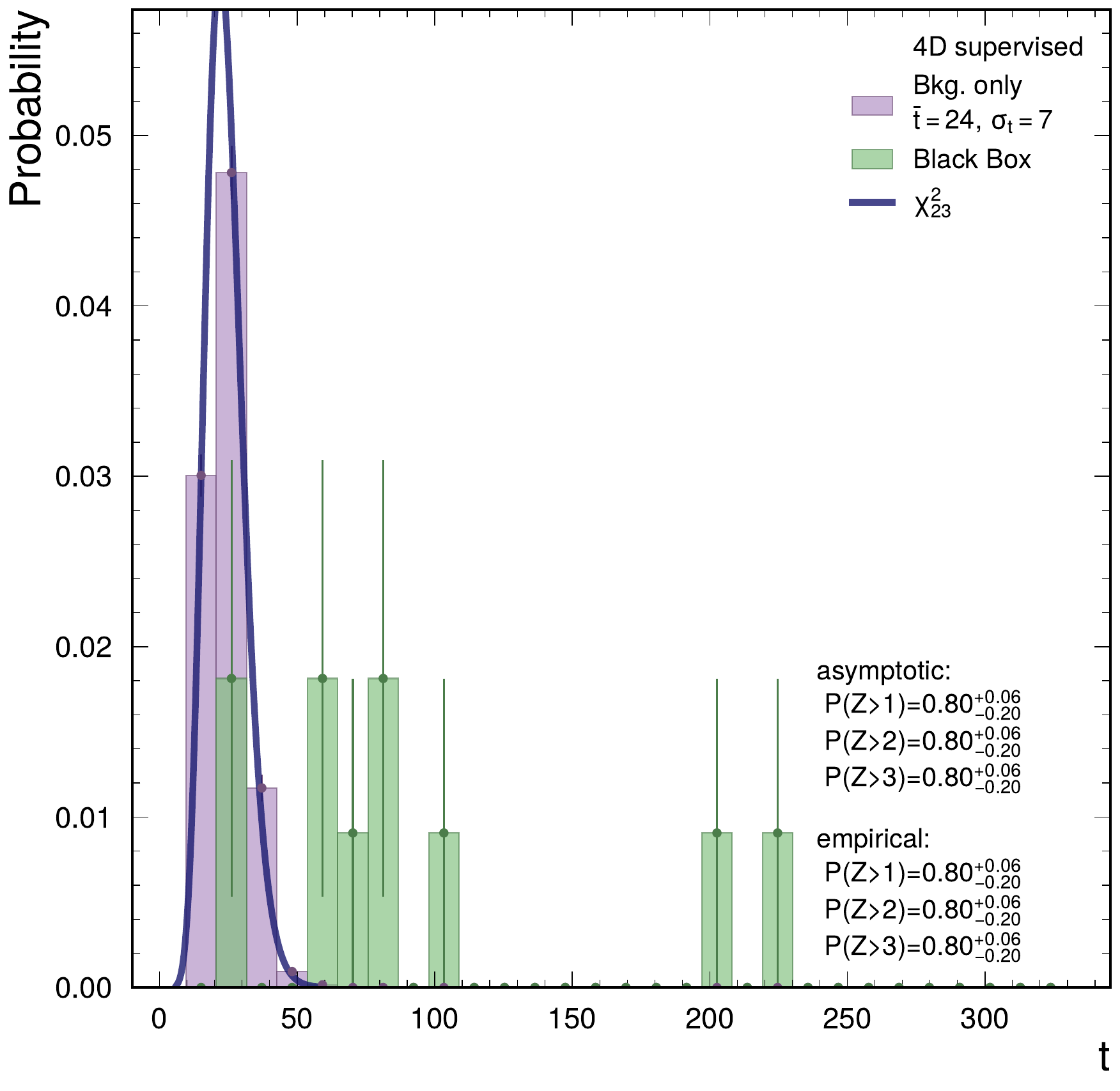}
    \caption{\textbf{Anomaly detection applied to the ``black box" dataset.} The outcome of the NPLM test statistic for ten splits of the black box (green histogram) are compared to the test distribution in absence of signals (purple histogram). For each of the splits we compute the NPLM test statistic 10 times and report the average number. The four panels report the results for different data embeddings: the original 57-dimensional representation (top left); the 6-dimensional representation given by the six highest $p_T$ within the 19 objects in the event (top right); the 6-dimensional neural embedding given by a variational autoencoder from \cite{ad_nmi}(bottom left); and the 4-dimensional neural embedding given by the Transformer-based architecture trained with supervised Sim-CLR loss (bottom right). The neural embedding based on contrastive-learning is the only representation that retains discovery power.}
    \label{fig:blackbox}
\end{figure*}

Finally, we apply the anomaly detection pipelines studied in this section to a ``black box" dataset provided in the AD challenge. The dataset consists of 1,001,000 events of unknown origin, corresponding to 10 times the expected yield of our experiments. The signal injection is expected to be 0.1\%. To match the experimental luminosity used in the signal benchmark study we run the NPLM test on 10 splits of the dataset separately and we compare the detection performance of our three baselines (57D source, 6D leading object $p_T$, and 6D VAE) to that of the best neural embedding, the 4-dimensional transformer-based model trained with supervised SimCLR loss. To account for the mild variance due to the finite size of the reference sample, we repeat the NPLM test $10$ times for each split, randomizing the sampling of the reference set. Figure~\ref{fig:blackbox} compares the NPLM test statistic outcome for each split (green histogram) with the distribution of the test in absence of signal (purple histogram). For each split we report the average observed value of the NPLM test over the ten runs obtained resampling the reference.
The comprehensive set of average values and corresponding standard deviations translated in $Z$-scores are reported in Table~\ref{tab:blackbox}. In the table we also provide an overall assessment of the statistical significance across all runs by computing a global significance. The latter is obtained by selecting the most significant observation among the runs and adjust it via Bonferroni correction.
The 4-dimensional Transformer is the only neural embedding that allow the discovery of a hidden signal in the ``black box" sample, significantly surpassing the detection performances of all the baselines considered in this study. The 4D contrastive embedding reaches a global $Z$-score $12.1\sigma$, opposed to $0,\,0.6$ and $1.4\sigma$ for the source, leading $p_T$ objects and Variatonal Autoencoder respectively.
To further inspect the properties of the blackbox dataset we produce a 2D UMAP (Uniform Manifold Approximation and Projection)~\cite{McInnes2018} for each split.
UMAP is a dimensionality reduction technique that helps visualize high-dimensional data in 2D or 3D while preserving its essential structure. It first constructs a graph by connecting each data point to its nearest neighbors, and then tries to find a 2D representation that preserves this local neighborhood structure as much as possible. Two points that are close to each other in the UMAP visualization, are therefore strongly connected in the original 4-dimensional space.
Figure~\ref{fig:umap} shows the 2D UMAP of the ``black box" data in split 9. The color code corresponds to the score output by NPLM, and only data points with NPLM score higher than 0.5 are considered. The most anomaly-like data points in mesh of pink and purple fall in localized regions of the domain, suggesting that the anomalies are part of the same local structure in the neural embedding. Additional plots for the remaining ``black box" splits are given in Appendix~\ref{app:3}.\\

\begin{table*}[htbp]
\renewcommand{\arraystretch}{1.5}
    \centering
    \makebox[\textwidth]{
    \begin{tabular}{c|cccccccccc|c}
    \toprule
         Embedding type         &  split 0&  split 1& split 2& split 3& split 4& split 5& split 6& split 7& split 8& split 9& global\\
         \hline
         57D source             & 
         $0.9^{0.7}_{0.6}$& $0^{0.7}_{0.6}$& $0.6^{0.7}_{0.6}$& $1.1^{0.6}_{0.6}$& $-1.5^{0.8}_{0.7}$& $-0.0^{0.4}_{0.4}$& $-1.3^{0.7}_{0.7}$& $0.3^{0.5}_{0.5}$& $0.7^{0.5}_{0.5}$& $0.9^{0.5}_{0.5}$& 0\\
         6D leading object $p_T$& 
         $-0.2^{0.4}_{0.4}$& $-0.5^{0.3}_{0.3}$& $0.4^{0.3}_{0.3}$& $-0.5^{0.4}_{0.4}$& $-0.6^{0.3}_{0.3}$& \boldsymbol{$0.7^{0.5}_{0.5}$}& $0.6^{0.4}_{0.4}$& $-0.2^{0.6}_{0.6}$& $-0.2^{0.4}_{0.4}$& $1.9^{0.5}_{0.5}$& $0.6^{0.8}_{0.6}$\\
         6D VAE                 & $1.4^{0.4}_{0.4}$& $0.6^{0.3}_{0.3}$& $0.7^{0.4}_{0.3}$& $-0.1^{0.3}_{0.3}$& $0.3^{0.3}_{0.3}$& $-0.7^{0.3}_{0.3}$& $-0.5^{0.4}_{0.4}$& \boldsymbol{$0.9^{0.2}_{0.2}$}& $1.2^{0.5}_{0.5}$& $2.4^{0.3}_{0.3}$&$1.4^{0.4}_{0.5}$\\
         \textbf{4D SimCLR sup.}        & \boldsymbol{$4.2^{0.2}_{0.2}$} & \boldsymbol{$5.3^{0.2}_{0.2}$} & \boldsymbol{$5.0^{0.1}_{0.1}$} & \boldsymbol{$11.3^{0.2}_{0.2}$} & \boldsymbol{$4.6^{0.3}_{0.2}$} & $0.6^{0.2}_{0.2}$ & \boldsymbol{$6.8^{0.2}_{0.2}$} & \boldsymbol{$0.9^{0.3}_{0.2}$} & \boldsymbol{$4.3^{0.2}_{0.2}$} & \boldsymbol{$12.3^{0.1}_{0.1}$}&\boldsymbol{$12.1^{0.1}_{0.1}$}\\
    \bottomrule
    \end{tabular}
    }
    \caption{\textbf{Anomaly detection applied to the ``black box" dataset.} For each black box split we report the outcome of the NPLM test applied to different data embeddings. The results are reported in terms of average observed $Z$-score, where the average is computed over 10 runs of the NPLM test with randomized resampling of the reference set. Error bands represent the 68\% confidence interval, computed as the standard deviation over the 10 runs. In the last columne we report the global significance across all runs. The latter is obtained by selecting the most significant observation among the runs and adjust it via Bonferroni correction.}
    \label{tab:blackbox}
\end{table*}

\section{Discussion}
\label{sec:discussion}
The results presented in Section~\ref{sec:experiments} show that contrastive-learning-based neural embeddings retain significantly higher anomaly detection power compared to the original 57-dimensional representation, or to a subset of variables, like the 6 highest $p_T$ values, that only allows for a partial view of the event. The contrastive learning scheme based on background supervision with SimCLR objective outperforms all other neural embedding strategies studied in this work, maintaining or even enhancing the sensitivity when the compression is more aggressive. 
As a function of the embedding dimensionality, we observe opposite trends in the accuracy of the background-only classification objective pursued as training task (Figure~\ref{fig:lin_eval}), and the anomalous signal detection objective pursued as downstream task (Figure~\ref{fig:pvals}). This experimental result remarks that the contrastive learning objective is not necessarily aligned with the anomaly detection objective, which is unknown until real data are analyzed.
Due to this reason, despite the increased computational demand, fine-tuning the neural embedding while optimizing the NPLM objective could potentially improve the detection performances. This is an interesting research direction that we leave to future work. 

Finally, our study shows that neural embeddings based on supervised contrastive learning—specifically, contrastive learning between background classes—significantly outperform self-supervised methods. While a substantial portion of this improvement can be attributed to the additional information provided by labels, our findings also underscore the need for greater attention to training details in self-supervised approaches, particularly in designing effective data augmentations. Furthermore, due to the absence of label information it is likely that self-supervised methods could benefit from larger training datasets. This could be readily achieved using real experimental data, which is produced in abundance at LHC experiments. We leave to future work a thorough exploration of this avenue.
Additionally, supervised contrastive learning could be further expanded to enhance sensitivity to specific families of signal hypotheses by adding a broad collection of labeled signal simulations to the training set. The resulting contrastive space  could be used to perform weakly-supervised anomaly detection strategies as demonstrated in~\cite{quak}.\\

\section{Conclusion and outlook}
\label{sec:conclusion}
We presented an end-to-end, signal-agnostic search strategy for inclusive final states at the LHC. By leveraging contrastive learning to build neural embeddings for anomaly detection, we significantly improve sensitivity to several unseen signals. Applying this method to the Anomaly Detection Challenge Blackbox dataset~\cite{delphesdataset} with an injected unknown signal in an unspecified final state, we demonstrate an eightfold improvement in signal sensitivity. This approach enables robust statistical anomaly detection searches for new physics beyond the Standard Model. 
The promising results obtained in our studies motivate further research aimed at making this pipeline viable for real experimental analyses. In the view of scaling the current study to real experimental scenarios characterized by large datasets, efficient self-supervised contrastive learning strategies leveraging real data for pre-training could represent a promising direction to reduce reliance on computationally expensive labeled simulations. Investigating the dependence of self-supervised performance on training sample size and designing optimized data augmentation techniques are critical directions for future research. Another key challenge of real analyses is the presence of systematic uncertainties affecting the modeling of the background processes.  
Previous work has demonstrated methods to incorporate such uncertainties into the NPLM test procedure using dedicated machine learning models~\cite{dAgnolo:2021aun}. Additionally, contrastive training schemes have shown promise in mitigating systematic effects at the embedding stage~\cite{Harris:2024sra}. However, understanding the extent to which systematic uncertainties propagate through contrastive-based neural embeddings and their impact on discovery sensitivity remains an open question for future exploration. Due to the ability to handle inclusive final states while preserving signal-independence, the analysis strategy presented in this work shows great potential for the statistical analysis of data collected by event-level anomaly detection triggers, such as the one used by the CMS experiment~\cite{CMS-DP-2024-059}, or for running quasi-online in real-time detector data acquisition systems~\cite{ARDINO2023167805}.

\section*{Acknowledgments}
G.G. and P.H. acknowledge the financial support of the National Science Foundation under Cooperative Agreement PHY-2019786 (The NSF AI Institute for Artificial Intelligence and Fundamental Interactions, http://iaifi.org/). K.G. and P.H. acknowledge the financial support National Science Foundation under Cooperative Agreement \#2117997 (A3D3, AI-accelerated algorithms for Data Driven Discovery) and NSF CSSI .Computations in this paper were partially run on the FASRC Cannon cluster supported by the FAS Division of Science Research Computing Group at Harvard University. T.\AA.~is supported by the Swiss National Science Foundation Grant No.~PZ00P2\_201594. L.X. is supported by the John Reed Fund from the MIT UROP office, and additionally, she was supported by MISTI France while performing research at CERN.

\bibliographystyle{naturemag}
\bibliography{references}

@article{Khot:2025kqg,
    author = "Khot, Ayush and Wang, Xiwei and Roy, Avik and Kindratenko, Volodymyr and Neubauer, Mark S.",
    title = "{Evidential Deep Learning for Uncertainty Quantification and Out-of-Distribution Detection in Jet Identification using Deep Neural Networks}",
    eprint = "2501.05656",
    archivePrefix = "arXiv",
    primaryClass = "hep-ex",
    month = "1",
    year = "2025"
}

@article{representations,
  author =       "Yoshua Bengio and Aaron Courville and Pascal Vincent",
  title =        "{Representation Learning: A Review and New Perspectives}",
  journal =      "IEEE Transactions on Pattern Analysis and Machine Intelligence",
  volume =       "35",
  number =       "8",
  pages =        "1798-1828",
  month =        "",
  year =         "2013",
  DOI =          "10.1109/TPAMI.2013.50"
}

@online{drawio,
    author    = "",
    title     = "drawio.com",
    url       = "https://www.drawio.com/",
    urldate   = {2024-08-24}
}

@inproceedings{transformer,
 author = {Vaswani, Ashish and Shazeer, Noam and Parmar, Niki and Uszkoreit, Jakob and Jones, Llion and Gomez, Aidan N and Kaiser, \L ukasz and Polosukhin, Illia},
 booktitle = {Advances in Neural Information Processing Systems},
 editor = {I. Guyon and U. Von Luxburg and S. Bengio and H. Wallach and R. Fergus and S. Vishwanathan and R. Garnett},
 pages = {},
 publisher = {Curran Associates, Inc.},
 title = {Attention is All you Need},
 url = {https://proceedings.neurips.cc/paper_files/paper/2017/file/3f5ee243547dee91fbd053c1c4a845aa-Paper.pdf},
 volume = {30},
 year = {2017}
}

@inproceedings{simsiam,
  author={Chen, Xinlei and He, Kaiming},
  booktitle={2021 IEEE/CVF Conference on Computer Vision and Pattern Recognition (CVPR)}, 
  title={Exploring Simple Siamese Representation Learning}, 
  year={2021},
  volume={},
  number={},
  pages={15745-15753},
  doi={10.1109/CVPR46437.2021.01549}
}

@inproceedings{simclr,
  author =       "Ting Chen and Simon Kornblith and Mohammed Norouzi and Geoffrey Hinton",
  title =        "A Simple Framework for Contrastive Learning of Visual Representations",
booktitle = 	 {Proceedings of the 37th International Conference on Machine Learning},
  volume =       "119",
  number =       "",
  pages =        "1597-1607",
  month =        "",
  year =         "2020",
  DOI =          "https://doi.org/10.48550/arXiv.2002.05709"
}

@inproceedings{supcon,
 author = {Khosla, Prannay and Teterwak, Piotr and Wang, Chen and Sarna, Aaron and Tian, Yonglong and Isola, Phillip and Maschinot, Aaron and Liu, Ce and Krishnan, Dilip},
 booktitle = {Advances in Neural Information Processing Systems},
 pages = {18661--18673},
 title = {Supervised Contrastive Learning},
 url = {https://proceedings.neurips.cc/paper_files/paper/2020/file/d89a66c7c80a29b1bdbab0f2a1a94af8-Paper.pdf},
 volume = {33},
 year = {2020}
}

@misc{adamw,
      title={Decoupled Weight Decay Regularization}, 
      author={Ilya Loshchilov and Frank Hutter},
      year={2019},
      doi = {https://doi.org/10.48550/arXiv.1711.05101},
}

@misc{lars,
      title={Large Batch Training of Convolutional Networks}, 
      author={Yang You and Igor Gitman and Boris Ginsburg},
      year={2017},
      doi = {https://doi.org/10.48550/arXiv.1708.03888},
}

@article{delphesdataset,
  author =       "Ekaterina Govorkova and Ema Puljak and Thea Aarrestad and Maurizio Pierini and Kinga Anna Wo/'zniak and Jennifer Ngadiuba",
  title =        "{LHC physics dataset for unsupervised New Physics detection at 40 MHz}",
  journal =      "Nature Sci Data",
  volume =       "9",
  number =       "118",
  pages =        "",
  year =         "2022",
  DOI =          "https://doi.org/10.1038/s41597-022-01187-8"
}

@article{pythia,
  author =       "Torbj{\"o}rn Sj{\"o}strand and Stefan Ask and Jesper R. Christiansen and Richard Corke and Nishita Desai and Philip Ilten and Stephen Mrenna and Stefan Prestel and Christine O. Rasmussen and Peter Z. Skands",
  title =        "{An introduction to PYTHIA 8.2}",
  journal =      "Computer Physics Communications",
  volume =       "191",
  number =       "",
  pages =        "159-177",
  year =         "2015",
  DOI =          "https://doi.org/10.1016/j.cpc.2015.01.024"
}

@article{delphes,
  author =       "The DELPHES 3 collaboration and J. de Favereau and C. Delaere and P. Demin and A. Giammanco and V. Lema\^itre and A. Mertens and M. Selvaggi",
  title =        "{DELHPES 3: a modular framework for fast simulation of a generic collider experiment}",
  journal =      "J. High Energ. Phys.",
  volume =       "2014",
  number =       "57",
  pages =        "",
  year =         "2014",
  DOI =          "https://doi.org/10.1007/JHEP02(2014)057"
}

@misc{falkon,
      title={Kernel methods through the roof: handling billions of points efficiently}, 
      author={Giacomo Meanti and Luigi Carratino and Lorenzo Rosasco and Alessandro Rudi},
      year={2020},
      doi={https://doi.org/10.48550/arXiv.2006.10350}
}

@online{pytorch_transformer,
    author    = "",
    title     = "Pytorch Transformer",
    year      = "",
    url       = "https://pytorch.org/docs/stable/generated/torch.nn.Transformer.html",
    urldate   = {2024-09-22}
}

@online{dino_github,
    author    = "Caron, Mathilde and Touvron, Hugo and Misra, Ishan and J\'egou, Herv\'e  and Mairal, Julien and Bojanowski, Piotr and Joulin, Armand",
    title     = "Self-Supervised Vision Transformers with DINO",
    year      = "",
    url       = "https://github.com/facebookresearch/dino/tree/main",
    urldate   = {2024-09-22}
}

@article{vicreg,
  title={Vicreg: Variance-invariance-covariance regularization for self-supervised learning},
  author={Bardes, Adrien and Ponce, Jean and LeCun, Yann},
  journal={arXiv preprint arXiv:2105.04906},
  year={2021}
}

@article{dAgnolo:2021aun,
    author = "d'Agnolo, Raffaele Tito and Grosso, Gaia and Pierini, Maurizio and Wulzer, Andrea and Zanetti, Marco",
    title = "{Learning new physics from an imperfect machine}",
    eprint = "2111.13633",
    archivePrefix = "arXiv",
    primaryClass = "hep-ph",
    doi = "10.1140/epjc/s10052-022-10226-y",
    journal = "Eur. Phys. J. C",
    volume = "82",
    number = "3",
    pages = "275",
    year = "2022"
}

@article{DAgnolo:2018cun,
    author = "D'Agnolo, Raffaele Tito and Wulzer, Andrea",
    title = "{Learning New Physics from a Machine}",
    eprint = "1806.02350",
    archivePrefix = "arXiv",
    primaryClass = "hep-ph",
    doi = "10.1103/PhysRevD.99.015014",
    journal = "Phys. Rev. D",
    volume = "99",
    number = "1",
    pages = "015014",
    year = "2019"
}

@article{Letizia:2022xbe,
    author = "Letizia, Marco and Losapio, Gianvito and Rando, Marco and Grosso, Gaia and Wulzer, Andrea and Pierini, Maurizio and Zanetti, Marco and Rosasco, Lorenzo",
    title = "{Learning new physics efficiently with nonparametric methods}",
    eprint = "2204.02317",
    archivePrefix = "arXiv",
    primaryClass = "hep-ph",
    doi = "10.1140/epjc/s10052-022-10830-y",
    journal = "Eur. Phys. J. C",
    volume = "82",
    number = "10",
    pages = "879",
    year = "2022"
}

@article{Grosso:2023scl,
    author = "Grosso, Gaia and Letizia, Marco and Pierini, Maurizio and Wulzer, Andrea",
    title = "{Goodness of fit by Neyman-Pearson testing}",
    eprint = "2305.14137",
    archivePrefix = "arXiv",
    primaryClass = "hep-ph",
    month = "5",
    year = "2023"
}

@article{Neyman:1933wgr,
    author = "Neyman, Jerzy and Pearson, Egon Sharpe",
    title = "{On the Problem of the Most Efficient Tests of Statistical Hypotheses}",
    doi = "10.1098/rsta.1933.0009",
    journal = "Phil. Trans. Roy. Soc. Lond. A",
    volume = "231",
    number = "694-706",
    pages = "289--337",
    year = "1933"
}

@article{Belis:2023mqs,
    author = "Belis, Vasilis and Odagiu, Patrick and Aarrestad, Thea Klaeboe",
    title = "{Machine learning for anomaly detection in particle physics}",
    eprint = "2312.14190",
    archivePrefix = "arXiv",
    primaryClass = "physics.data-an",
    doi = "10.1016/j.revip.2024.100091",
    journal = "Rev. Phys.",
    volume = "12",
    pages = "100091",
    year = "2024"
}

@article{ATLAS:2020iwa,
    author = "{ATLAS Collaboration}",
    collaboration = "ATLAS",
    title = "{Dijet resonance search with weak supervision using $\sqrt{s}=13$ TeV $pp$ collisions in the ATLAS detector}",
    eprint = "2005.02983",
    archivePrefix = "arXiv",
    primaryClass = "hep-ex",
    reportNumber = "CERN-EP-2020-062",
    doi = "10.1103/PhysRevLett.125.131801",
    journal = "Phys. Rev. Lett.",
    volume = "125",
    number = "13",
    pages = "131801",
    year = "2020"
}

@article{CMS:2024lwn,
    collaboration = "CMS",
    title = "{Model-agnostic search for dijet resonances with anomalous jet substructure in proton-proton collisions at $\sqrt{s}$ = 13 TeV}",
    reportNumber = "CMS-PAS-EXO-22-026",
    year = "2024"
}

@article{backtoroots,
    author = {Finke, Thorben and Hein, Marie and Kasieczka, Gregor and Kr\"amer, Michael and M\"uck, Alexander and Prangchaikul, Parada and Quadfasel, Tobias and Shih, David and Sommerhalder, Manuel},
    title = "{Back To The Roots: Tree-Based Algorithms for Weakly Supervised Anomaly Detection}",
    eprint = "2309.13111",
    archivePrefix = "arXiv",
    primaryClass = "hep-ph",
    reportNumber = "TTK-23-26",
    month = "9",
    year = "2023"
}

@article{Grosso:2024nho,
    author = "Grosso, Gaia",
    title = "{Anomaly-aware summary statistic from data batches}",
    eprint = "2407.01249",
    archivePrefix = "arXiv",
    primaryClass = "hep-ex",
    doi = "10.1007/JHEP12(2024)093",
    journal = "JHEP",
    volume = "12",
    pages = "093",
    year = "2024"
}

@article{CGV-107,
url = {http://dx.doi.org/10.1561/0600000107},
year = {2023},
volume = {15},
journal = {Foundations and Trends® in Computer Graphics and Vision},
title = {An Introduction to Neural Data Compression},
doi = {10.1561/0600000107},
issn = {1572-2740},
number = {2},
pages = {113-200},
author = {Yibo Yang and Stephan Mandt and Lucas Theis}
}

@article{Harris:2024sra,
    author = "Harris, Philip and Kagan, Michael and Krupa, Jeffrey and Maier, Benedikt and Woodward, Nathaniel",
    title = "{Re-Simulation-based Self-Supervised Learning for Pre-Training Foundation Models}",
    eprint = "2403.07066",
    archivePrefix = "arXiv",
    primaryClass = "hep-ph",
    month = "3",
    year = "2024"
}

@article{ad_nmi,
	author = {Govorkova, Ekaterina and Puljak, Ema and Aarrestad, Thea and James, Thomas and Loncar, Vladimir and Pierini, Maurizio and Pol, Adrian Alan and Ghielmetti, Nicol{\`o} and Graczyk, Maksymilian and Summers, Sioni and Ngadiuba, Jennifer and Nguyen, Thong Q. and Duarte, Javier and Wu, Zhenbin},
	journal = {Nature Machine Intelligence},
	number = {2},
	pages = {154--161},
	title = {Autoencoders on field-programmable gate arrays for real-time, unsupervised new physics detection at 40 MHz at the Large Hadron Collider},
	volume = {4},
	year = {2022}}

@article{ATLAS:2014sxa,
    collaboration = "ATLAS",
    title = "{A general search for new phenomena with the ATLAS detector in pp collisions at $\sqrt{s}=8$ TeV}",
    reportNumber = "ATLAS-CONF-2014-006",
    month = "3",
    year = "2014"
}

@article{CMS:2020zjg,
    author = "Sirunyan, Albert M and others",
    collaboration = "CMS",
    title = "{MUSiC: a model-unspecific search for new physics in proton\textendash{}proton collisions at $\sqrt{s} = 13\,\text {TeV} $}",
    eprint = "2010.02984",
    archivePrefix = "arXiv",
    primaryClass = "hep-ex",
    reportNumber = "CMS-EXO-19-008, CERN-EP-2020-171",
    doi = "10.1140/epjc/s10052-021-09236-z",
    journal = "Eur. Phys. J. C",
    volume = "81",
    number = "7",
    pages = "629",
    year = "2021"
}

@article{Bardhan:2025icr,
    author = "Bardhan, Jai and Agrawal, Radhikesh and Tilak, Abhiram and Neeraj, Cyrin and Mitra, Subhadip",
    title = "{HEP-JEPA: A foundation model for collider physics using joint embedding predictive architecture}",
    eprint = "2502.03933",
    archivePrefix = "arXiv",
    primaryClass = "cs.LG",
    month = "2",
    year = "2025"
}

@inproceedings{Wildridge:2024yeg,
    author = "Wildridge, Andrew J. and Rodgers, Jack P. and Colbert, Ethan M. and yao, Yao and Jung, Andreas W. and Liu, Miaoyuan",
    title = "{Bumblebee: Foundation Model for Particle Physics Discovery}",
    booktitle = "{38th conference on Neural Information Processing Systems}",
    eprint = "2412.07867",
    archivePrefix = "arXiv",
    primaryClass = "hep-ex",
    month = "12",
    year = "2024"
}

@article{Birk:2024knn,
    author = "Birk, Joschka and Hallin, Anna and Kasieczka, Gregor",
    title = "{OmniJet-\ensuremath{\alpha}: the first cross-task foundation model for particle physics}",
    eprint = "2403.05618",
    archivePrefix = "arXiv",
    primaryClass = "hep-ph",
    doi = "10.1088/2632-2153/ad66ad",
    journal = "Mach. Learn. Sci. Tech.",
    volume = "5",
    number = "3",
    pages = "035031",
    year = "2024"
}

@article{Golling:2024abg,
    author = "Golling, Tobias and Heinrich, Lukas and Kagan, Michael and Klein, Samuel and Leigh, Matthew and Osadchy, Margarita and Raine, John Andrew",
    title = "{Masked particle modeling on sets: towards self-supervised high energy physics foundation models}",
    eprint = "2401.13537",
    archivePrefix = "arXiv",
    primaryClass = "hep-ph",
    doi = "10.1088/2632-2153/ad64a8",
    journal = "Mach. Learn. Sci. Tech.",
    volume = "5",
    number = "3",
    pages = "035074",
    year = "2024"
}

@article{Dillon:2023zac,
    author = "Dillon, Barry M. and Favaro, Luigi and Feiden, Friedrich and Modak, Tanmoy and Plehn, Tilman",
    title = "{Anomalies, representations, and self-supervision}",
    eprint = "2301.04660",
    archivePrefix = "arXiv",
    primaryClass = "hep-ph",
    doi = "10.21468/SciPostPhysCore.7.3.056",
    journal = "SciPost Phys. Core",
    volume = "7",
    pages = "056",
    year = "2024"
}

@misc{openai2024gpt4technicalreport,
      title={GPT-4 Technical Report}, 
      author={OpenAI and Josh Achiam and Steven Adler and Sandhini Agarwal and Lama Ahmad and Ilge Akkaya and Florencia Leoni Aleman and Diogo Almeida and Janko Altenschmidt and Sam Altman and Shyamal Anadkat and Red Avila and Igor Babuschkin and Suchir Balaji and Valerie Balcom and Paul Baltescu and Haiming Bao and Mohammad Bavarian and Jeff Belgum and Irwan Bello and Jake Berdine and Gabriel Bernadett-Shapiro and Christopher Berner and Lenny Bogdonoff and Oleg Boiko and Madelaine Boyd and Anna-Luisa Brakman and Greg Brockman and Tim Brooks and Miles Brundage and Kevin Button and Trevor Cai and Rosie Campbell and Andrew Cann and Brittany Carey and Chelsea Carlson and Rory Carmichael and Brooke Chan and Che Chang and Fotis Chantzis and Derek Chen and Sully Chen and Ruby Chen and Jason Chen and Mark Chen and Ben Chess and Chester Cho and Casey Chu and Hyung Won Chung and Dave Cummings and Jeremiah Currier and Yunxing Dai and Cory Decareaux and Thomas Degry and Noah Deutsch and Damien Deville and Arka Dhar and David Dohan and Steve Dowling and Sheila Dunning and Adrien Ecoffet and Atty Eleti and Tyna Eloundou and David Farhi and Liam Fedus and Niko Felix and Simón Posada Fishman and Juston Forte and Isabella Fulford and Leo Gao and Elie Georges and Christian Gibson and Vik Goel and Tarun Gogineni and Gabriel Goh and Rapha Gontijo-Lopes and Jonathan Gordon and Morgan Grafstein and Scott Gray and Ryan Greene and Joshua Gross and Shixiang Shane Gu and Yufei Guo and Chris Hallacy and Jesse Han and Jeff Harris and Yuchen He and Mike Heaton and Johannes Heidecke and Chris Hesse and Alan Hickey and Wade Hickey and Peter Hoeschele and Brandon Houghton and Kenny Hsu and Shengli Hu and Xin Hu and Joost Huizinga and Shantanu Jain and Shawn Jain and Joanne Jang and Angela Jiang and Roger Jiang and Haozhun Jin and Denny Jin and Shino Jomoto and Billie Jonn and Heewoo Jun and Tomer Kaftan and Łukasz Kaiser and Ali Kamali and Ingmar Kanitscheider and Nitish Shirish Keskar and Tabarak Khan and Logan Kilpatrick and Jong Wook Kim and Christina Kim and Yongjik Kim and Jan Hendrik Kirchner and Jamie Kiros and Matt Knight and Daniel Kokotajlo and Łukasz Kondraciuk and Andrew Kondrich and Aris Konstantinidis and Kyle Kosic and Gretchen Krueger and Vishal Kuo and Michael Lampe and Ikai Lan and Teddy Lee and Jan Leike and Jade Leung and Daniel Levy and Chak Ming Li and Rachel Lim and Molly Lin and Stephanie Lin and Mateusz Litwin and Theresa Lopez and Ryan Lowe and Patricia Lue and Anna Makanju and Kim Malfacini and Sam Manning and Todor Markov and Yaniv Markovski and Bianca Martin and Katie Mayer and Andrew Mayne and Bob McGrew and Scott Mayer McKinney and Christine McLeavey and Paul McMillan and Jake McNeil and David Medina and Aalok Mehta and Jacob Menick and Luke Metz and Andrey Mishchenko and Pamela Mishkin and Vinnie Monaco and Evan Morikawa and Daniel Mossing and Tong Mu and Mira Murati and Oleg Murk and David Mély and Ashvin Nair and Reiichiro Nakano and Rajeev Nayak and Arvind Neelakantan and Richard Ngo and Hyeonwoo Noh and Long Ouyang and Cullen O'Keefe and Jakub Pachocki and Alex Paino and Joe Palermo and Ashley Pantuliano and Giambattista Parascandolo and Joel Parish and Emy Parparita and Alex Passos and Mikhail Pavlov and Andrew Peng and Adam Perelman and Filipe de Avila Belbute Peres and Michael Petrov and Henrique Ponde de Oliveira Pinto and Michael and Pokorny and Michelle Pokrass and Vitchyr H. Pong and Tolly Powell and Alethea Power and Boris Power and Elizabeth Proehl and Raul Puri and Alec Radford and Jack Rae and Aditya Ramesh and Cameron Raymond and Francis Real and Kendra Rimbach and Carl Ross and Bob Rotsted and Henri Roussez and Nick Ryder and Mario Saltarelli and Ted Sanders and Shibani Santurkar and Girish Sastry and Heather Schmidt and David Schnurr and John Schulman and Daniel Selsam and Kyla Sheppard and Toki Sherbakov and Jessica Shieh and Sarah Shoker and Pranav Shyam and Szymon Sidor and Eric Sigler and Maddie Simens and Jordan Sitkin and Katarina Slama and Ian Sohl and Benjamin Sokolowsky and Yang Song and Natalie Staudacher and Felipe Petroski Such and Natalie Summers and Ilya Sutskever and Jie Tang and Nikolas Tezak and Madeleine B. Thompson and Phil Tillet and Amin Tootoonchian and Elizabeth Tseng and Preston Tuggle and Nick Turley and Jerry Tworek and Juan Felipe Cerón Uribe and Andrea Vallone and Arun Vijayvergiya and Chelsea Voss and Carroll Wainwright and Justin Jay Wang and Alvin Wang and Ben Wang and Jonathan Ward and Jason Wei and CJ Weinmann and Akila Welihinda and Peter Welinder and Jiayi Weng and Lilian Weng and Matt Wiethoff and Dave Willner and Clemens Winter and Samuel Wolrich and Hannah Wong and Lauren Workman and Sherwin Wu and Jeff Wu and Michael Wu and Kai Xiao and Tao Xu and Sarah Yoo and Kevin Yu and Qiming Yuan and Wojciech Zaremba and Rowan Zellers and Chong Zhang and Marvin Zhang and Shengjia Zhao and Tianhao Zheng and Juntang Zhuang and William Zhuk and Barret Zoph},
      year={2024},
      eprint={2303.08774},
      archivePrefix={arXiv},
      primaryClass={cs.CL},
      url={https://arxiv.org/abs/2303.08774}, 
}

@article{Leigh:2024ked,
    author = "Leigh, Matthew and Klein, Samuel and Charton, Fran\c{c}ois and Golling, Tobias and Heinrich, Lukas and Kagan, Michael and Ochoa, In\^es and Osadchy, Margarita",
    title = "{Is Tokenization Needed for Masked Particle Modelling?}",
    eprint = "2409.12589",
    archivePrefix = "arXiv",
    primaryClass = "hep-ph",
    month = "9",
    year = "2024"
}

@article{Mikuni:2024qsr,
    author = "Mikuni, Vinicius and Nachman, Benjamin",
    title = "{OmniLearn: A Method to Simultaneously Facilitate All Jet Physics Tasks}",
    eprint = "2404.16091",
    archivePrefix = "arXiv",
    primaryClass = "hep-ph",
    month = "4",
    year = "2024"
}

@article{Park:2022zov,
    author = "Park, Sang Eon and Harris, Philip and Ostdiek, Bryan",
    title = "{Neural embedding: learning the embedding of the manifold of physics data}",
    eprint = "2208.05484",
    archivePrefix = "arXiv",
    primaryClass = "hep-ph",
    doi = "10.1007/JHEP07(2023)108",
    journal = "JHEP",
    volume = "07",
    pages = "108",
    year = "2023"
}

@article{Li:2024htp,
    author = "Li, Congqiao and others",
    title = "{Accelerating Resonance Searches via Signature-Oriented Pre-training}",
    eprint = "2405.12972",
    archivePrefix = "arXiv",
    primaryClass = "hep-ph",
    reportNumber = "FERMILAB-PUB-24-0699-V",
    month = "5",
    year = "2024"
}

@article{McInnes2018, 
doi = {10.21105/joss.00861}, 
url = {https://doi.org/10.21105/joss.00861},
year = {2018}, 
publisher = {The Open Journal}, 
volume = {3}, 
number = {29}, 
pages = {861}, 
author = {Leland McInnes and John Healy and Nathaniel Saul and Lukas Großberger}, 
title = {UMAP: Uniform Manifold Approximation and Projection}, 
journal = {Journal of Open Source Software} }

@article{CMS-DP-2024-059,
      collaboration = "CMS",
      title         = "{2024 Data Collected with AXOL1TL Anomaly Detection at the
                       CMS Level-1 Trigger}",
      year          = "2024",
      url           = "https://cds.cern.ch/record/2904695",
}

@article{ARDINO2023167805,
	author = {Rocco Ardino and Christian Deldicque and Marc Dobson and Sabrina Giorgetti and Gaia Grosso and Thomas James and Emilio Meschi and Dinyar Rabady and Attila Racz and Hannes Sakulin and Petr Zejdl},
	journal = {Nuclear Instruments and Methods in Physics Research Section A: Accelerators, Spectrometers, Detectors and Associated Equipment},
	pages = {167805},
	title = {A 40 MHz Level-1 trigger scouting system for the CMS Phase-2 upgrade},
	volume = {1047},
	year = {2023}}

@article{quak,
	author = {Park, Sang Eon and Rankin, Dylan and Udrescu, Silviu-Marian and Yunus, Mikaeel and Harris, Philip},
	journal = {Journal of High Energy Physics},
	number = {6},
	pages = {30},
	title = {Quasi anomalous knowledge: searching for new physics with embedded knowledge},
	volume = {2021},
	year = {2021}}

@article{PhysRevD.106.056005,
  title = {Self-supervised anomaly detection for new physics},
  author = {Dillon, Barry M. and Mastandrea, Radha and Nachman, Benjamin},
  journal = {Phys. Rev. D},
  volume = {106},
  issue = {5},
  pages = {056005},
  numpages = {12},
  year = {2022},
  month = {Sep},
  publisher = {American Physical Society},
  doi = {10.1103/PhysRevD.106.056005},
  url = {https://link.aps.org/doi/10.1103/PhysRevD.106.056005}
}

@article{D0:2000dnz,
    author = "Abazov, V. M. and others",
    collaboration = "D0",
    title = "{A Quasi Model Independent Search for New Physics at Large Transverse Momentum}",
    eprint = "hep-ex/0011067",
    archivePrefix = "arXiv",
    reportNumber = "FERMILAB-PUB-00-302-E",
    doi = "10.1103/PhysRevD.64.012004",
    journal = "Phys. Rev. D",
    volume = "64",
    pages = "012004",
    year = "2001"
}

@article{D0:2001mmn,
    author = "Abbott, B. and others",
    collaboration = "D0",
    title = "{A Quasi Model Independent Search for New High $p_T$ Physics at D0}",
    eprint = "hep-ex/0011071",
    archivePrefix = "arXiv",
    reportNumber = "FERMILAB-PUB-00-304-E",
    doi = "10.1103/PhysRevLett.86.3712",
    journal = "Phys. Rev. Lett.",
    volume = "86",
    pages = "3712--3717",
    year = "2001"
}

@article{H1:2004rlm,
    author = "Aktas, A. and others",
    collaboration = "H1",
    title = "{A General search for new phenomena in ep scattering at HERA}",
    eprint = "hep-ex/0408044",
    archivePrefix = "arXiv",
    reportNumber = "DESY-04-140",
    doi = "10.1016/j.physletb.2004.09.057",
    journal = "Phys. Lett. B",
    volume = "602",
    pages = "14--30",
    year = "2004"
}

@article{CDF:2007iou,
    author = "Aaltonen, T. and others",
    collaboration = "CDF",
    title = "{Model-Independent and Quasi-Model-Independent Search for New Physics at CDF}",
    eprint = "0712.1311",
    archivePrefix = "arXiv",
    primaryClass = "hep-ex",
    reportNumber = "FERMILAB-PUB-07-657-E",
    doi = "10.1103/PhysRevD.78.012002",
    journal = "Phys. Rev. D",
    volume = "78",
    pages = "012002",
    year = "2008"
}

@article{CDF:2008voc,
    author = "Aaltonen, T. and others",
    collaboration = "CDF",
    title = "{Global Search for New Physics with 2.0 fb$^{-1}$ at CDF}",
    eprint = "0809.3781",
    archivePrefix = "arXiv",
    primaryClass = "hep-ex",
    reportNumber = "FERMILAB-PUB-08-400-E",
    doi = "10.1103/PhysRevD.79.011101",
    journal = "Phys. Rev. D",
    volume = "79",
    pages = "011101",
    year = "2009"
}

@article{H1:2008aak,
    author = "Aaron, F. D. and others",
    collaboration = "H1",
    title = "{A General Search for New Phenomena at HERA}",
    eprint = "0901.0507",
    archivePrefix = "arXiv",
    primaryClass = "hep-ex",
    reportNumber = "DESY-08-173",
    doi = "10.1016/j.physletb.2009.03.034",
    journal = "Phys. Lett. B",
    volume = "674",
    pages = "257--268",
    year = "2009"
}

@article{D0:2011ccx,
    author = "Abazov, Victor Mukhamedovich and others",
    collaboration = "D0",
    title = "{Model independent search for new phenomena in $p \bar{p}$ collisions at $\sqrt{s}=1.96$ TeV}",
    eprint = "1108.5362",
    archivePrefix = "arXiv",
    primaryClass = "hep-ex",
    reportNumber = "FERMILAB-PUB-11-400-E",
    doi = "10.1103/PhysRevD.85.092015",
    journal = "Phys. Rev. D",
    volume = "85",
    pages = "092015",
    year = "2012"
}

@article{ATLAS:2023azi,
    author = "Aad, Georges and others",
    collaboration = "ATLAS",
    title = "{Anomaly detection search for new resonances decaying into a Higgs boson and a generic new particle $X$ in hadronic final states using $\sqrt{s} = 13$ TeV $pp$ collisions with the ATLAS detector}",
    eprint = "2306.03637",
    archivePrefix = "arXiv",
    primaryClass = "hep-ex",
    reportNumber = "CERN-EP-2023-045",
    doi = "10.1103/PhysRevD.108.052009",
    journal = "Phys. Rev. D",
    volume = "108",
    pages = "052009",
    year = "2023"
}

@article{ATLAS:2023ixc,
    author = "Aad, Georges and others",
    collaboration = "ATLAS",
    title = "{Search for New Phenomena in Two-Body Invariant Mass Distributions Using Unsupervised Machine Learning for Anomaly Detection at s=13\,\,TeV with the ATLAS Detector}",
    eprint = "2307.01612",
    archivePrefix = "arXiv",
    primaryClass = "hep-ex",
    reportNumber = "CERN-EP-2023-112",
    doi = "10.1103/PhysRevLett.132.081801",
    journal = "Phys. Rev. Lett.",
    volume = "132",
    number = "8",
    pages = "081801",
    year = "2024"
}

@article{ATLAS:2025obc,
    author = "Aad, Georges and others",
    collaboration = "ATLAS",
    title = "{Weakly supervised anomaly detection for resonant new physics in the dijet final state using proton-proton collisions at $\sqrt{s}=13$ TeV with the ATLAS detector}",
    eprint = "2502.09770",
    archivePrefix = "arXiv",
    primaryClass = "hep-ex",
    reportNumber = "CERN-EP-2025-002",
    month = "2",
    year = "2025"
}

@article{Freytsis:2023cjr,
    author = "Freytsis, Marat and Perelstein, Maxim and San, Yik Chuen",
    title = "{Anomaly detection in the presence of irrelevant features}",
    eprint = "2310.13057",
    archivePrefix = "arXiv",
    primaryClass = "hep-ph",
    doi = "10.1007/JHEP02(2024)220",
    journal = "JHEP",
    volume = "02",
    pages = "220",
    year = "2024"
}

@article{Dillon:2021gag,
    author = "Dillon, Barry M. and Kasieczka, Gregor and Olischlager, Hans and Plehn, Tilman and Sorrenson, Peter and Vogel, Lorenz",
    title = "{Symmetries, safety, and self-supervision}",
    eprint = "2108.04253",
    archivePrefix = "arXiv",
    primaryClass = "hep-ph",
    doi = "10.21468/SciPostPhys.12.6.188",
    journal = "SciPost Phys.",
    volume = "12",
    number = "6",
    pages = "188",
    year = "2022"
}

@article{Favaro:2023xdl,
    author = {Favaro, Luigi and Kr\"amer, Michael and Modak, Tanmoy and Plehn, Tilman and R\"uschkamp, Jan},
    title = "{Semi-visible jets, energy-based models, and self-supervision}",
    eprint = "2312.03067",
    archivePrefix = "arXiv",
    primaryClass = "hep-ph",
    doi = "10.21468/SciPostPhys.18.2.042",
    journal = "SciPost Phys.",
    volume = "18",
    number = "2",
    pages = "042",
    year = "2025"
}

@article{deFavereau:2013fsa,
    author = "de Favereau, J. and Delaere, C. and Demin, P. and Giammanco, A. and Lema\^\i{}tre, V. and Mertens, A. and Selvaggi, M.",
    collaboration = "DELPHES 3",
    title = "{DELPHES 3, A modular framework for fast simulation of a generic collider experiment}",
    eprint = "1307.6346",
    archivePrefix = "arXiv",
    primaryClass = "hep-ex",
    doi = "10.1007/JHEP02(2014)057",
    journal = "JHEP",
    volume = "02",
    pages = "057",
    year = "2014"
}

@article{Sjostrand:2014zea,
    author = {Sj\"ostrand, Torbj\"orn and Ask, Stefan and Christiansen, Jesper R. and Corke, Richard and Desai, Nishita and Ilten, Philip and Mrenna, Stephen and Prestel, Stefan and Rasmussen, Christine O. and Skands, Peter Z.},
    title = "{An introduction to PYTHIA 8.2}",
    eprint = "1410.3012",
    archivePrefix = "arXiv",
    primaryClass = "hep-ph",
    reportNumber = "LU-TP-14-36, MCNET-14-22, CERN-PH-TH-2014-190, FERMILAB-PUB-14-316-CD, DESY-14-178, SLAC-PUB-16122",
    doi = "10.1016/j.cpc.2015.01.024",
    journal = "Comput. Phys. Commun.",
    volume = "191",
    pages = "159--177",
    year = "2015"
}


\appendix
\subsection{Median $Z$-score for different signal injections}\label{app:1}
\begin{figure*}[htbp]
    \centering
    \includegraphics[width=0.45\linewidth]{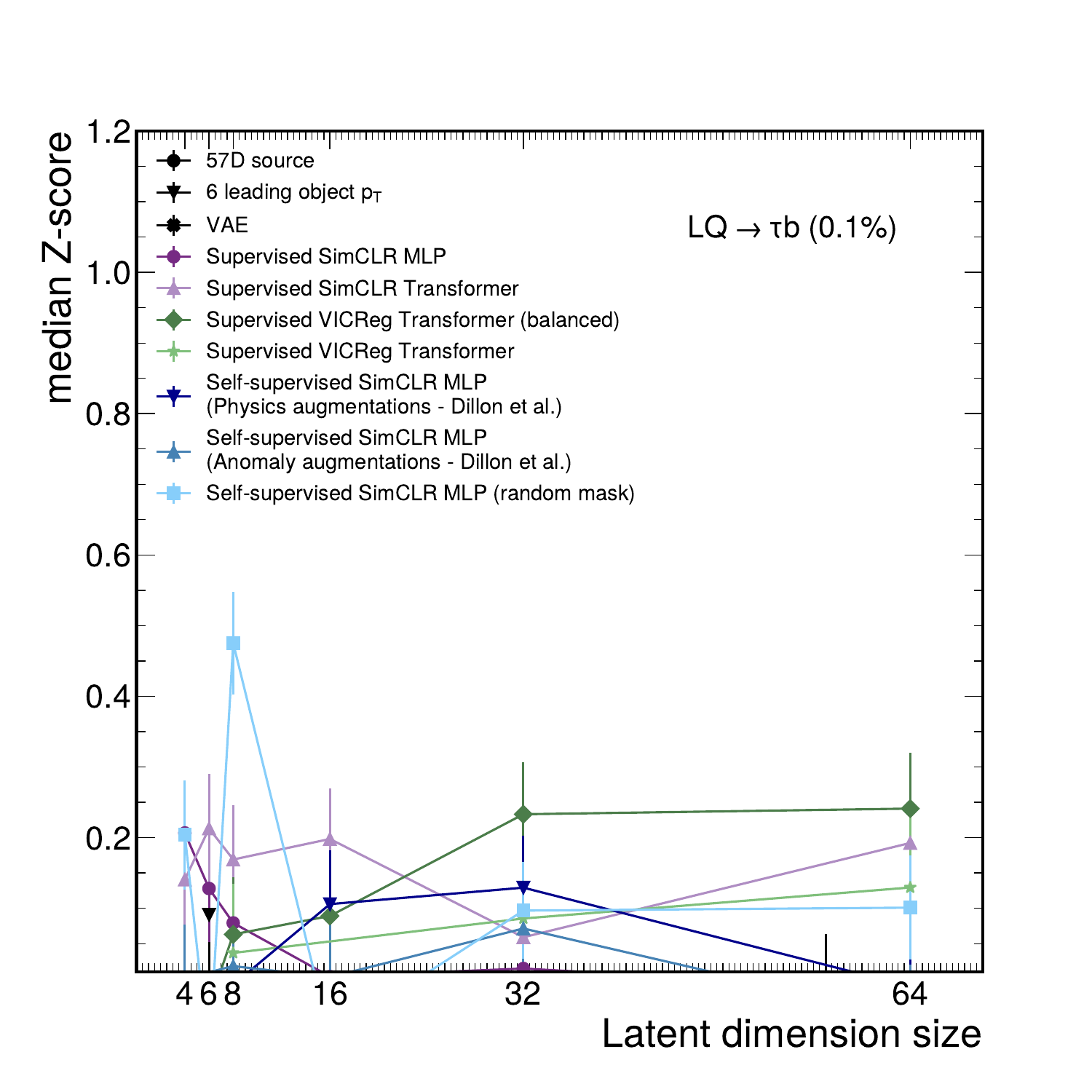}
    \includegraphics[width=0.45\linewidth]{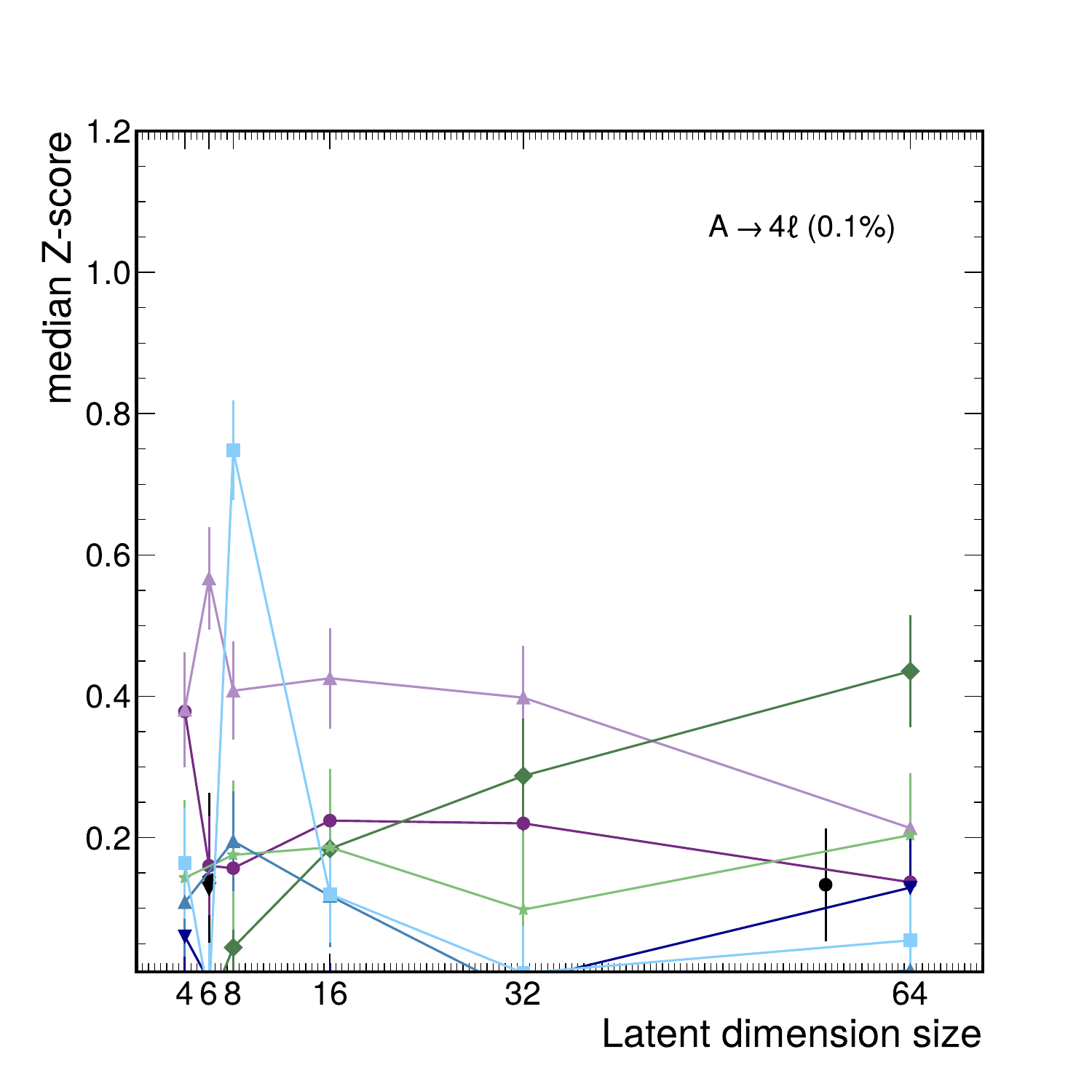}\\
    \hspace{0.02cm}
    \includegraphics[width=0.45\linewidth]{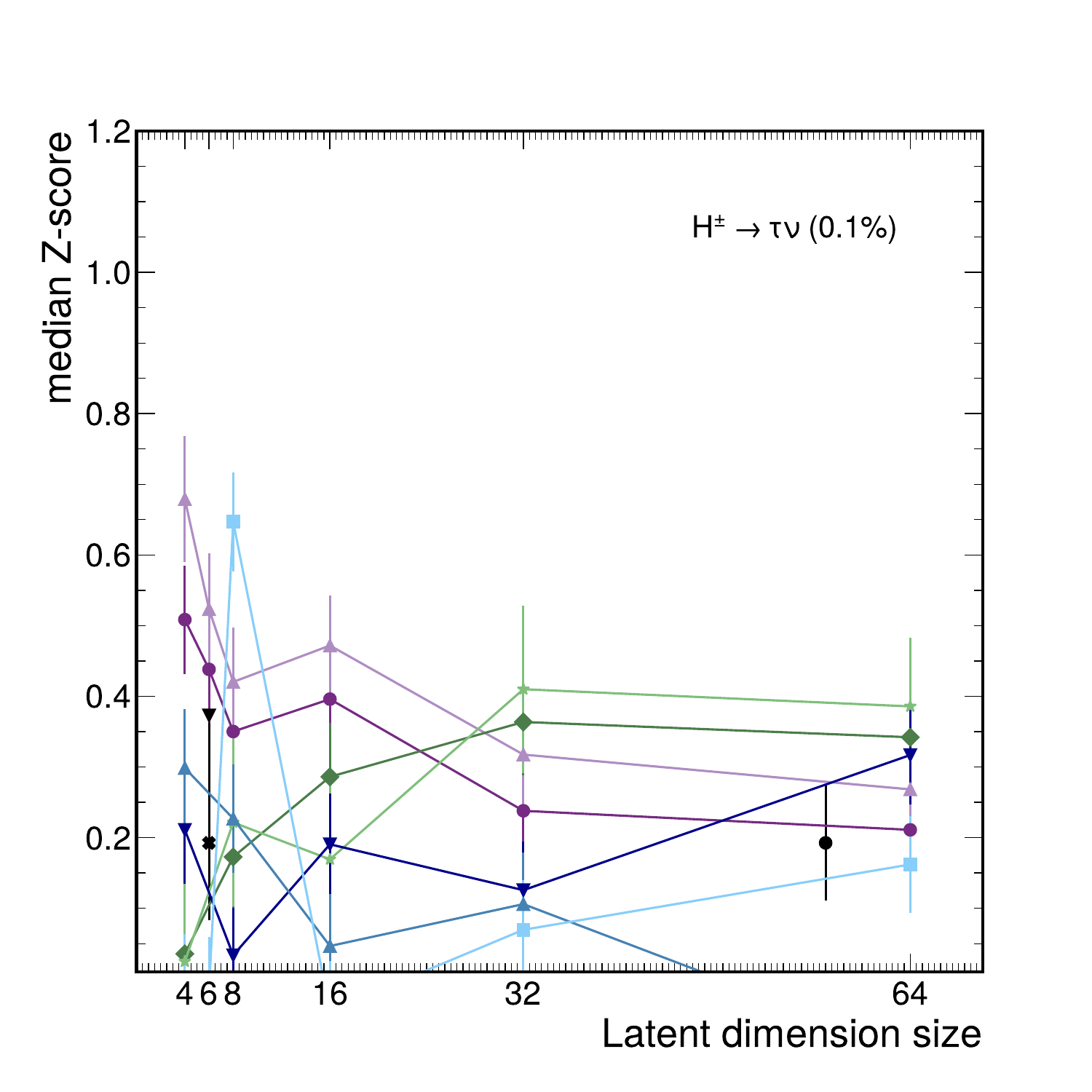}
    \includegraphics[width=0.45\linewidth]{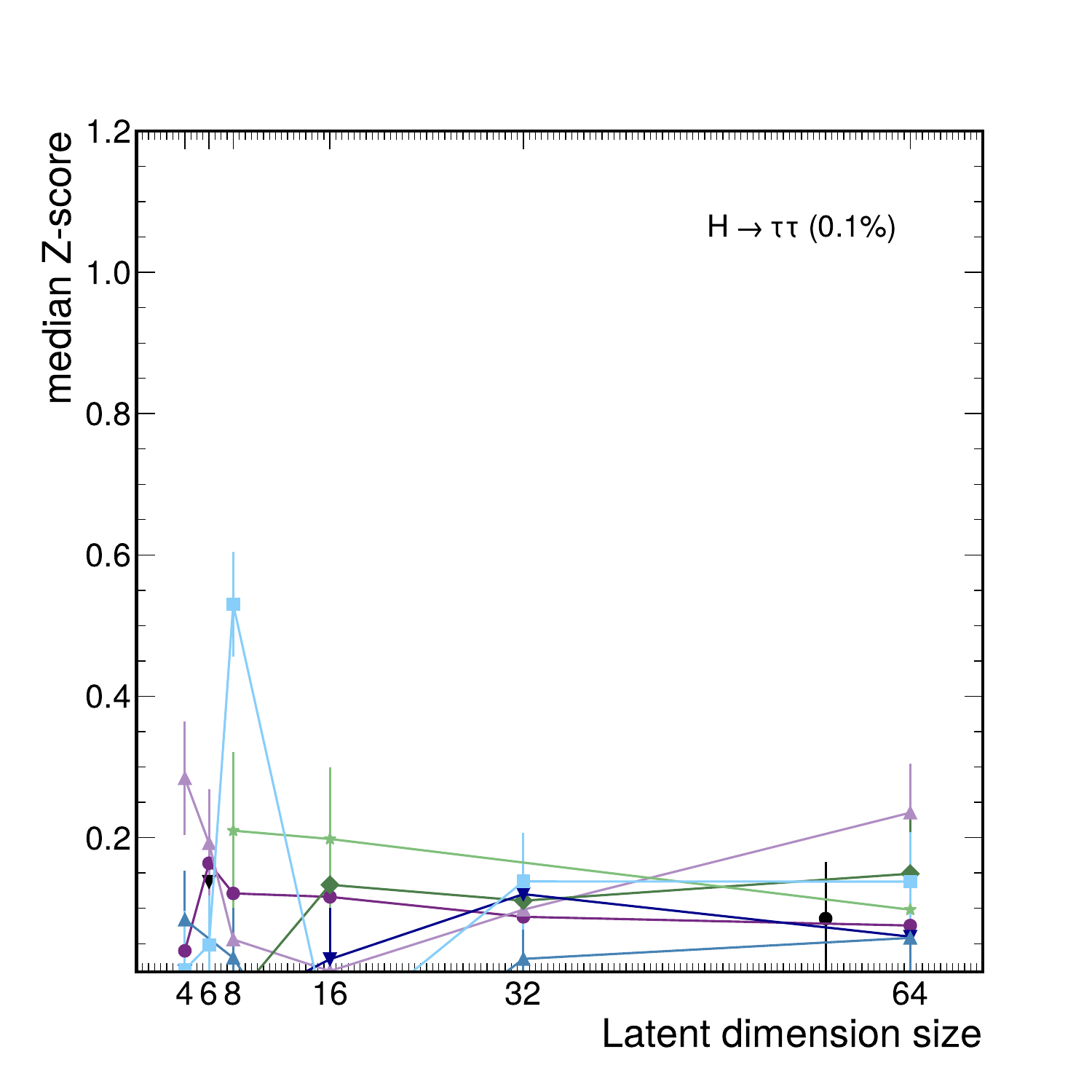}
    
    \caption{The median observed $Z$-score as a function of the feature embedding size after injecting a signal into the SM background pseudo-dataset, corresponding to $0.1\%$ of the total integral. Results are shown for the $LQ \rightarrow \tau b$ (upper left), $A \rightarrow 4\ell$ (upper right), $H^{\pm} \rightarrow \tau \nu$ (bottom left), and $H \rightarrow \tau \tau$ (bottom right) signals. The median $Z$-score is presented for different embedding models: the MLP- and Transformer-based models trained with supervision and SimCLR loss (dark and light purple, respectively),the Transformer-based model trained with supervision and VICReg loss using either balanced background classes (dark green) or background classes weighted to match the composition expected in data (light green), the MLP-based model trained self-supervised with a SimCLR loss with 50\% random masking (light blue), physics-inspired augmentations from~\cite{Dillon:2021gag} (medium blue), or physics- and anomaly-inspired augmentations from~\cite{Dillon:2021gag} (dark blue).
    Additionally, we compare these results with three baseline embeddings: 57D source (black circles), the 6D leading object $p_T$ (black triangles), and the 6D VAE (black crosses). The typical $3\sigma$ level for evidence and $5\sigma$ level for discovery are reported in dashed lines. Where empirical Z-scores cannot be computed ($Z>3\sigma$) we rely on the asymptotic formula.}
    \label{fig:zscores1}
\end{figure*}

\begin{figure*}[htbp]
    \centering
    \includegraphics[width=0.45\linewidth]{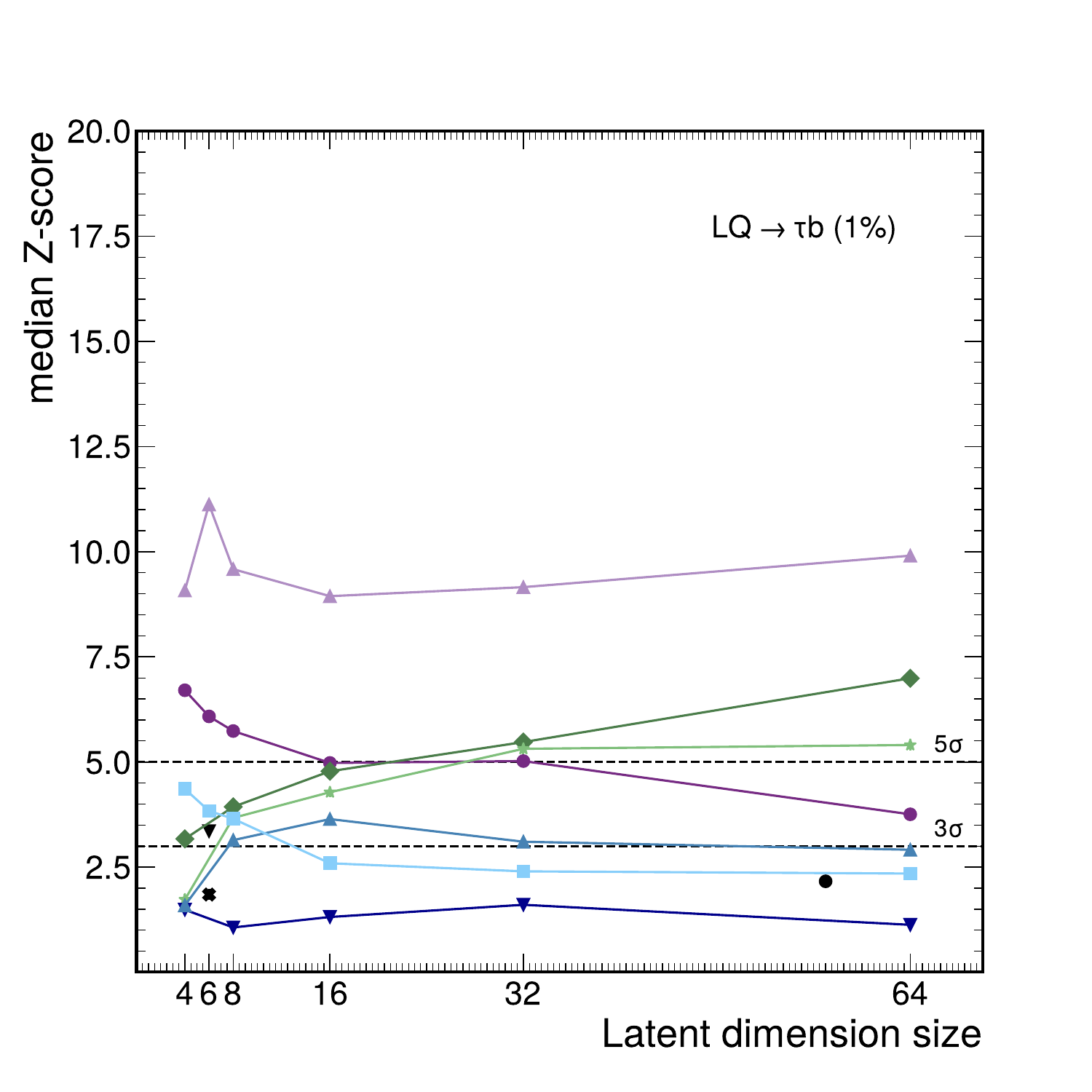}
    \includegraphics[width=0.45\linewidth]{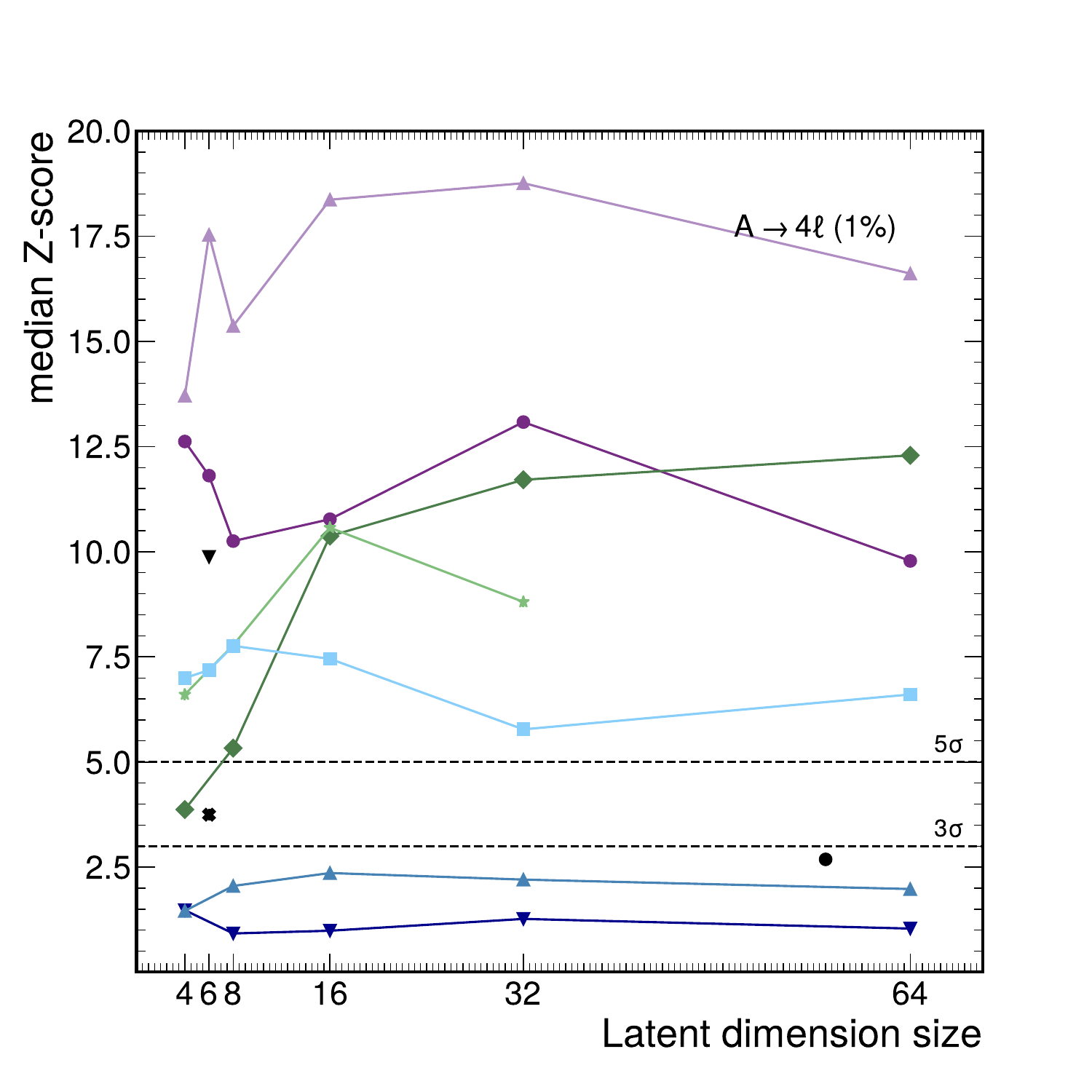}\\
    \hspace{0.02cm}
    \includegraphics[width=0.45\linewidth]{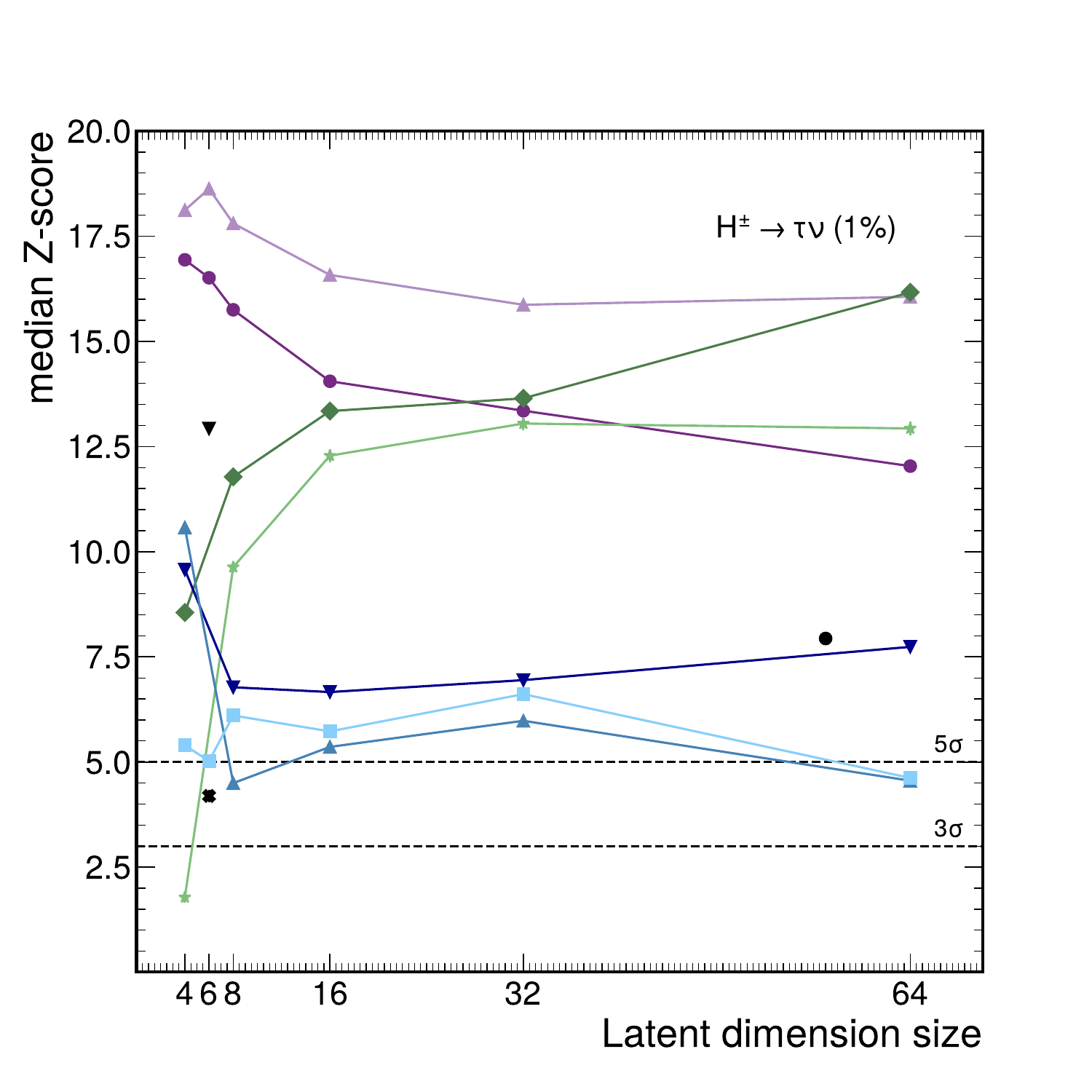}
    \includegraphics[width=0.45\linewidth]{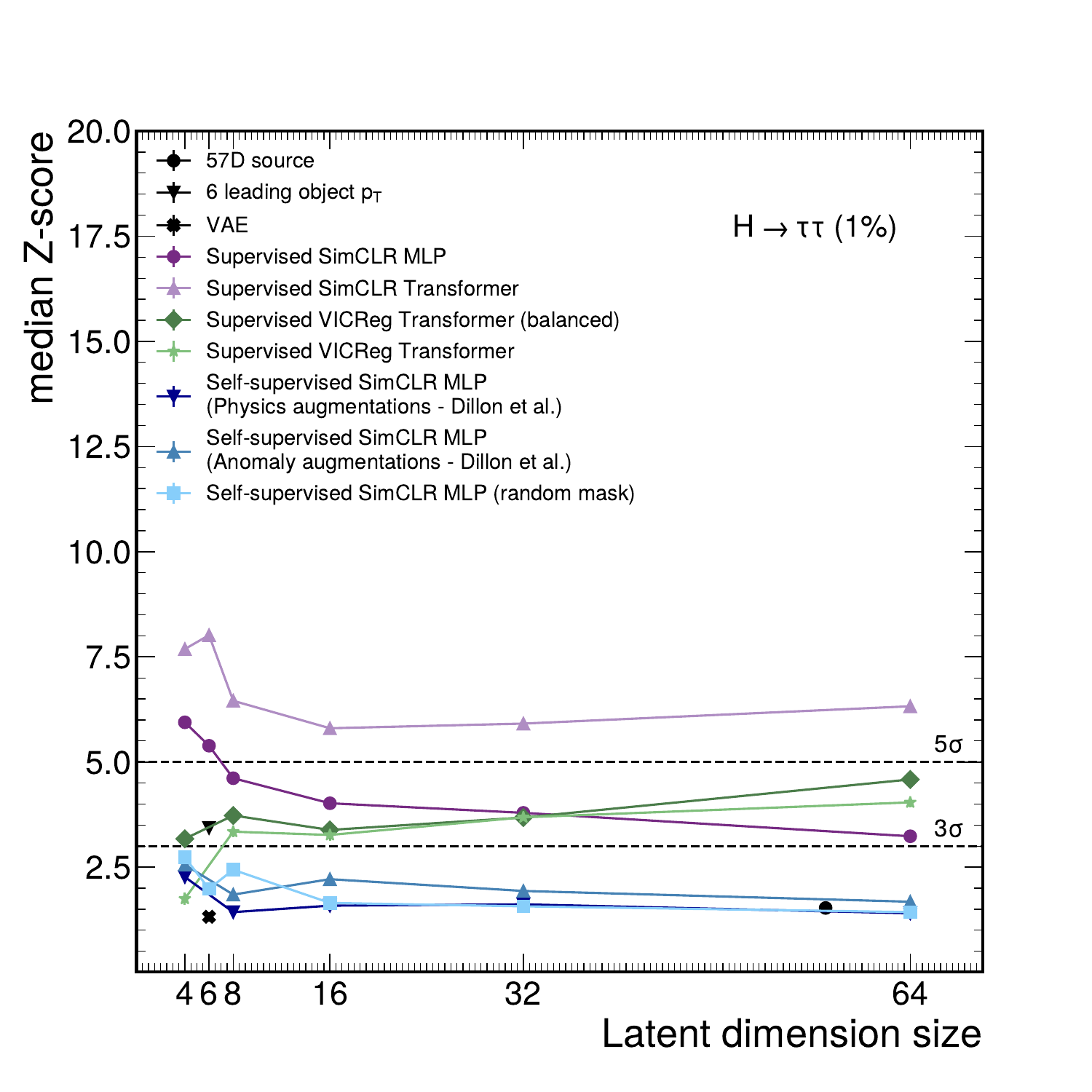}
    
    \caption{The median observed $Z$-score as a function of the feature embedding size after injecting a signal into the SM background pseudo-dataset, corresponding to $1\%$ of the total integral. Results are shown for the $LQ \rightarrow \tau b$ (upper left), $A \rightarrow 4\ell$ (upper right), $H^{\pm} \rightarrow \tau \nu$ (bottom left), and $H \rightarrow \tau \tau$ (bottom right) signals. The median $Z$-score is presented for different embedding models: the MLP- and Transformer-based models trained with supervision and SimCLR loss (dark and light purple, respectively),the Transformer-based model trained with supervision and VICReg loss using either balanced background classes (dark green) or background classes weighted to match the composition expected in data (light green), the MLP-based model trained self-supervised with a SimCLR loss with 50\% random masking (light blue), physics-inspired augmentations from~\cite{Dillon:2021gag} (medium blue), or physics- and anomaly-inspired augmentations from~\cite{Dillon:2021gag} (dark blue).
    Additionally, we compare these results with three baseline embeddings: 57D source (black circles), the 6D leading object $p_T$ (black triangles), and the 6D VAE (black crosses). The typical $3\sigma$ level for evidence and $5\sigma$ level for discovery are reported in dashed lines. Where empirical Z-scores cannot be computed ($Z>3\sigma$) we rely on the asymptotic formula.}
    \label{fig:zscores2}
\end{figure*}

In this appendix we report additional examples of observed median $Z$-score as the ones showed in Figure~\ref{fig:pvals} in the main body of this paper. Figure~\ref{fig:zscores1} is obtained from a $0.1\%$ signal injection whereas Figure~\ref{fig:zscores2} from $1\%$ signal injection.

\subsection{Discovery powers for $0.5$ and $1 \%$ signal injections}
\begin{figure*}[htbp]
    \centering
    \includegraphics[width=0.45\linewidth]{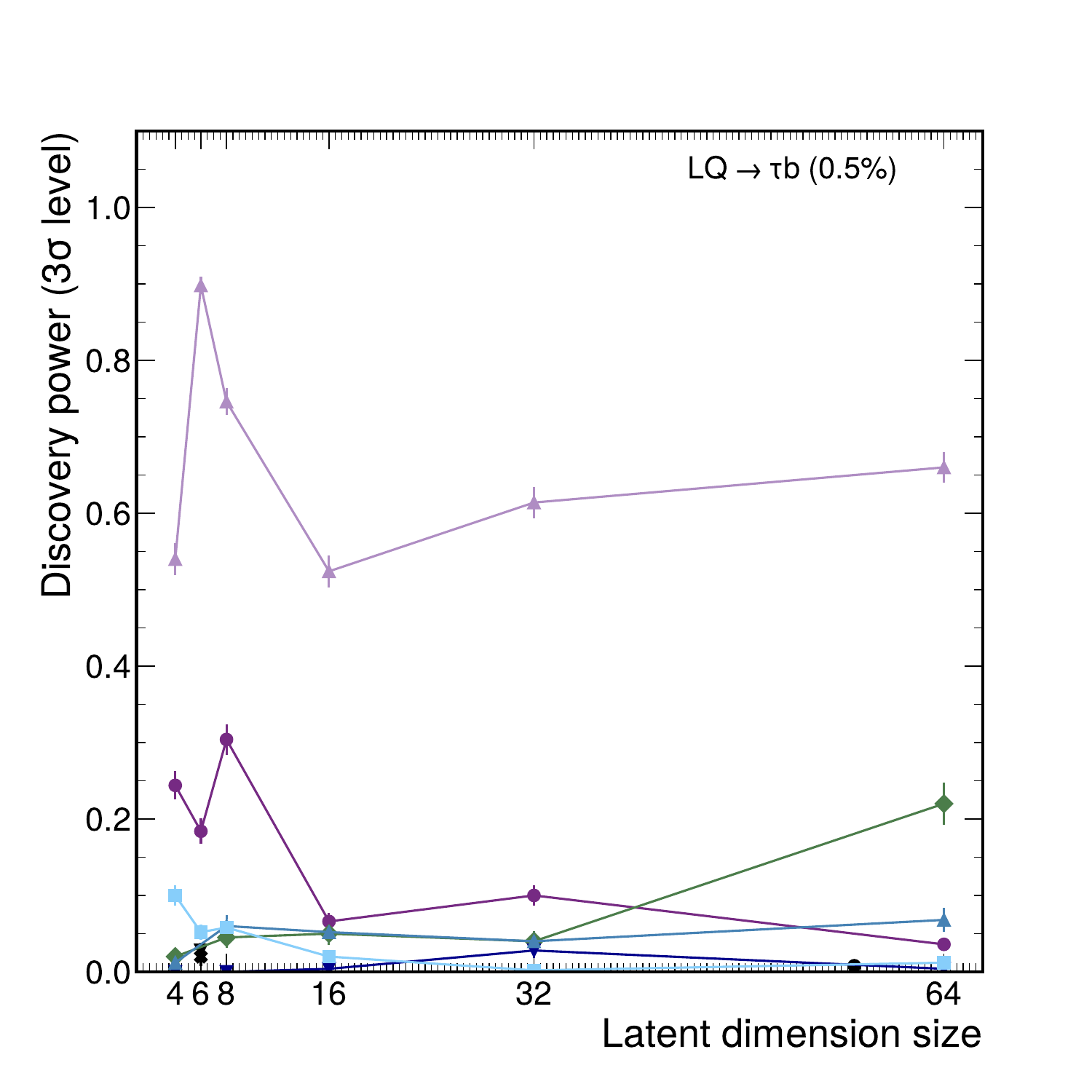}
    \includegraphics[width=0.45\linewidth]{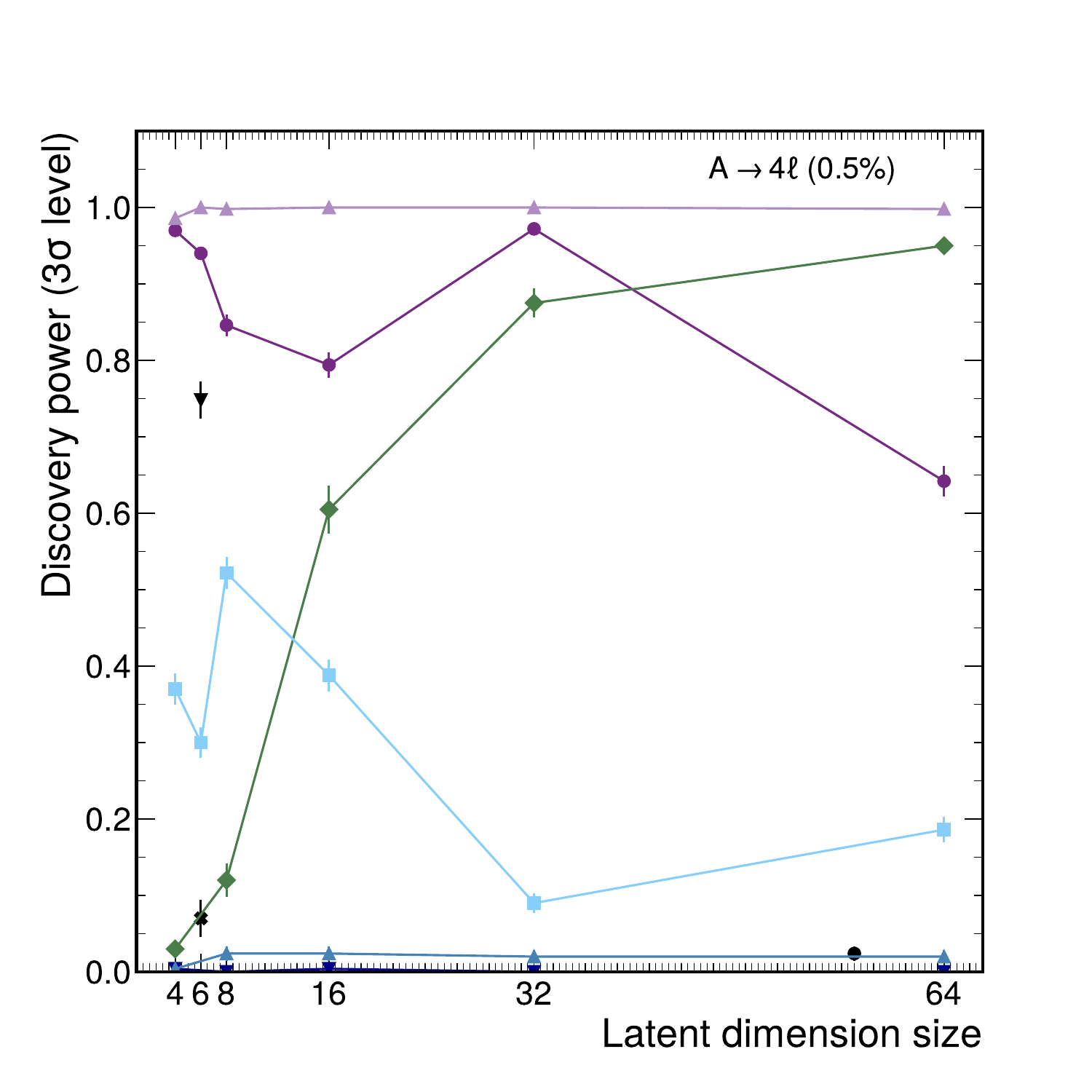}
    \includegraphics[width=0.45\linewidth]{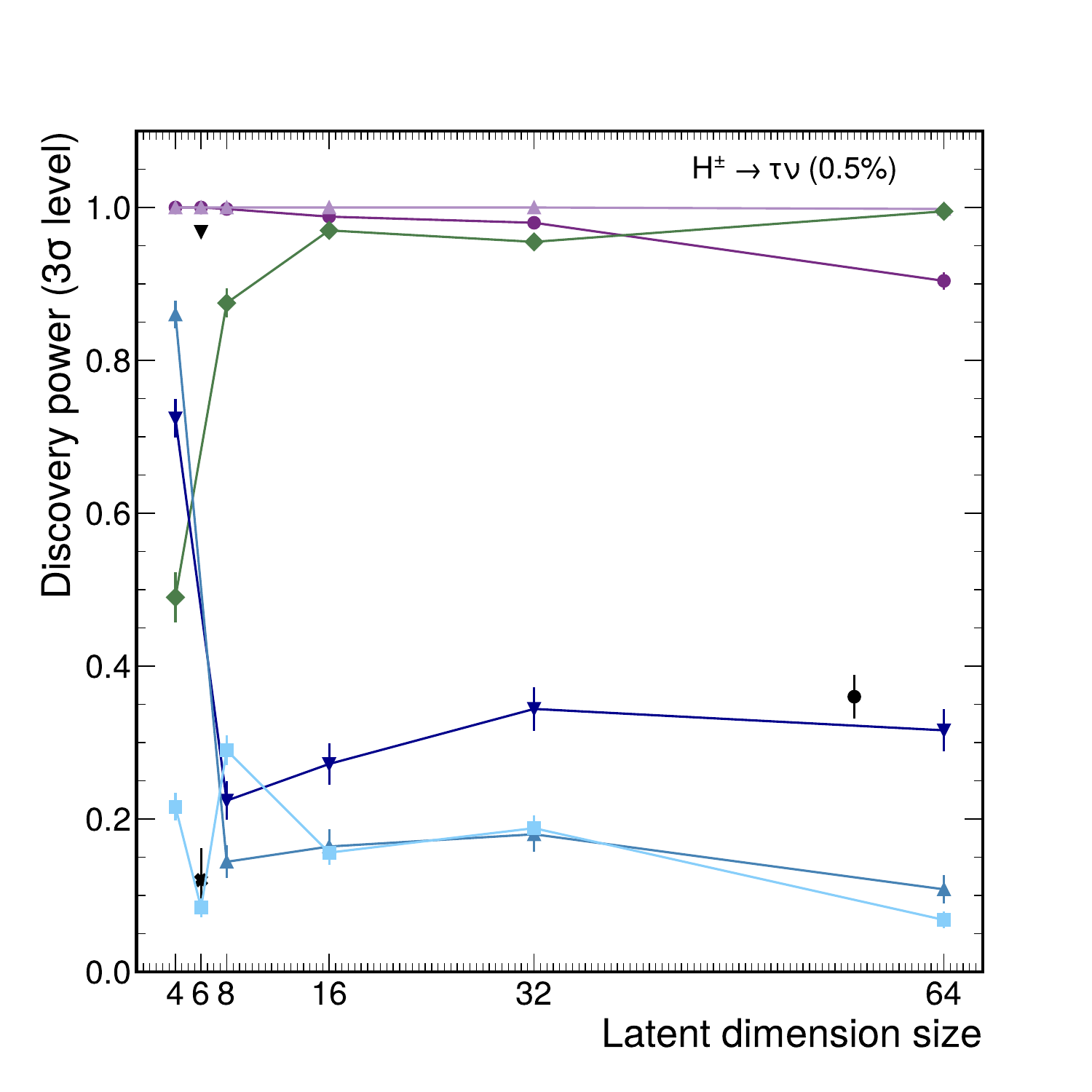}
    \includegraphics[width=0.45\linewidth]{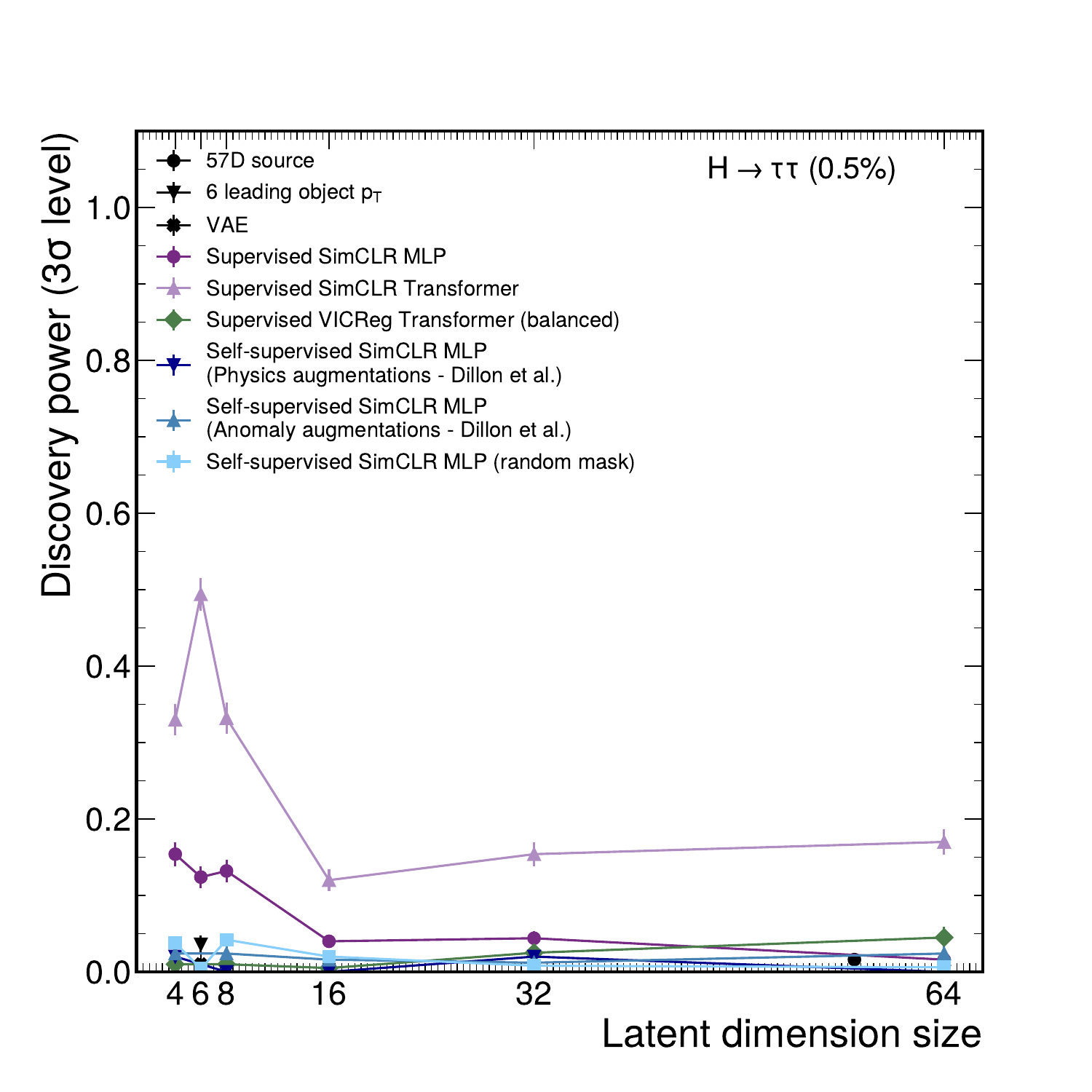}
    \caption{The empirical discovery power ($3\sigma$) as a function of the feature embedding size after injecting a signal into the SM background pseudodataset, corresponding to $0.5\%$ of the total integral. Results are shown for the $LQ \rightarrow \tau b$ (upper left), $A \rightarrow 4\ell$ (upper right), $H^{\pm} \rightarrow \tau \nu$ (bottom left), and $H \rightarrow \tau \tau$ (bottom right) signals. The discovery power is presented for different models: the MLP- and Transformer-based models trained with supervision and SimCLR loss (dark and light purple, respectively),the Transformer-based model trained with supervision and VICReg loss using balanced background classes (dark green), the MLP-based model trained self-supervised with a SimCLR loss with 50\% random masking (light blue), physics-inspired augmentations from~\cite{Dillon:2021gag} (medium blue), or physics- and anomaly-inspired augmentations from~\cite{Dillon:2021gag} (dark blue). Additionally, we compare these results with three baseline embeddings: 57D source (black circles), the 6D leading object $p_T$ (black triangles), and the 6D VAE (black crosses). The typical $3\sigma$ level for evidence and $5\sigma$ level for discovery are reported in dashed lines. }
    \label{fig:pvals2}
\end{figure*}

\begin{figure*}[htbp]
    \centering
    \includegraphics[width=0.45\linewidth]{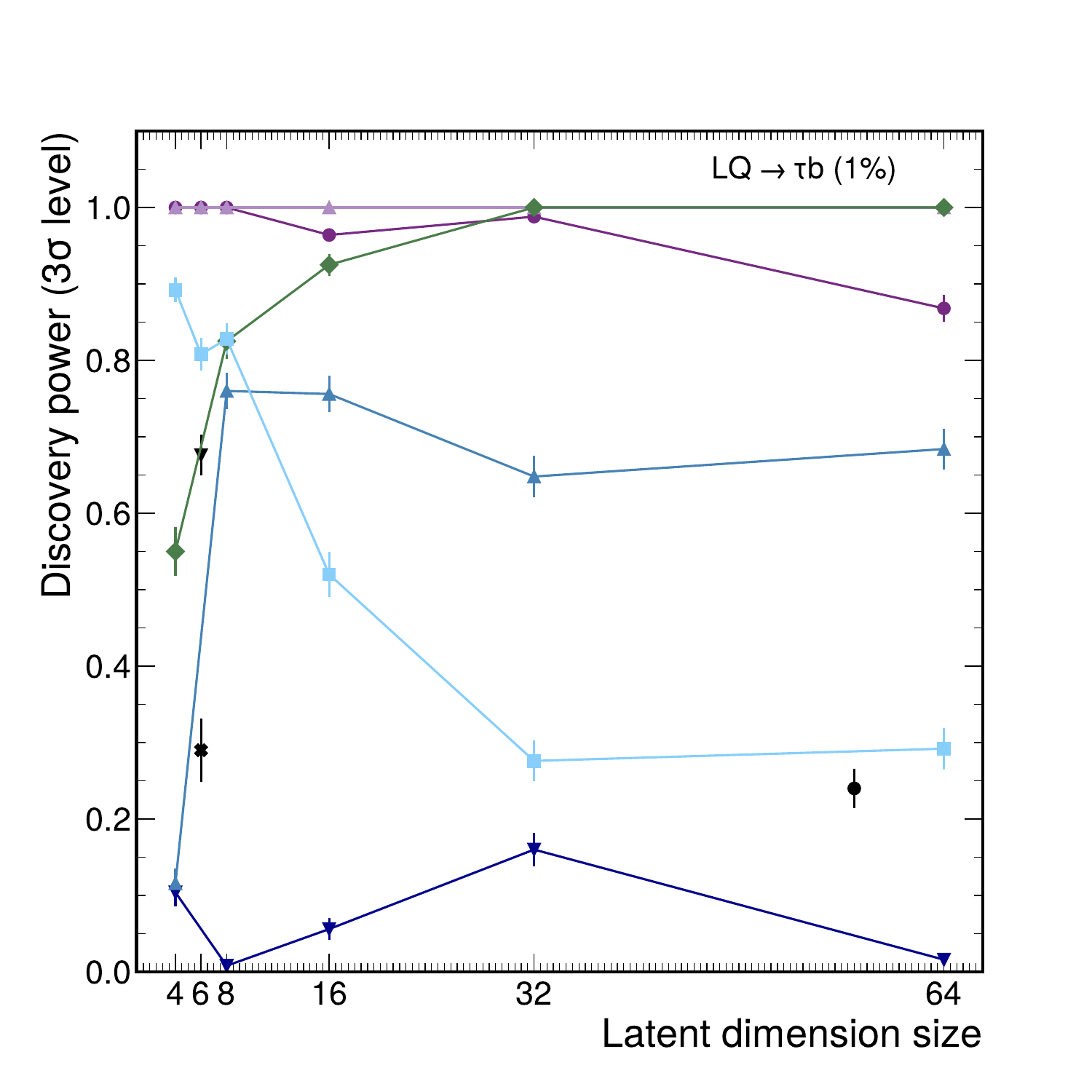}
    \includegraphics[width=0.45\linewidth]{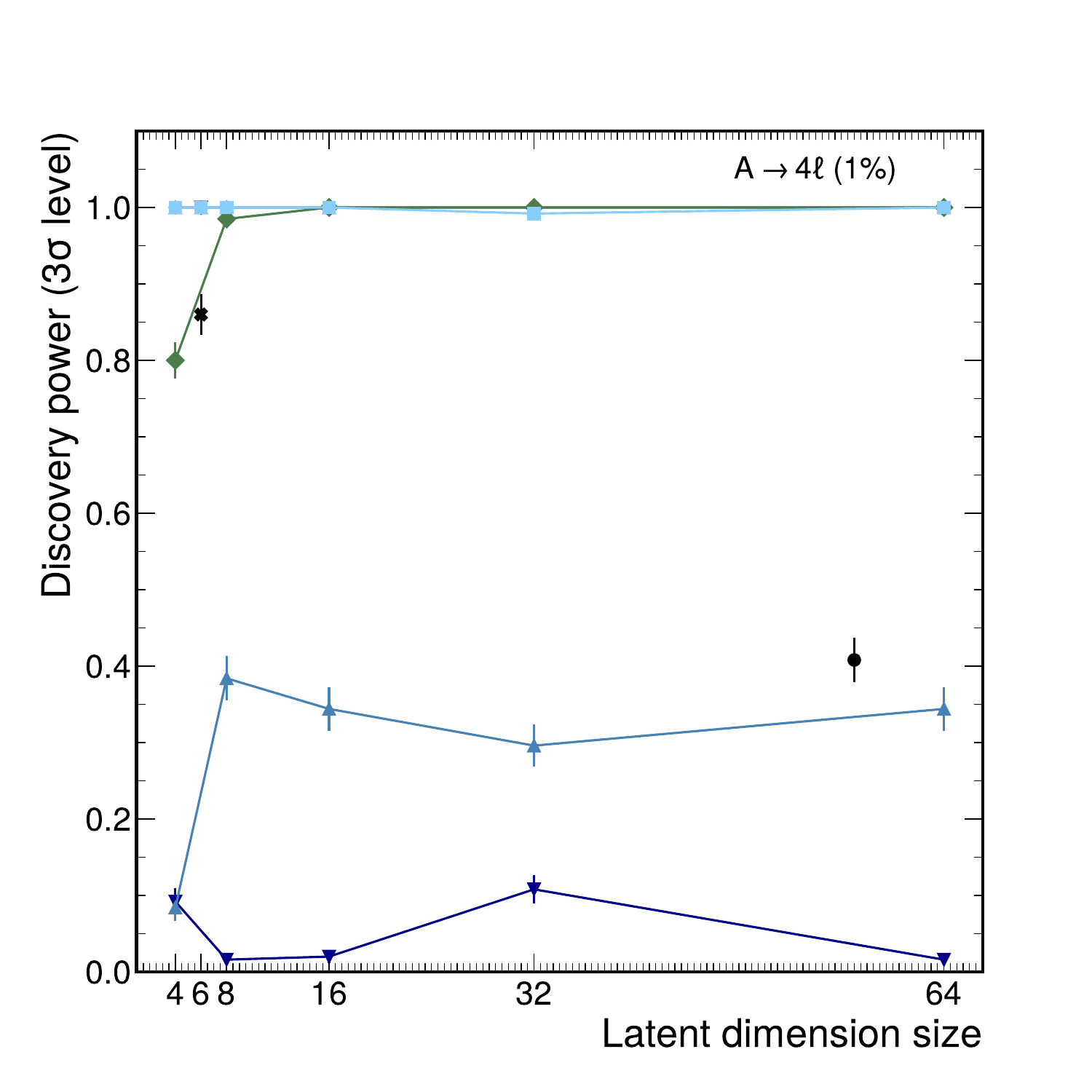}
    \includegraphics[width=0.45\linewidth]{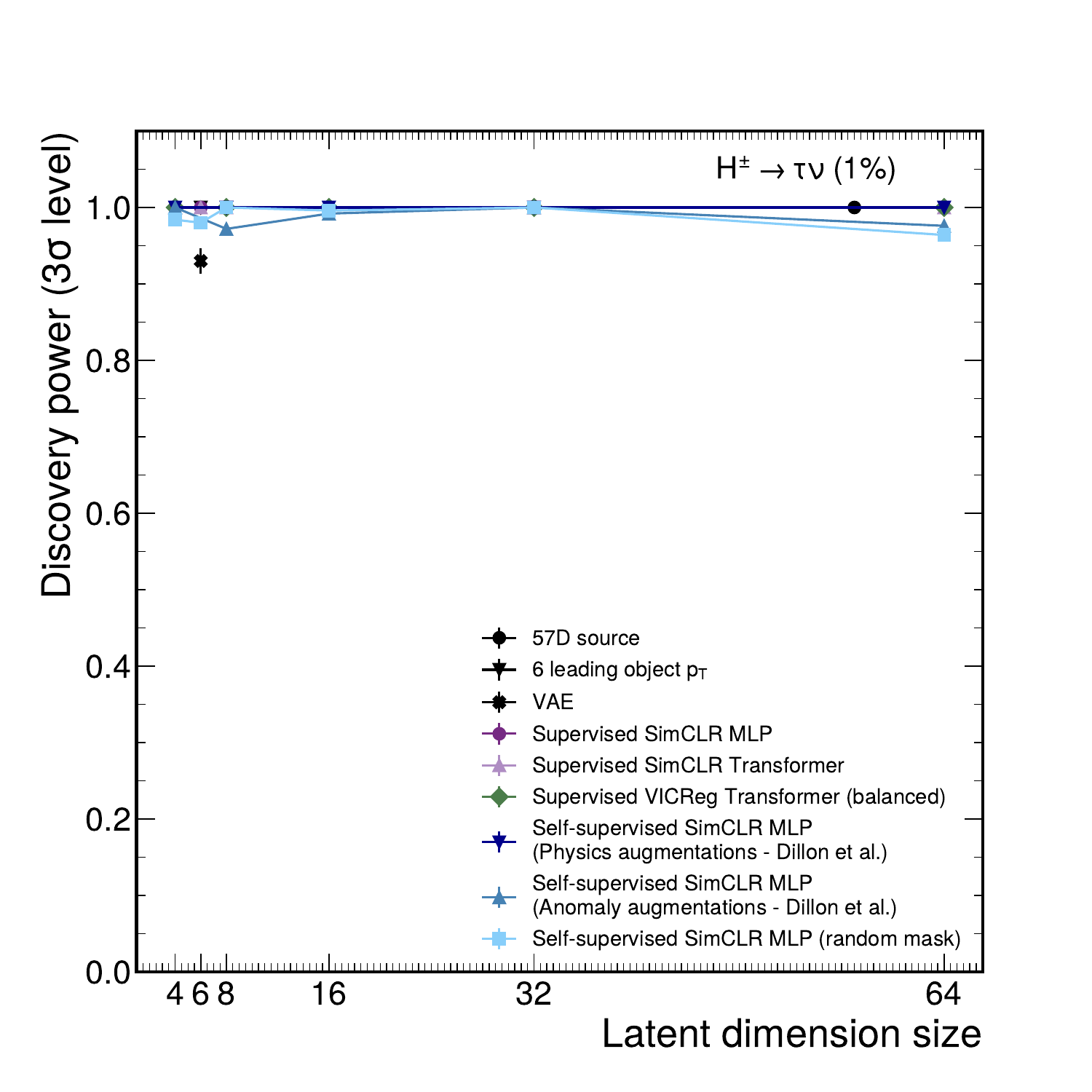}
    \includegraphics[width=0.45\linewidth]{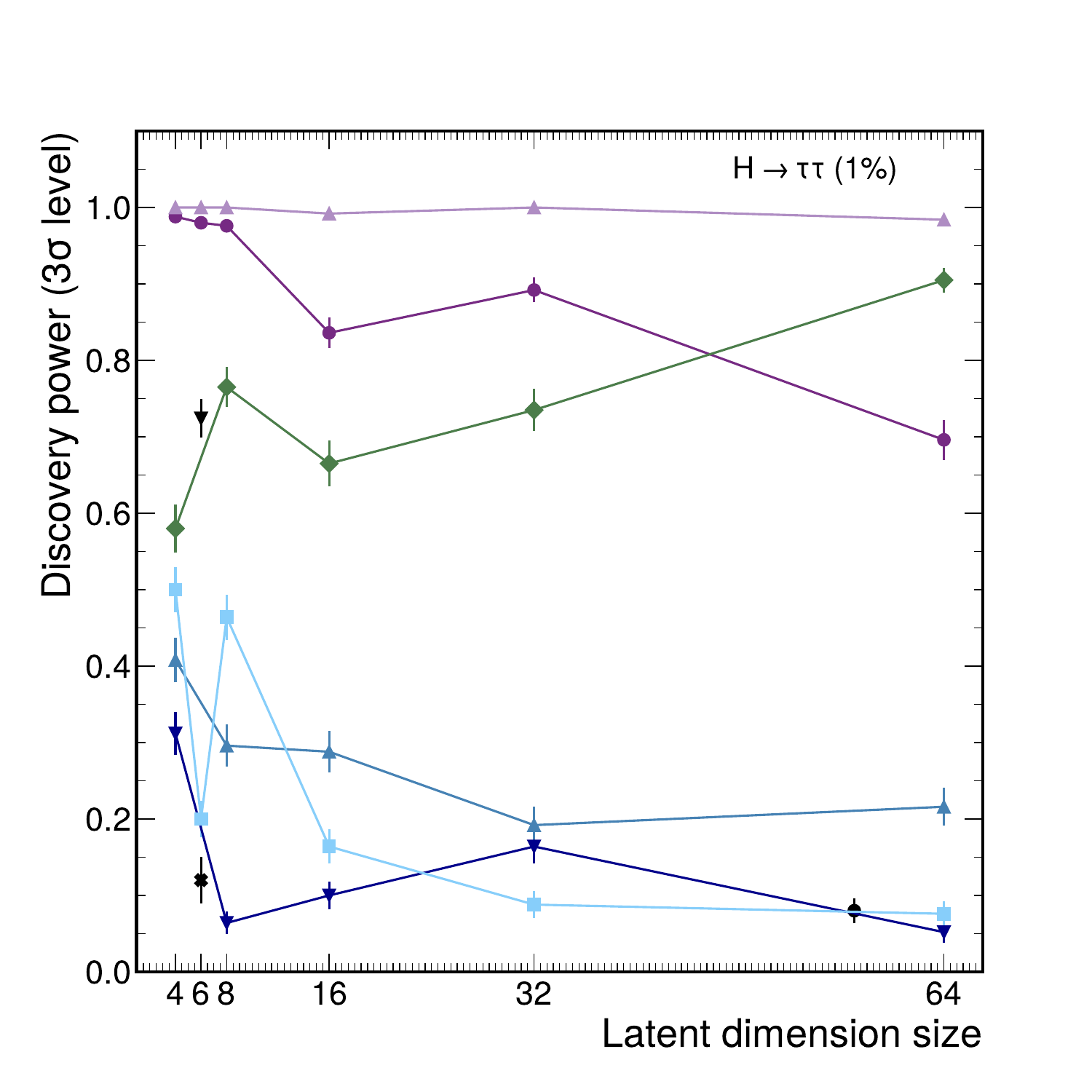}
    \caption{The empirical discovery power ($3\sigma$) as a function of the feature embedding size after injecting a signal into the SM background pseudodataset, corresponding to $1\%$ of the total integral. Results are shown for the $LQ \rightarrow \tau b$ (upper left), $A \rightarrow 4\ell$ (upper right), $H^{\pm} \rightarrow \tau \nu$ (bottom left), and $H \rightarrow \tau \tau$ (bottom right) signals. The discovery power is presented for different models: the MLP- and Transformer-based models trained with supervision and SimCLR loss (dark and light purple, respectively),the Transformer-based model trained with supervision and VICReg loss using balanced background classes (dark green), the MLP-based model trained self-supervised with a SimCLR loss with 50\% random masking (light blue), physics-inspired augmentations from~\cite{Dillon:2021gag} (medium blue), or physics- and anomaly-inspired augmentations from~\cite{Dillon:2021gag} (dark blue). Additionally, we compare these results with three baseline embeddings: 57D source (black circles), the 6D leading object $p_T$ (black triangles), and the 6D VAE (black crosses). The typical $3\sigma$ level for evidence and $5\sigma$ level for discovery are reported in dashed lines.}
    \label{fig:pvals1}
\end{figure*}

Figures~\ref{fig:pvals2} and~\ref{fig:pvals1} report the statistical power for signal discovery at $3\sigma$ level obtained using the NPLM test over $0.5$ and $1 \%$ signal injections respectively. The four panels in each figure report the power curves for the four signal benchmarks considered in this work: $LQ \rightarrow \tau b$ (upper left), $A \rightarrow 4\ell$ (upper right), $H^{\pm} \rightarrow \tau \nu$ (bottom left), and $H \rightarrow \tau \tau$ (bottom right) signals. Different curves represent the power of different embeddings as a function of the dimensionality of the embedding. The statistical power at $3\sigma$ represent the probability of observing a $p$-value lower or equal to $1.35 \cdot 10^{-3}$, namely a $Z$-score greater or equal to 3.

\subsection{UMAPs}\label{app:3}
\begin{figure*}[htbp]
    \centering
\includegraphics[width=0.32\linewidth]{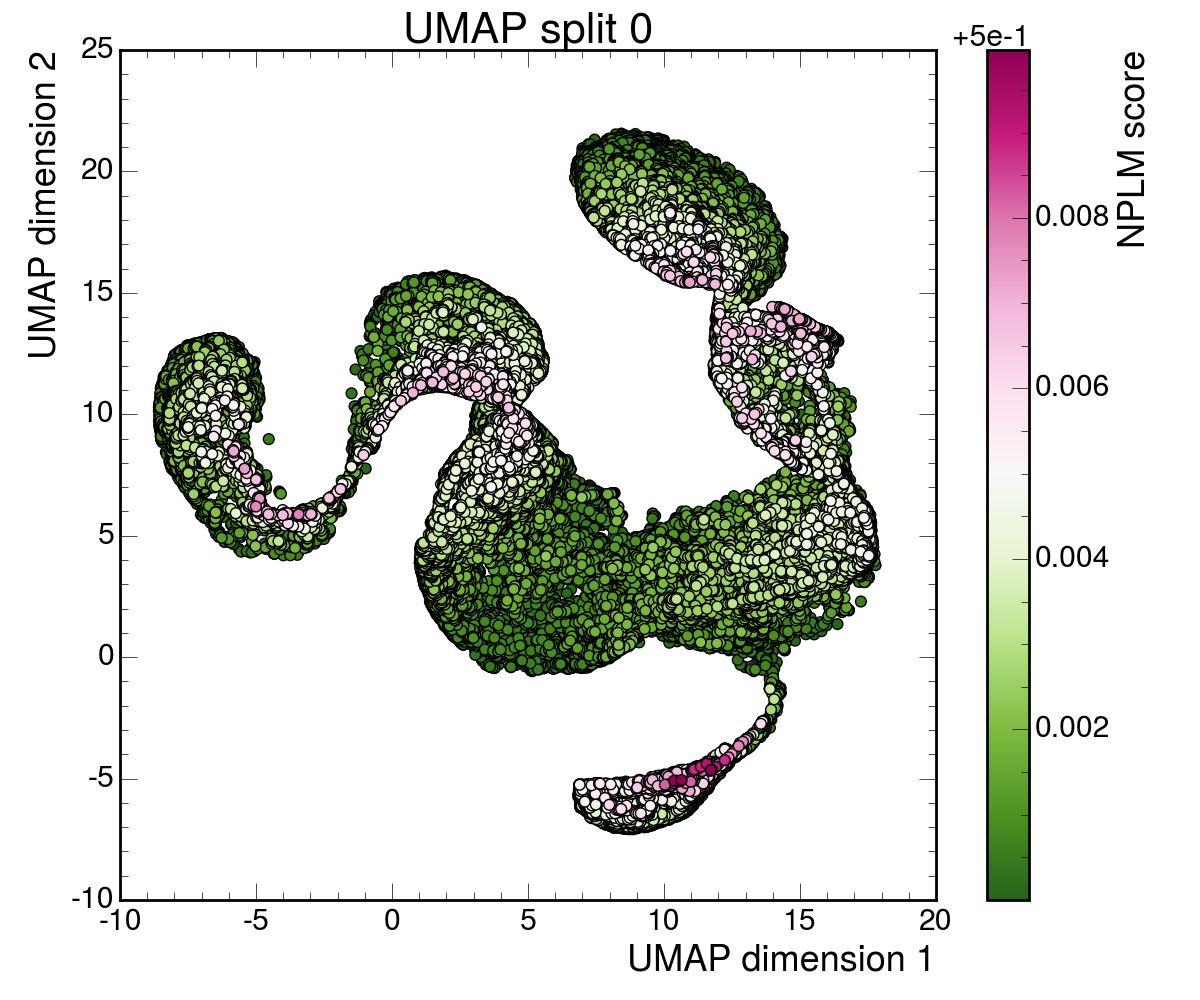}
\includegraphics[width=0.32\linewidth]{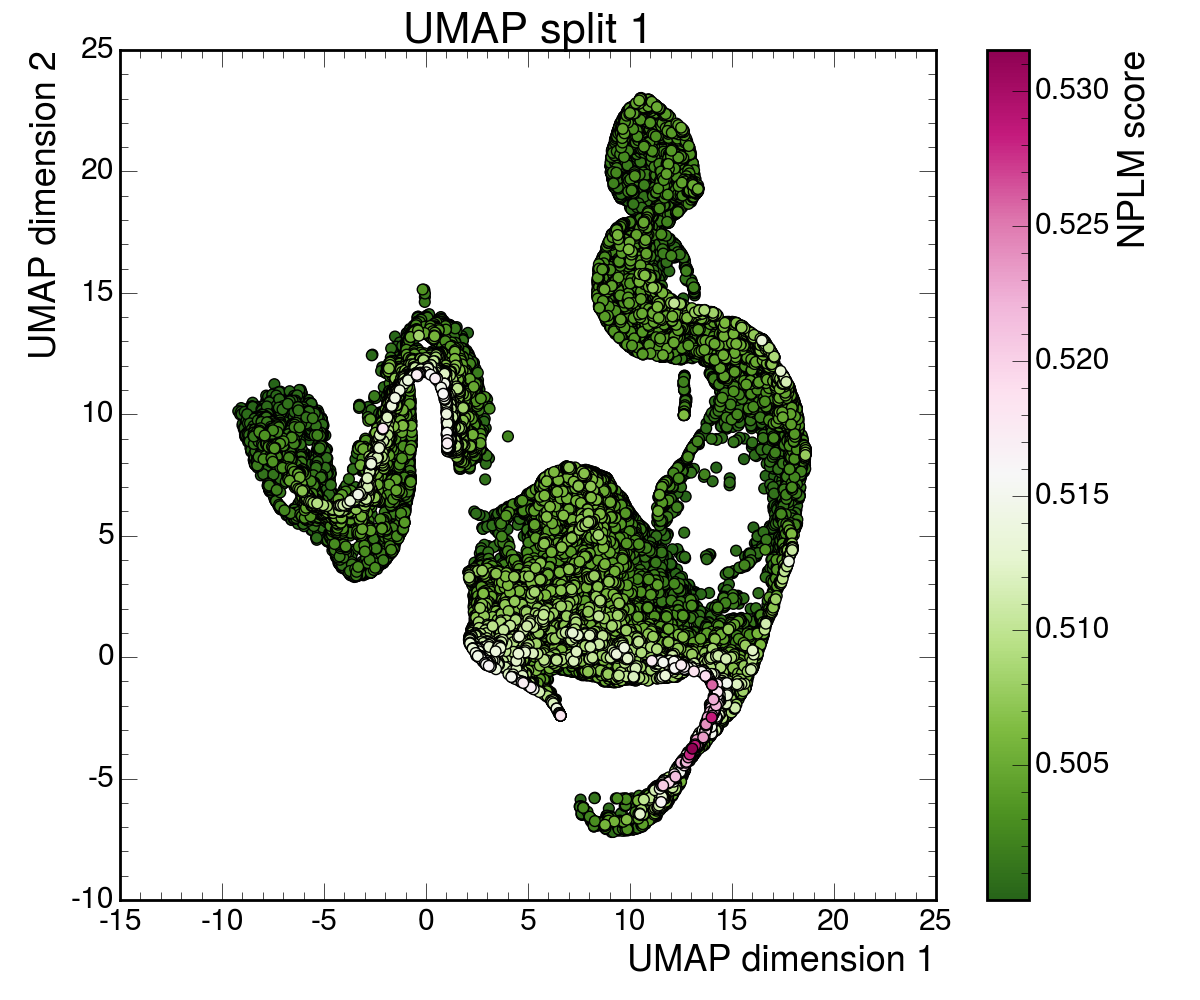}
\includegraphics[width=0.32\linewidth]{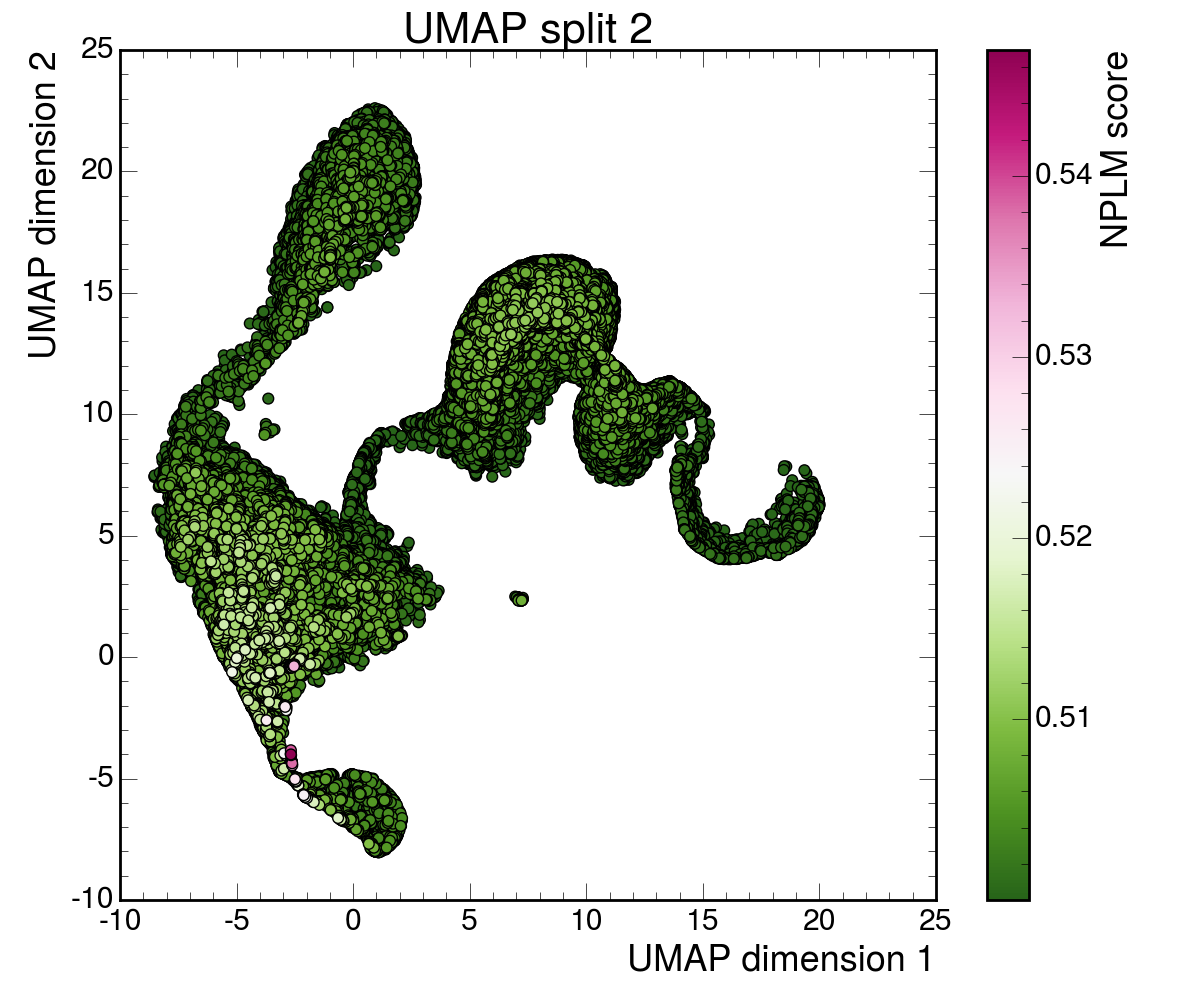}\\
\includegraphics[width=0.32\linewidth]{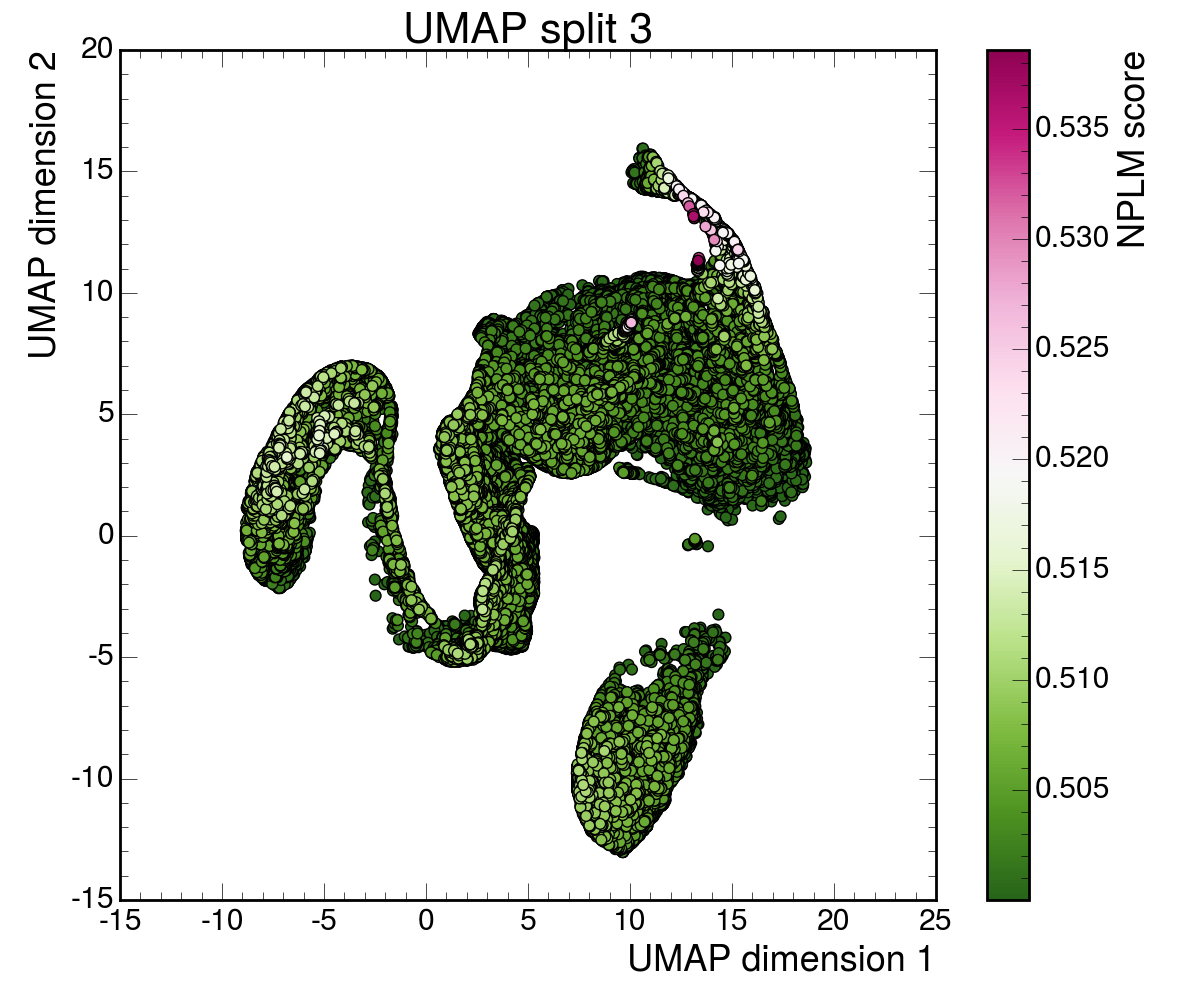}
\includegraphics[width=0.32\linewidth]{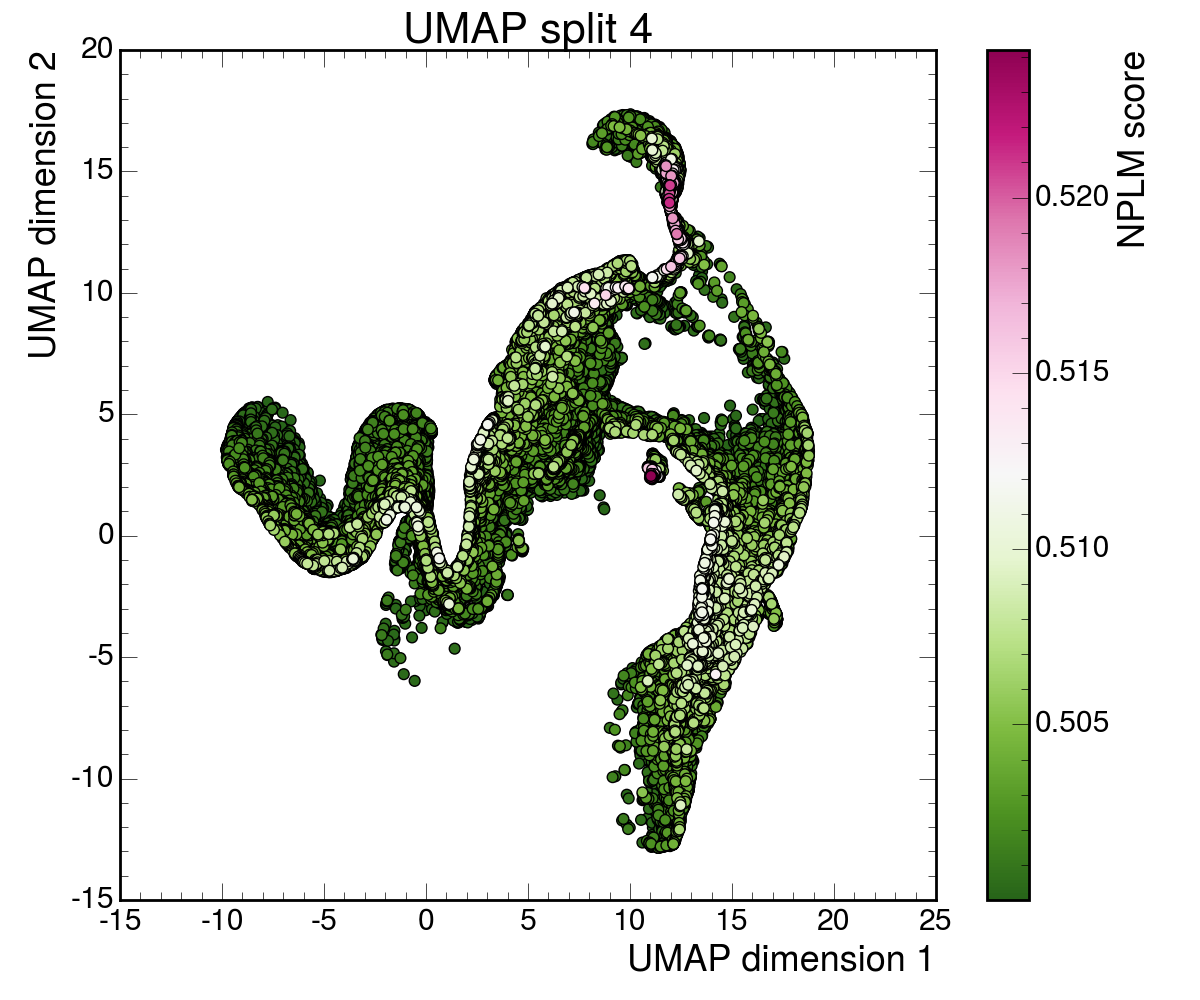}
\includegraphics[width=0.32\linewidth]{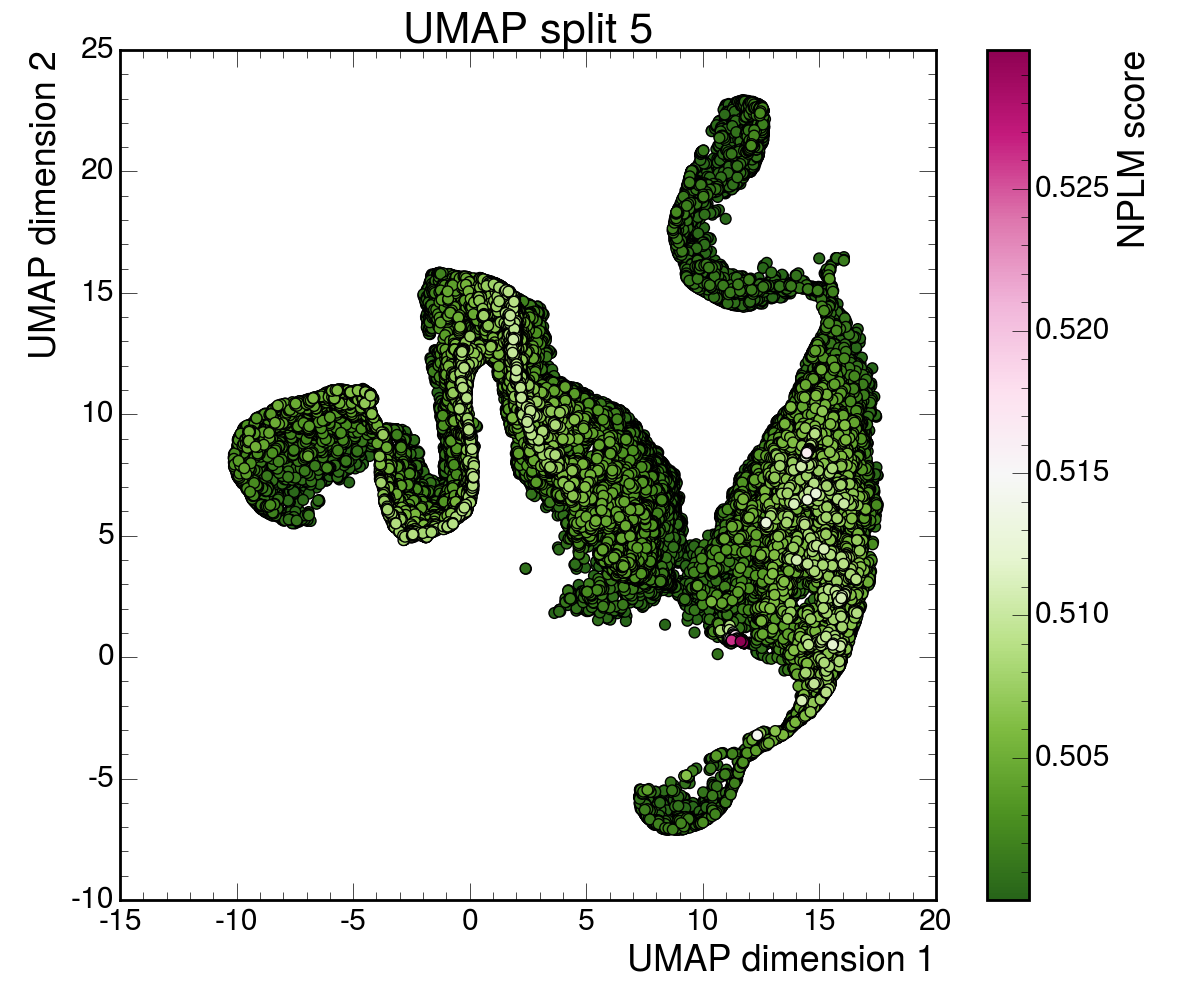}\\
\includegraphics[width=0.32\linewidth]{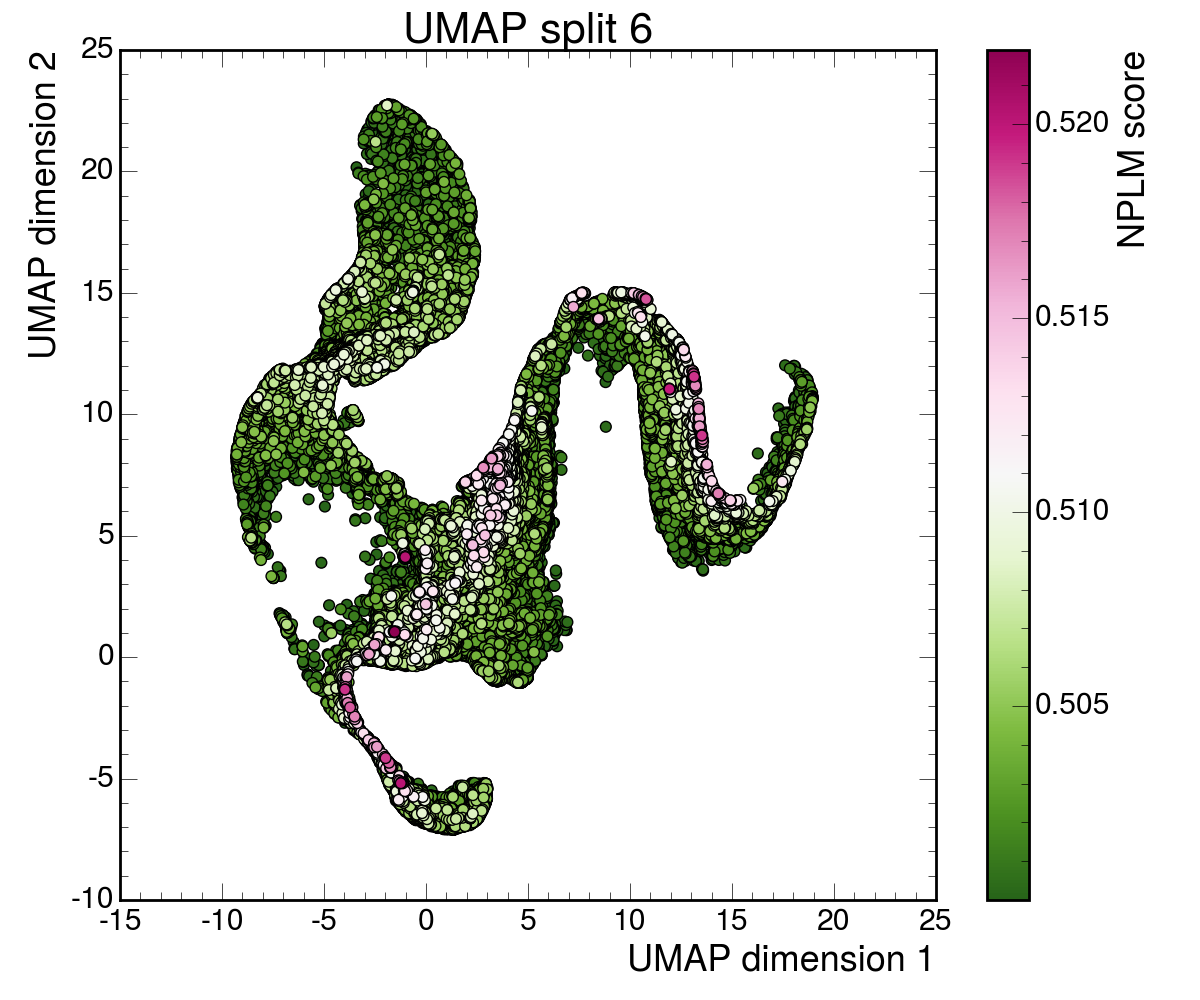}
\includegraphics[width=0.32\linewidth]{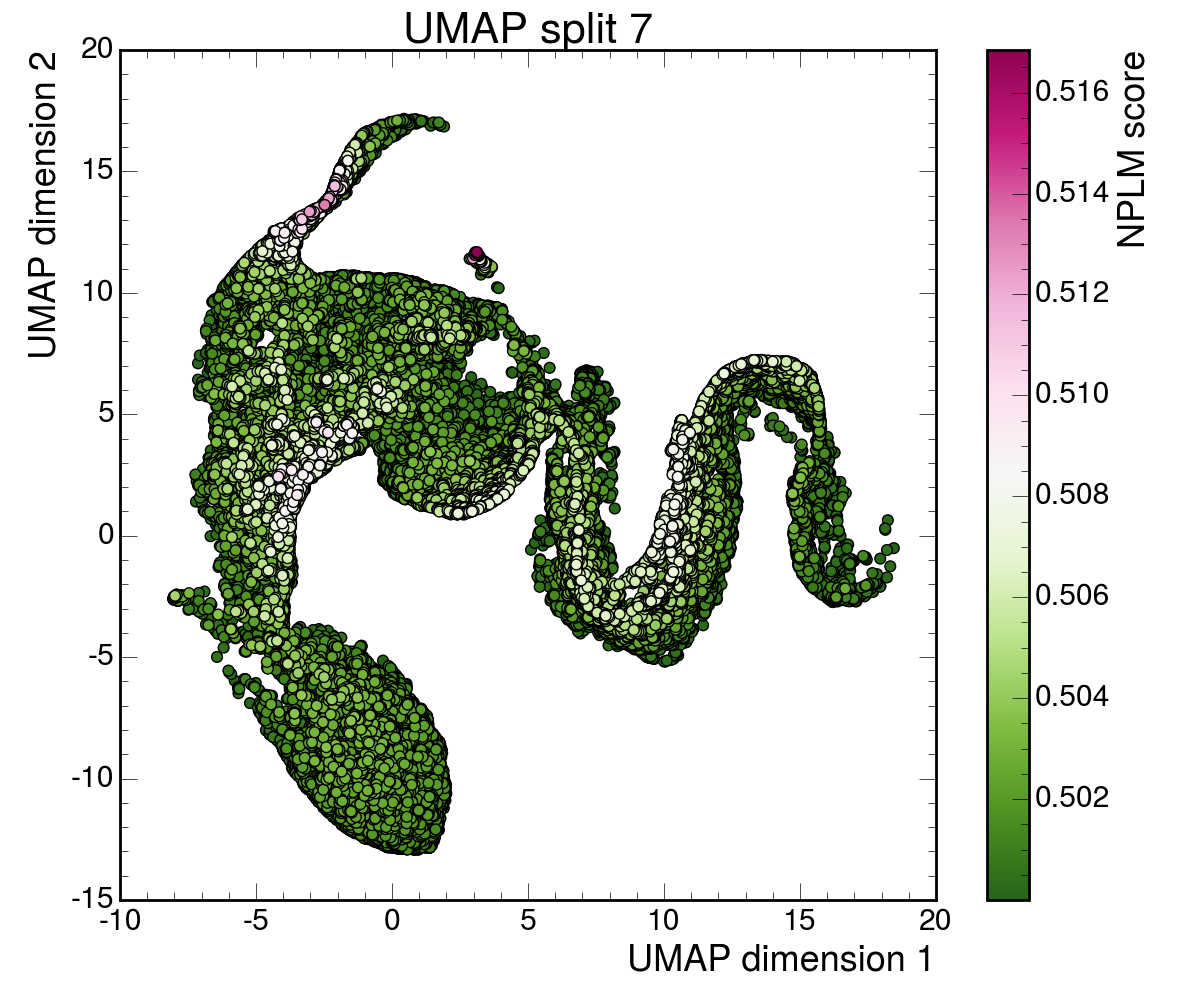}
\includegraphics[width=0.32\linewidth]{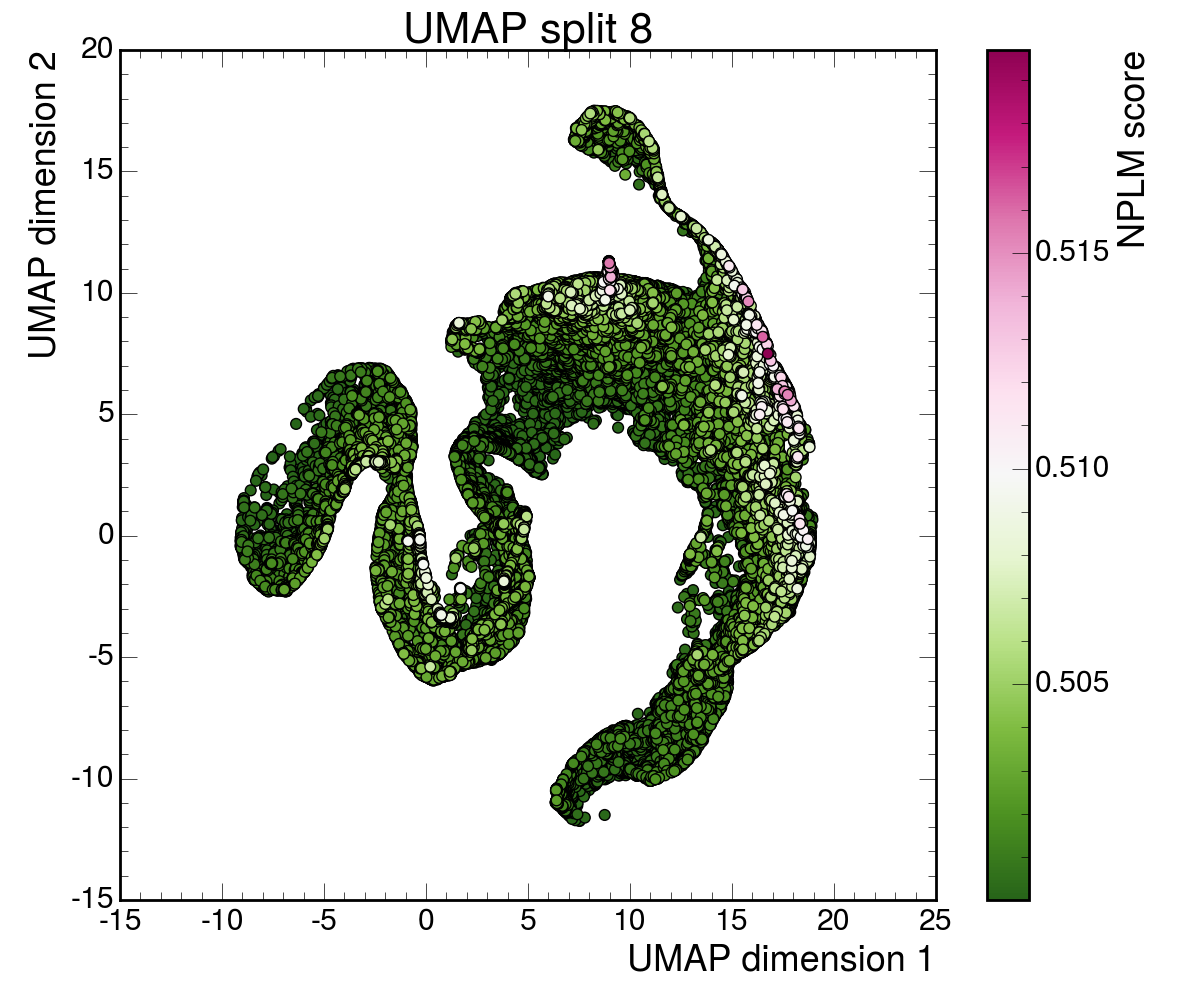}
\caption{\textbf{UMAP visualization of a ``black box" batch.} We inspect each split by means of a two-dimensional UMAP, where the color code represents the sigmoid activated score output by NPLM. The plots only show data with score grater than 0.5. Data points are ordered in score so that higher score data points are on top and always visible.}
\label{fig:umap1}
\end{figure*}

To complement Figure~\ref{fig:umap}, we show in Figure~\ref{fig:umap1} the two-dimensional UMAP of the various ``black box" data splits embedded in the 4D contrastive space used in the studies described in Section~\ref{sec:experiments}. The plots show data with NPLM score grater than 0.5. Data points are ordered by score so that higher score data points are on top and always well visible.
\end{document}